\documentclass[12pt]{article}

\usepackage{amssymb}
\usepackage{amsfonts}
\usepackage{amsmath}
\usepackage[nohead]{geometry}
\usepackage[singlespacing]{setspace}
\usepackage[bottom]{footmisc}
\usepackage{indentfirst}
\usepackage{endnotes}
\usepackage{graphicx}
\usepackage{rotating}
\usepackage{multirow}
\usepackage{epsfig,graphics}
\usepackage{xspace}
\usepackage[active]{srcltx}
\usepackage[retainorgcmds]{IEEEtrantools}
\usepackage{lscape,psfrag,color,dcolumn,tabularx}
\usepackage{booktabs} 
\usepackage{caption} 
\usepackage{bbm}
\usepackage{enumitem}
\usepackage{units}
\usepackage{subcaption}
\usepackage{bm}
\usepackage{accents}
\usepackage{yfonts}
\usepackage{epstopdf}

\newtheorem{theorem}{Theorem}

\newtheorem{definition}[theorem]{Definition}

\newtheorem{lemma}{Lemma}

\newtheorem{proposition}{Proposition}

\newtheorem{assumption}{Assumption}
\newcommand{\ud}{\, \mathrm{d}}

\newenvironment{proof}[1][Proof]{\noindent\textbf{#1.} }{\ \rule{0.5em}{0.5em}}

\makeatletter
\def\@biblabel#1{\hspace*{-\labelsep}}
\makeatother
\graphicspath{{images/}}
\usepackage{booktabs}
\usepackage{float}

\usepackage[round, sort, numbers, authoryear]{natbib}
\usepackage[colorlinks,pagebackref=true]{hyperref}
\definecolor{darkblue}{rgb}{0.0,0.0,0.3}
\hypersetup{citecolor = darkblue}
\hypersetup{urlcolor = darkblue}

\renewcommand*{\backref}[1]{}
\renewcommand*{\backrefalt}[4]{%
  \ifcase #1 %
  \or
    [#2]%
  \else
    [#2]%
  \fi%
}

\begin{document}
\title{Insider Trading with Penalties}
\author{Sylvain Carr\'e \footnote{Swiss Finance Institute and Ecole Polytechnique F\'ed\'erale de Lausanne, e-mail: sylvain.carre@epfl.ch }\quad \quad \quad \quad Pierre Collin-Dufresne  \footnote{Swiss Finance Institute and Ecole Polytechnique F\'ed\'erale de Lausanne, e-mail: pierre.collin-dufresne@epfl.ch} \quad \quad \quad \quad Franck Gabriel \footnote{ Department of Mathematics, Ecole Polytechnique F\'ed\'erale de Lausanne, e-mail: franck.gabriel@epfl.ch. }$\ \,$\footnote{Supported by the ERC CG CRITICAL, held by Pr. Hairer.} }

\maketitle

\begin{abstract}
We consider a one-period Kyle (1985) framework
where the insider can be subject to a penalty if she trades.
We establish existence and uniqueness of equilibrium for virtually any
penalty function when noise is uniform.
In equilibrium, the demand of the insider and the price functions are in general non-linear
and remain analytically tractable because the \emph{expected} price function is linear.

We use this result to investigate the trade off between price efficiency 
and ``fairness'': we consider a regulator that wants to
minimise post-trade standard deviation for a given level of uninformed
traders' losses. The minimisation is over the function space of penalties;
for each possible penalty, our existence and uniqueness theorem allows
to define unambiguously the post-trade standard deviation and the uninformed traders'
losses that prevail in equilibrium.

Optimal penalties are characterized in closed-form. They must increase
quickly with the magnitude of the insider's order for small orders and become
flat for large orders:
in cases where the fundamental realizes at very high or very low values,
the insider finds it optimal to trade despite the high penalty.
Although such trades --if they occur-- are costly for liquidity traders,
they signal extreme events and therefore incorporate
a lot of information into prices.

We generalize this result in two directions by imposing a budget constraint
on the regulator and considering the cases of either non-pecuniary or pecuniary penalties.
In the first case, we establish that optimal penalties are a subset of the previously optimal penalties:
the patterns of equilibrium trade volumes and prices is unchanged.
In the second case, we also fully characterize the constrained efficient points and penalties and show
that new patterns emerge in the demand schedules of the insider trader and the associated price functions.

\end{abstract}

\newpage
\section{Introduction}
\label{sec:introduction}

This paper derives and uses analytical results
about a one-period \cite{Kyle85} model with non-Gaussian noises and penalties
associated with insider trading. The natural benchmark for such a framework is a one-period Kyle model with non-Gaussian noise and without penalties.
Characterizing the benchmark equilibrium is useful in order to 
study the more general case with penalties.

\cite{RochetVila} have studied this problem under the assumption that the informed trader
is able to observe the noise trader's demand and can therefore condition its order to this demand.
In that context, they show existence and uniqueness of equilibrium regardless of
the distributional assumptions on the noises. Unfortunately, their approach does not
allow in general to construct explicitly the equilibrium. Furthermore, it does not seem
possible to replicate their result in the presence of penalties.\footnote{In \cite{RochetVila}, the proof of existence
and uniqueness relies on the fact that equilibrium price functions are \emph{optimal price functions}
in the sense that they minimise the expected insider gains' functional. This property holds true
because they can write the chain of equivalences $(X,P)$ equilibrium $\leftrightarrow$ $\mathbb{E}[X\lvert d]=0$
$\leftrightarrow$ $P$ optimal price function, where $d$ is the aggregate order. With a penalty $C$,
the central link breaks down, because then the first order condition of the insider's program 
combined with the price efficiency condition yields
$p'(d)\mathbb{E}[X\lvert d]=\mathbb{E}[C'(x)\lvert d]$.} But allowing for penalties on insider trading is crucial
for our purpose. Indeed, we want to understand how a regulator can trade off efficiently between
information incorporation and protection of liquidity traders, and how he can slide along
the efficient choices depending on the weights he attributes  to these two conflicting objectives.

\cite{BVH} 
study different models of market making with one or several strategic agents
without the assumption of normality and without hypothetizing that strategic agents
can observe the noise traders' demand. A one-period Kyle model
with one strategic trader and
non-Gaussian noise is a particular instance of their analysis. Their results imply
that in that case, when the distribution of
the noise equals in law a linear transformation of the fundamental,
a linear equilibrium exists. In this equilibrium, the demand of the IT has the same distribution
as the demand of the NT, what we call a \emph{mimicking} property.\footnote{We provide a discussion
in Appendix \ref{sec:discussionBVH}.}
\cite{BVH} only focus on linear equilibria. When adding an arbitrary penalty function to the model,
one can no longer expect to have linear equilibria, and there is \emph{a priori} no method to construct
equilibria explicitly in a systematic manner.

We use a setup
that remains tractable after the introduction of any penalty function --- at the cost
of a distributional assumption. We find that uniform noise has the property that
even though the IT demand function and the price function are non-linear, the expected price
function is linear, \emph{whatever the penalty} $C$. The equilibrium demand then simply appears
as the maximiser of a known objective. 
We obtain uniqueness
of the equilibrium among virtually all strategies for any penalty $C$ as a simple corollary of
our analysis.  By contrast, proving
a general uniqueness result for the one-period Kyle model with Gaussian noise \emph{without penalty}
was an extremely involved mathematical problem which was only addressed almost thirty
years after the \cite{Kyle85} seminal paper (\cite{KBL}, \cite{McLennan}).

\section{The Model}

\noindent As in the one-period version of \cite{Kyle85}, the model features
a risk-neutral insider trader (IT), noise traders (NT) and competitive market makers (MM).
Agents are trading an asset with fundamental value $v$. The IT perfectly
observes $v$ and places an order $X(v)$. NT have a stochastic demand $u$ independent
of $v$. MM observes the total demand $X(v)+u$ and executes orders at a price $P$ such that 
she breaks even on average. 

The first difference of our model with \cite{Kyle85} is that we consider 
uniform -- instead of Gaussian -- noises: 
\begin{eqnarray*}
u &\sim& U(-1,1),\\
v &\sim& U(-1,1),\\
u&\perp& v.
\end{eqnarray*}

The choice of $[-1,1]$ as the support is for clarity and without loss of generality;
one could equivalently assume $u\sim U(-a,a)$ and $v\sim U(b,c)$ with $a>0$ and $b<c$:
see Appendix \ref{sec:normalizationsupports}.

\

The second difference is that a regulator may decide to penalize
trades of size $x$ by a cost $C(x)$. We interpret $C$ as a product $C=\alpha \tilde{C}$:
$\alpha$ is the exogenous probability that the regulator starts
and successfully completes an investigation, while $\tilde{C}(x)$ is the cost imposed 
to the IT conditional on the investigation being successful and the order of the IT being $x$.
Success of the investigation means that 
the regulator correctly identifies the order of the informed trader $x$ and gathers sufficient
evidence to enforce payment of the corresponding fine. In other cases, the IT can not be
constrained to pay any fine. Under these assumptions, the regulator never makes
type 1 errors (never convicts a trader that didn't use insider information) but can make type 2
errors (not convicting a trader that did use insider information).

\subsection{The Insider Trader's Problem}
\label{sec:sec1}
\subsubsection{Benchmark Equilibrium without Penalties}
\noindent In the absence of penalties, the IT solves
\begin{equation}
\label{eq:maximisationITbenchmark}
\max_{x\in I} \, x\mathbb{E}_u[v-P(x+u)]
\end{equation}
taking the price function $P$ of the MM as given. The MM breaks even on average:
\begin{equation}
\label{eq:breakevenMMbenchmark}
P(d) = \mathbb{E}[v\lvert X(v)+u=d].
\end{equation}
An equilibrium is a pair $(X,P)$ that satisfies \eqref{eq:maximisationITbenchmark}
and \eqref{eq:breakevenMMbenchmark}.

\noindent From the discussion in section \ref{sec:introduction}, $(X,P)$ defined by
\begin{eqnarray}
X(v) &=& v\\
P(x+u)&=& \frac{x+u}{2}
\end{eqnarray}
is an equilibrium of the one-period Kyle model without penalty.
We refer to it as the \emph{(linear) mimicking equilibrium.} Indeed,
$X(v)$ and $u$ are equal in distribution. Notice that the image of $X$
is $[-1,1]$. We will prove later
that this equilibrium is unique among all equilibria featuring a non-decreasing
demand whose image lies in $[-1,1]$.

\

\noindent With penalties, the optimal demand is no longer mimicking
the random demand $u$. One intuitive interpretation is that while
mimicking $u$ allows the IT to best conceal herself from the market maker,
she can't hide from the regulator (in case investigation is open and succeeds).
This leads to a lower demand than in the case without penalties. We now
define formally the equilibrium with penalties.

\subsubsection{One-Period Kyle Model with Penalties}
\label{sec:sec1subsec1}
\noindent The IT solves
\begin{equation}
\label{eq:maximisationIT}
\max_{x\in I} \, x\mathbb{E}_u[v-P(x+u)]-C(x),
\end{equation}
taking the price function $P$ of the MM as given. The MM breaks even on average:
\begin{equation}
\label{eq:breakevenMM}
P(d) = \mathbb{E}[v\lvert X(v)+u=d].
\end{equation}
This game involving the IT and the MM is denoted $\mathcal{K}(C)$.
An equilibrium of $\mathcal{K}(C)$  is a pair $(X,P)$ such that $X$ solves \eqref{eq:maximisationIT} and
$P$ satisfies \eqref{eq:breakevenMM}.

\

\noindent The interval $I\subset\mathbb{R}$ in the maximisation program \eqref{eq:maximisationIT}
is the set of admissible insider's demand. 
In order to be able to prove uniqueness, we make the following assumption:
\begin{assumption}
\label{assumption:i}
$I = [-1,1].$
\end{assumption}

\noindent The bounds of $I$ are
those that obtain in the linear mimicking equilibrium when there is no penalty function.
They are therefore natural: a demand function $X$ whose image is not contained in
$[-1,1]$ would imply that for some values of the fundamental $v$, the magnitude of the IT order
is \emph{higher} when there is a penalty, compared to the linear equilibrium without penalty.\footnote{At
present, we do not know whether an equilibrium featuring such a demand function can exist.}

\

\noindent To conclude this section, we state two remarks and introduce some notation.

(i) The data of a strategy $X$
implies a pricing function $P$ via equation \eqref{eq:breakevenMM}. 
That is, if $X$ is part of an equilibrium, then the pricing function must
be given by $P$. We denote the pricing function associated with a demand schedule $X$
by $P(X)$.

(ii) In the IT's maximisation program \eqref{eq:maximisationIT}, the pricing function $P$
only intervenes through the \emph{expected price function}, denoted $\hat{P}$ and defined
by
\begin{equation}
\label{eq:expectedpricefunction}
\hat{P}(x) = \mathbb{E}_u[P(x+u)].
\end{equation}
\noindent $\hat{P}$ represents the price that the risk-neutral IT will
face on average if she places an order $x$. The program \eqref{eq:maximisationIT}
can be rewritten in terms of the expected price function only:
\begin{equation}
\label{eq:maximisationIThat}
\max_{x\in I} \, x(v-\hat{P}(x))-C(x).
\end{equation}

\subsubsection{Out-of-Equilibrium Pricing}
\label{sec:sec1subsec2}
\noindent The noise $u$ we consider has bounded support. Moreover, 
the discussion above indicates that the equilibrium demand functions $X$
we will consider satisfy $\lvert X\lvert\leq 1$. This means that the aggregate order,
$d=X(v)+u$ belongs to a bounded set $D$.
The conditional expectation in \eqref{eq:breakevenMM}
is not defined for values of $d\notin D$, meaning that we must make
an assumption on the out-of-equilibrium pricing of the MM:

\begin{assumption}
\label{assumption:outofequilibriumpricing}
For any equilibrium $(X,P)$ of $\mathcal{K}(C)$ we consider, with $X$ non-decreasing and
$X([-1,1])\subset [-1,1]$, we always impose the following out-of-equilibrium pricing (letting $x_M=X(1)$):
\begin{eqnarray*}
P(d)&=&1 \quad \, \, \, \, \text{for} \quad d>1+x_M\\
P(d)&=&-1 \quad \text{for}\quad d<-1-x_M.
\end{eqnarray*}
\end{assumption}

This assumption states that when the MM observes
a positive aggregate order larger than its maximal possible equilibrium size,
she prices the asset as if it had realized at its maximal value, $v=1$. 
Similarly, when the aggregate order is negative with a magnitude larger than
the maximal possible equilibrium size, the MM prices as if $v=-1$.
When constructing equilibria, we do not always recall that Assumption \ref{assumption:outofequilibriumpricing}
is used to define the out-of-equilibrium pricing. When verifying that $(X,P)$ is an equilibrium,
one must not only check that $X(v)$ maximises the IT's program \eqref{eq:maximisationIT}
among all $x$ in the candidate support $[-x_M,x_M]$, but also among values of $x$ in $I\setminus [-x_M;x_M]$.
For these values of $x$, the aggregate order $d=x+u$ realizes in the out-of-equiibrium
region with positive probability, in which case Assumption \ref{assumption:outofequilibriumpricing}
defines the price $P(d)$.

Finally, notice that Assumption \ref{assumption:outofequilibriumpricing} fully characterizes out-of-equilibrium pricing:
indeed, any $d\in [-1-x_M,1+x_M]$ belongs to the support of $u+X(v)$, since $u$ is $U(-1,1)$ and $x_M\leq 1$.

\subsubsection{A first example}
\label{sec:sec1subsec3}
\noindent We now present an example of an equilibrium of $\mathcal{K}(C)$.
More illustrations can be found in section \ref{sec:examplesequilibria}, where
we discuss the intuitions behind some typical behaviours of the equilibrium demand and price 
functions in the presence of penalties.

\

\noindent Let $K\in \left(0,\frac{1}{2}\right)$ and 
\begin{equation*}
C(x) = K\mathbb{I}_{x\neq 0}.
\end{equation*}
Under this penalty function, the insider trader undergoes an expected sanction
of $K$ if she trades. This example is particularly important because we will see
that such penalty functions are among the optimal regulations.

\

\noindent We will show that $(X,P(X))$ is an equilibrium, where 
\begin{equation}
X(v)=v\mathbb{I}_{\lvert v\lvert >\sqrt{2K}}.
\end{equation}
As we will see, the price function $P(X)$ is  non-linear but the expected price function $\hat{P}$ satisfies $\hat{P}(x)=\frac{x}{2}$. Hence, the IT maximises under the same expected price function as in the linear mimicking equilibrium.
Facing an expected price identical to the one without penalties, the
IT only trades when its previously optimal strategy --- the linear mimicking demand --- allows
her to recoup the penalty $K$ on average. Without penalties, the profit of the IT when she
observes a fundamental $v$ is $\frac{v^2}{2}$. With a constant penalty upon trading equal to $K$, the
IT does not trade as long as $\frac{v^2}{2}<K$. When $\frac{v^2}{2}>K$, the IT considers $K$ as a sunk cost
and optimizes as if there was no penalty, thus selecting $X(v)=v$. Notice that the demand function
is non-linear and exhibits a jump at $\pm\sqrt{2K}$.

\subsubsection{Indistinguishable Equilibria}
\label{sec:sec1subsec4}
\noindent In the equilibrium of the example above, the IT would earn the same profit upon observation
of $v=\pm\sqrt{2K}$ by selecting $X(v)=0$ or $X(v)=v$: zero in both cases. In general, when the
penalty function exhibits jumps, we should expect the existence of such indifference points. At these points,
the IT can achieve a given profit by placing a small order and undergoing a small expected sanction
or by placing a larger order, associated with a larger expected penalty. However, as long as
the set of $v$ such that the maximisation program of the IT \eqref{eq:maximisationIT} admits several solutions
has measure zero, these indifference points will almost surely not be reached. The equilibrium will
therefore be independent of the choice of the maximiser $X(v)$, in the sense that any \emph{ex post} model observable
is almost surely the same --- \emph{e.g.} demand of the IT $X(v)$, observed price $P(d)$ --- and any
\emph{ex ante} model quantity --- such as the IT expected profit or the expected penalty collected from the IT ---
is the same. In that case, we wish to consider that any choice of maximiser $X$ induces the \emph{same}
equilibrium. We formalize this by introducing an equivalence relation between equilibria that
we call indistinguishability.

\

\noindent Assume that
$X$ and $X'$ are two solutions of the IT's maximisation program \eqref{eq:maximisationIT}
and agree outside of a countable set.
In that case,
$P(X)=P(X')$. This means that if $(X,P)$ is an equilibrium
then so is $(X',P)$.
This leads us to the following definition.

\begin{definition}
\label{def:equivalenceclass}
Let $(X,P)$ and $(X',P')$ be two equilibria of $\mathcal{K}(C)$.
We say that $(X,P)$ and $(X',P')$ are {\em indistinguishable} 
if $X$ and $X'$ agree outside of a countable set. 
Indistinguishability defines an equivalence relation over the set of equilibria
of $\mathcal{K}(C)$. 
\end{definition}

\noindent From now on, we identify an equilibrium
of $\mathcal{K}(C)$ to its equivalence class. Definition \ref{def:equivalenceclass}
is useful because we will see that maximisers of \eqref{eq:maximisationIT}
\emph{have to} agree outside of a countable set, and so the equilibria they induce
will belong to the same equivalence class.

\subsection{The regulator's problem}
\label{sec:theregulatorsproblem}
\noindent In our model, the regulator is concerned about two quantities:\footnote{
In section \ref{sec:pecuniary}, the regulator additionally needs to take care of
the expected fine she collects for budget reasons.}
(i) the post-trade standard deviation of the fundamental, $\sigma(v\lvert d)$
and (ii) the P\&\,L of the uninformed traders:
\begin{equation}
\label{eq:definitionlossNT}
g(u,v)=u(v-P(X(v)+u)).
\end{equation}

Quantity (i) matters because one would like to have informative prices:
when (i) is small, the residual uncertainty about $v$ is also small.
Quantity (ii) captures the willingness of the regulator to have liquid markets.
In a liquid market, agents who have to trade for non-fundamental reasons
do not experience high losses. This corresponds to a situation
where $g$ is not too negative. The core issue is that
improving upon criterion (i) generally causes
criterion (ii) to worsen.

\noindent Let
\begin{equation}
\label{eq:var}
S=\mathbb{E}[\sigma(v\lvert d)]
\end{equation}
be the expectation of the post-trade standard deviation of $v$ and
\begin{equation}
\label{eq:expectedlossesNT}
G=\mathbb{E}[g(u,v)]
\end{equation}
denote the expected P\&\,L of the NT. 

\

\noindent The objective of the regulator can now be stated as the characterization
of the \emph{efficient frontier}, with the following definition:

\begin{definition}
\label{def:dominatedpoints}
(i) A point $(G,S)$ is implementable if it is the outcome of an equilibrium of $\mathcal{K}(C)$
for some admissible penalty $C$.

\

\noindent (ii) An implementable point $(G,S)$ is dominated 
by $(G',S')$ if $(G',S')$ is implementable and $G'\geq G$, $S'\leq S$
with at least one strict inequality.

\

\noindent (iii) The set of implementable non-dominated points is called the efficient frontier.

\

\noindent In section \ref{sec:pecuniary}, we will need the following refinement of (ii):

\

\noindent (ii') An implementable point $(G,S)$ belonging to some subset of the plane $H$ is dominated in $H$
by $(G',S')$ if $(G',S')$ is implementable, $G'\geq G$, $S'\leq S$
with at least one strict inequality and $(G',S')\in H$.
\end{definition}

\noindent Points outside the efficient frontier are irrelevant from the regulator's perspective,
as she can improve upon one of his objectives without harming the other one. By contrast,
any point belonging to the efficient frontier could be picked by a regulator for a suitable
weighting\footnote{Not necessarily linear.} of the objectives. Our goal is to characterize the efficient frontier
and the penalties that implement it.

\subsection{Admissible penalty functions}

\noindent We do not impose any restriction on the penalty function,
except that it only depends 
in a non-decreasing manner on the magnitude
of the order of the insider trader, and that
there is no sanction when she does not trade.

\begin{definition}
$C:[-1;1]\rightarrow \mathbb{R}_+$ is a penalty function if it is symmetric and non-decreasing, left-continuous over $[0;1]$
and satisfies $C(0)=0$. 
The set of penalty functions is denoted $\mathcal{C}$.
\end{definition}

\noindent The class $\mathcal{C}$ is very general and defined
by economically relevant requirements.
In particular, it would be unnatural and perhaps politically hard to implement to impose
a higher sanction on a smaller trade. The left-continuity assumption
simply makes sure that the supremum of the possible profits is attainable.

\section{Existence and uniqueness of equilibrium for $\mathcal{K}(C)$}
\label{sec:exuniq}

\noindent In this section, we set out to prove the following Theorem:
\begin{theorem}
\label{theorem:existenceuniqueness}
For any $C\in\mathcal{C}$, the Kyle game $\mathcal{K}(C)$ with penalty function $C$
admits a unique equilibrium $(X(C),P(C))$.\footnote{Recall that equilibria
are identified with their equivalence class, see Definition \ref{def:equivalenceclass}
.}
In general, $X$ and $P$ are non-linear.
\end{theorem}

\noindent One consequence of this result is that for each $C\in\mathcal{C}$, the regulator's
quantities of interest are defined unambiguously as the outcomes of the unique equilibrium
in $\mathcal{K}(C)$. In particular, the efficient frontier is defined unambiguously.
\subsection{ Analysis of the  expected price function }

\subsubsection{Under uniform noises, the {\em expected} price function is linear regardless of the IT demand}

\noindent Lemma \ref{lemma:cs} contains the key observation at the root of our analysis. Recall that for any odd non-decreasing function $X : [-1,1] \to [-1,1]$, we denote by $P(X)$ the pricing function associated with $X$ (equation \eqref{eq:breakevenMM}) and given $P=P(X)$, $\hat{P}$ is the expected price function (equation \eqref{eq:expectedpricefunction}): $\hat{P}(x)$ is the price that the IT will face on average if she places an order $x$. 

\begin{lemma}
\label{lemma:cs}
Let  $X : [-1,1] \to [-1,1]$  be an odd non-decreasing function, and $x_M = X(1)$.  The expected price function  $\hat{P}$ is linear on $[-x_M,x_M]$: 
\begin{align*}
\hat{P}(x) = \frac{x}{2}.
\end{align*} 
\end{lemma}

Lemma \ref{lemma:cs} is crucial because it makes the surprising
statement that the expected price function that must
prevail in equilibrium is $\hat{P}(x)=x/2$ without requiring any knowledge:
neither the form of $C$ nor guesses about $X$ or $P$ are needed.

In turn, this implies that the equilibrium demand of the IT, $X(v)$,
must be a maximiser of 
\begin{equation}
\label{eq:profitf}
\psi_C(.,v): x \mapsto x\left(v-\frac{x}{2}\right)-C(x),
\end{equation}
as the IT maximises its expected profit knowing that $\hat{P}(x)=x/2$.
This demand $X$ induces some price function $P$. Remark that,
again by Lemma \ref{lemma:cs}, the expected price indeed satisfies
$\hat{P}(x)=x/2$. This indicates that $(X,P(X))$ is an equilibrium of
$\mathcal{K}(C).$ 

While this discussion provides the intuition on how we construct
the equilibrium of $\mathcal{K}(C)$ for an arbitrary $C$, several
technical issues must be addressed in order to make the argument formal.
One must check that any selection of maximiser is non-decreasing 
and take care of the out-of-equilibrium pricing: notice in particular
that Lemma \ref{lemma:cs} only characterizes $\hat{P}$ over
$[-x_M,x_M]$, while we need to compute the IT's expected profit
for all admissible demands $x$. Additional results are also required
to establish uniqueness of he equilibrium of $\mathcal{K}(C)$.
The main step in that direction is to show that $\psi_C(.,v)$ admits a 
unique maximiser except for a countable number of values of $v$
(section \ref{sec:unicitymaximiser}).

We now provide the proof of this lemma. Section \ref{sec:intuititionlemmacs} clarifies 
the main intuitions. 

\

\begin{proof}[Proof of Lemma \ref{lemma:cs}]
We use the notation $p(.)$ for a density and $p(.\lvert .)$ for a conditional density.
Write
\begin{eqnarray*}
p(v\lvert d)&\propto& p(d\lvert v)p(v)\\
&\propto& \mathbb{I}_{X(v)\in [d-1;d+1]}\mathbb{I}_{v\in [-1;1]}.
\end{eqnarray*}
That is, for $-1-x_M\leq d\leq 1+x_M$, $v\lvert d$ is uniform over
\begin{eqnarray}
\label{eq:vsachantd}
\{v\in [-1;1] \lvert X(v) \in [d-1;d+1]\}&=&\{v\in [-1;1] \lvert X(v) \in [d-1;d+1]\cap[-x_M;x_M]\}\nonumber\\
&=& [(X^{-1}_{\ell}\left((d-1)\vee(-x_M)\right);X^{-1}_r\left((d+1)\wedge x_M\right))]\nonumber\\
&& \,
\end{eqnarray}
where
\begin{eqnarray*}
X^{-1}_{\ell}(x)&=&\inf\{v\lvert X(v)\geq x\}\\
X^{-1}_r(x)&=&\sup\{v\lvert X(v)\leq x\}.
\end{eqnarray*}
$X^{-1}_{\ell}$ and $X^{-1}_r$ only disagree when there is $v$ such that $X(v)=x$ and $X$ is locally constant at $v$, i.e. they agree outside of a countable set. Then, letting $P=P(X)$, 
\begin{equation*}
P(d)=\frac{1}{2}\left(X^{-1}_{\ell}\left((d-1)\vee(-x_M)\right)+X^{-1}_r\left((d+1)\wedge x_M\right)\right).
\end{equation*}
Now since
\begin{equation*}
\hat{P}(x)=\frac{1}{2}\int_{x-1}^{x+1}P(z) \ud z,
\end{equation*}
by differentiation it is enough to show that $P(x+1)-P(x-1)=1$ a.e.. Using the expression of $P$ found above, we obtain that for $-x_M\leq x\leq x_M$:
\begin{eqnarray*}
2(P(x+1)-P(x-1))&=&X^{-1}_{\ell}\left(x\vee(-x_M)\right)+X^{-1}_r\left((x+2)\wedge x_M\right)\\
&-&X^{-1}_{\ell}\left((x-2)\vee(-x_M)\right)-X^{-1}_r\left(x\wedge x_M\right)\\
&=&X^{-1}_r(x_M)-X^{-1}_{\ell}(-x_M)\\
&=&2
\end{eqnarray*}
a.e.. This is because $X^{-1}_{\ell}=X^{-1}_r$ a.e., $X^{-1}_r(x_M)=1$, and $X^{-1}_{\ell}(-x_M)=-1$. 
\end{proof}

\

\noindent Having identified $\hat{P}$, we know that
the insider trader's problem is to maximise $\psi_C(.,v)$
as defined in \eqref{eq:profitf}. Because we will use this function
throughout the paper, we repeat its definition here:

\begin{definition}
The insider's expected profit (under the correct expected price function $\hat{P}(x)=x/2$)
for a demand $x$ when the fundamental value is $v$ is
\begin{equation}
\label{eq:profitpsi}
\psi_C(x,v)=x\left(v-\frac{x}{2}\right)-C(x).
\end{equation}
\end{definition}
Notice that $\psi_C$ is an ``expected'' profit because we interpret
$C$ as an average cost ---an investigation may not be started or not succeed---
while $\hat{P}(x)$ is an expected price because the realization of $u$ is random
and the realized price is $P(x+u)$.

\subsubsection{Intuition}
\label{sec:intuititionlemmacs}
\noindent In order to isolate the intuition
behind Lemma \ref{lemma:cs}, let us consider the case
where $X$ is continuous and strictly increasing.

Assume that the market maker observes 
an aggregate order $d>0$. Since the demand of
the noise traders $u$ takes values
in $[-1,1]$, the possible demands of the IT 
$X(v)$ consistent with the observation of $d$
are exactly the admissible demands such that $d-1\leq X(v) \leq d+1$.
Because admissible demands satisfy $X(v)\leq 1$ and $d+1>1$,
the information obtained by the market maker when she observes $d$
is that $X(v)\geq d-1$. Thus, she knows that $v\geq X^{-1}(d-1)$.  
Intuitively, the fact that the aggregate order is positive rules out
extreme negative values of $v$ and the MM deduces a lower
bound on $v$, $X^{-1}(d-1)$.

Moreover, due to the uniform noise assumption,
all values of $v$ above this lower bound are equally likely. Therefore, the 
price $P(d)$ is given by the midpoint of the interval $\left[X^{-1}(d-1),1\right]$. 

In a similar manner, when $d<0$, the price $P(d)$ is given by the midpoint of the interval $\left[-1,X^{-1}(d+1)\right]$. 

\begin{figure}[tbhp]
\begin{center}
\includegraphics[width=300pt]{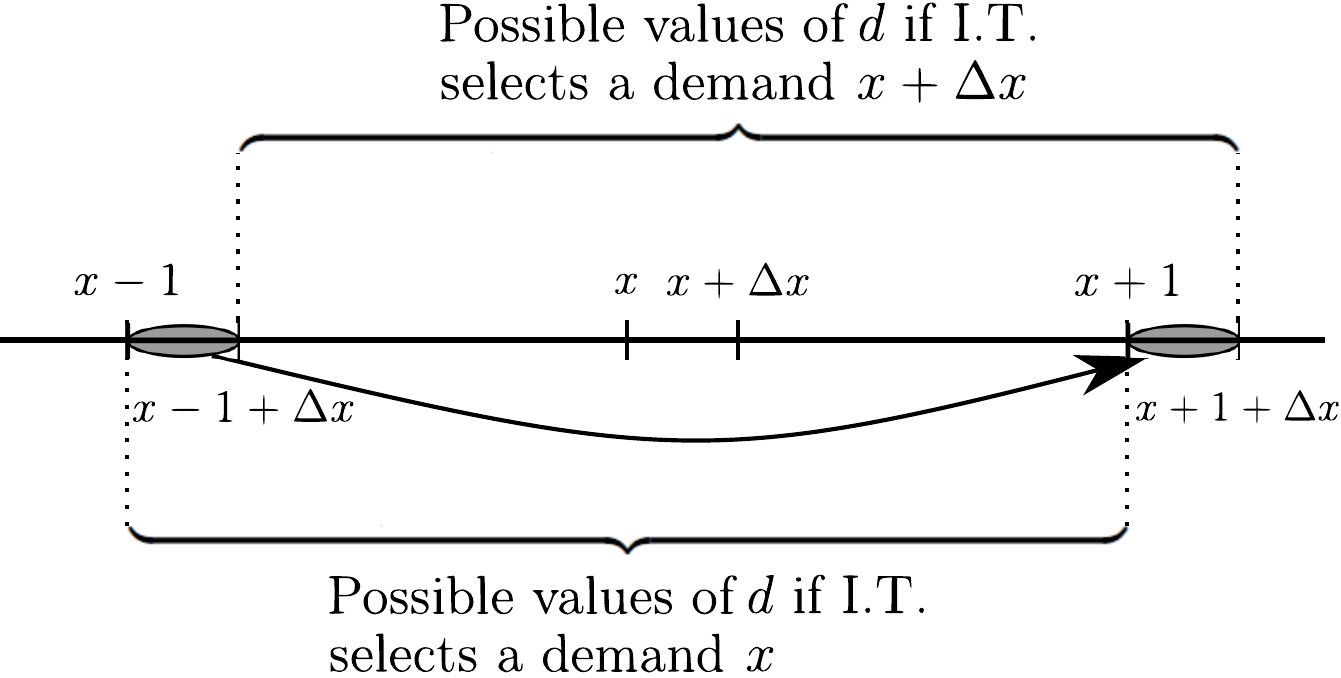}
\end{center}
\caption{Marginal expected price impact of an increase in $x$}
\label{fig:marginalpriceimpact}
\end{figure}

Now, assume that the IT wants to place an order $x$.
The IT is only concerned by the expected price impact,
$\hat{P}(x)$, which is a uniform average of the $P(d)$
over $d\in [x-1,x+1]$, the set of possible aggregate demands
given an IT demand $x$. If, instead, the IT decides to place an order
$x+\Delta x$, the set of possible aggregate demands $d$ is
$d\in [x-1+\Delta x,x+1+\Delta x]$:
see Figure \ref{fig:marginalpriceimpact}.

\

Thus, the only contribution to the marginal increase in expected price
$\hat{P}(x+\Delta x)-\hat{P}(x)$ is due to the fact that the weight
that was attributed to the interval $[x-1,x-1+\Delta x]$ is now
attributed to the interval $[x+1,x+1+\Delta x]$. Crucially,
this weight is the same due to the uniform noise assumption.
Considering a vanishing $\Delta x$, one concludes that the
marginal impact of increasing demand on expected price is
proportional to $P(x+1)-P(x-1)$.

\begin{figure}[tbhp]
\begin{center}
\includegraphics[width=450pt]{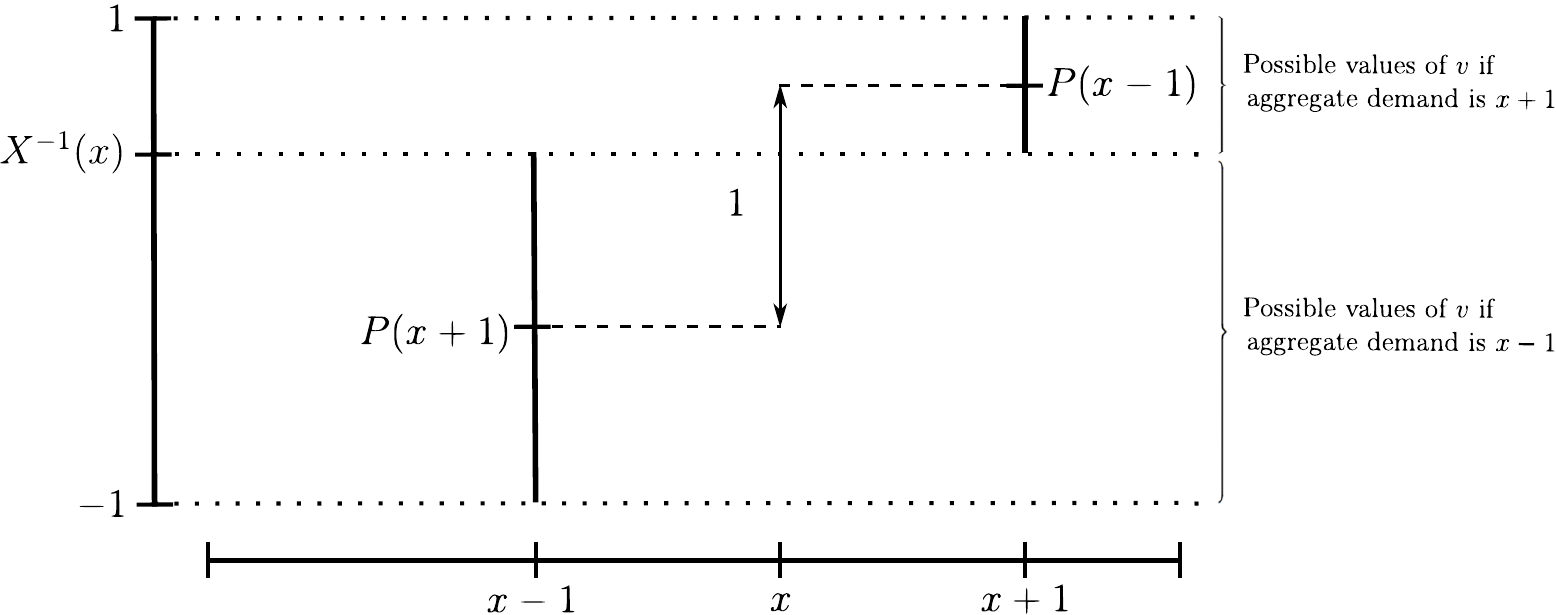}
\end{center}
\caption{The marginal expected price impact is constant}
\label{fig:marginalpriceimpact2}
\end{figure}

We have seen above that $P(x+1)$ is the midpoint
of $\left[X^{-1}((x+1)-1),1\right]=\left[X^{-1}(x),1\right]$, and that $P(x-1)$ is the midpoint
of $\left[-1,X^{-1}((x-1)+1)\right]=\left[-1,X^{-1}(x)\right]$. Therefore, the marginal impact on the expected price is proportional to the distance between these two midpoints: 
\begin{align*}
\frac{d}{dx} \hat{P}(x) \propto P(x+1)-P(x-1) = \frac{1+X^{-1}(x)}{2} - \frac{X^{-1}(x) - 1}{2} = 1.
\end{align*}
Figure \ref{fig:marginalpriceimpact2} provides an illustration of this result. This shows that the expected price function is linear. Notice that the arguments above rely heavily on the uniform noise assumption: with other noises,
one cannot expect in general to have a linear expected price function.

\

\subsection{Candidate optimal demands are unique up to changes on a countable set}
\label{sec:unicitymaximiser}
\noindent In this section, we set out to
obtain an unambiguous definition of the strategy $X$ that
will be our maximiser.

\begin{definition}
Let $V$, $I$ be two intervals of $\mathbb{R}$. A correspondence
$\mathcal{X}:V\rightarrow \mathcal{P}(I)\setminus \emptyset$ is {\em non-decreasing} if 
for any $v_1<v_2$ in $V$, $\sup \mathcal{X}(v_1)\leq\inf \mathcal{X}(v_2)$.
\end{definition}
\noindent Notice that if $\mathcal{X}$ is a one-to-one mapping,
then we recover the usual notion of a non-decreasing function.

\begin{lemma}
\label{lemma:correspondnondecrease}
Let $\mathcal{X}:V\rightarrow \mathcal{P}(I)\setminus \emptyset$ be a non-decreasing correspondence. Then for all $v$ in $V$ except on a countable set, $\mathcal{X}(v)$ is a singleton.
\end{lemma}

\begin{proof}
The argument is the same as for the proof that a non-decreasing
function has at most a countable number of discontinuities.
\end{proof}

\

\noindent For a given penalty $C\in\mathcal{C}$, let
$\mathcal{X}_C$ be the correspondence mapping $v\in [-1;1]$
to the set of maximisers of the insider trader's profit function
when she observes a realization $v$ of the fundamental:

\begin{equation*}
\mathcal{X}_C(v) = \underset{x}{\arg \max} \, \, \psi_C(x,v).
\end{equation*}

\noindent Recall that $\psi_C$ is defined in \eqref{eq:profitpsi}.

\begin{lemma}
\label{lemma:xincreasing}
For any $v \in [-1,1]$, $\mathcal{X}_C(v) \neq \emptyset$, and $\mathcal{X}_C$ is a non-decreasing correspondence.
\end{lemma}

\begin{proof}
First, let us show that $\mathcal{X}_C(v)$ is never empty.
Let $v \in [-1,1]$, the function $\psi_C(.,v)$
has a finite upper bound  as $C\geq 0$. Let $M=\sup_x \psi_C(x,v)<\infty$ and $(x_n)$
such that $\psi_C(x_n,v)\rightarrow M$.  There is an extraction of $(x_n)$,
still denoted $(x_n)$, such that $x_n$ converges to $x$ and either (i) $(x_n)$ is increasing or (ii)
$(x_n)$ is decreasing. By symmetry, we can assume without loss
of generality that $x> 0$ or $x=0$ and the case $(ii)$ holds.
Let us first consider case (i). Since $C$ is
left-continuous and $x\mapsto x\left(v-\frac{x}{2}\right)$ is continuous,
$\psi_C(x_n,v)$ converges to $\psi_C(x,v)$: therefore $\psi_C(x,v)=M$ and $x\in 
\mathcal{X}_C(v)$. Let us now consider case (ii). Since $C$ is non decreasing, it has a right limit at $x$ denoted by $C(x^+)$ which is greater than $C(x)$. Taking the limit in the definition of $\psi_C(x_n,v)$, the value of $\psi_C(x_n,v)$ converges to $x\left(v-\frac{x}{2}\right)-C(x^+) \leq x\left(v-\frac{x}{2}\right)-C(x)$. Using the fact that $\psi_C(x_n,v)$ converges to $M$, we conclude that $C(x^+) = C(x)$ and $\psi_C(x,v) = M$.

Now, let us show that $\mathcal{X}_C$ is a non-decreasing correspondence.
Let $v_1<v_2$ in $[-1;1]$ and $x_1^*\in \mathcal{X}_C(v_1)$ and $x_2^* \in \mathcal{X}_{C}(v_2)$. For any $x \in [-1,1]$: 
\begin{align*}
\psi_C(x,v_2) = \psi_C(x,v_1) + (v_2-v_1)x.
\end{align*}
Using the fact that $x_1^* \in \mathcal{X}_C(v_1)$ and $v_1<v_2$, for any $x < x_1^*$, 
\begin{align*}
\psi_C(x,v_2) < \psi_C(x_1^*,v_1)+(v_2-v_1) x_1^* =  \psi_C(x_1^*,v_2).
\end{align*}
By definition, $\psi_C(x_2^*,v_2) \geq \psi_C(x_1^*,v_2)$, thus $x_2^* \geq x_1^*$. Since this inequality holds for any $x_1^*\in \mathcal{X}_C(v_1)$ and $x_2^* \in \mathcal{X}_{C}(v_2)$, we get that $\sup \mathcal{X}_C(v_1)\leq\inf \mathcal{X}_C(v_2)$: the correspondence $\mathcal{X}_{C}$ is non-decreasing. 
\end{proof}

\

\noindent The combination of Lemmas \ref{lemma:correspondnondecrease} and 
\ref{lemma:xincreasing} ensures that the maximiser of the IT's expected profit
is unique except for a countable number of values of $v$: 

\begin{lemma}
\label{lemma:maximiserunique}
There exists a non-decreasing function $X_{C}$ such that for all $v \in [-1,1]$ except on a countable set, 
\begin{align*}
\mathcal{X}_C(v) = \left\{ X_{C}(v)\right\}. 
\end{align*}
All such $X_C$ agree outside of a countable set. 
\end{lemma}

As we identify equilibria 
in a same equivalence class, as introduced in Definition \ref{def:equivalenceclass},
we do not need to specify which particular $X_C$ we consider:
we can unambiguously talk about ``a maximiser'' of the expected profit.
We are now ready
to derive the main result of this section.

\subsection{Existence and uniqueness of the equilibrium of $\mathcal{K}(C)$}
\noindent We recast the statement of Theorem \ref{theorem:existenceuniqueness}
by indicating what the equilibrium optimal demand is:

\

\noindent\textit{Let
$C\in\mathcal{C}$ and $X_C(v)$ be a maximiser of $x\mapsto x\left(v-\frac{x}{2}\right)-C(x)$.
Then $(X_C,P(X_C))$ is an equilibrium of $\mathcal{K}(C)$. This is the unique equilibrium among the pairs $(X,P)$ such that $X:[-1,1]\rightarrow[-1,1]$ is non-decreasing.}

\

\begin{proof}[Proof of Theorem \ref{theorem:existenceuniqueness}] From Lemma \ref{lemma:cs}, $\hat{P}_C(x)=\frac{x}{2}$ for $-x_M\leq x\leq x_M$.
Since $X_C(v)$ is a maximiser of $x\left(v-\frac{x}{2}\right)-C(x)$, $x=X(v)$ is an optimal
response to the expected price function $\hat{P}$ among all $x\in [-x_M,x_M]$. 
To confirm that $(X_C,P(X_C))$ is an equilibrium,
we need to check what happens if the IT makes a choice outside of the candidate support
$[-x_M,x_M]$, knowing that the out-of-equilibrium pricing is defined by Assumption \ref{assumption:outofequilibriumpricing}. Consider for instance the case $x\in (x_M,1]$,
as the case $x\in [-1,-x_M]$ is identical by symmetry.
Then
\begin{eqnarray}
\label{eq:ptildeoutofeq}
\hat{P}_C(x)&=&\frac{1}{2}\int_{x-1}^{x+1} P_C(z) \ud z\nonumber\\
&=&\frac{1}{2}(x-x_M)+\frac{1}{2}\int_{x_M-1}^{x_M+1} P_C(z) \ud z-\frac{1}{2}\int_{x_M-1}^{x-1} P_C(z) \ud z\nonumber\\
&=&\frac{1}{2}(x-x_M)+\hat{P}_C(x_M)-\frac{1}{2}\int_{x_M-1}^{x-1} P_C(z) \ud z\nonumber\\
&=&\frac{1}{2}(x-x_M)+\frac{x_M}{2}-\frac{1}{2}\int_{x_M-1}^{x-1} P_C(z) \ud z\nonumber\\
&=&\frac{x}{2}.
\end{eqnarray}
This is because when $z\in [x_M-1,x-1]$, $z-1<x-2\leq -1\leq -x_M$
and $z+1\geq x_M$ so from \eqref{eq:vsachantd}, 
$v\lvert z$ is uniform over $[-1,1]$ and $P_C(z)=0$.

As $X(v)$ maximises $x\mapsto x\left(v-\frac{x}{2}\right)-C(x)$,
and $\hat{P}_C(x)= \frac{x}{2}$ for $x\in (x_M,1]$, $X(v)$ maximises
$x\mapsto x\left(v-\hat{P}_C(x)\right)-C(x)$ over $[-1,1]$: $(X_C,P_C)$ is an equilibrium.

We now prove uniqueness. Let
$X':[-1,1]\rightarrow [-x_M',x_M']$ be a non-decreasing strategy of the IT. By Lemma \ref{lemma:cs},
the expected price $\hat{P}'$
associated with $X'$ is $\frac{x}{2}$ for $x\in [-x_M',x_M']$. But the computation
of $\hat{P}'$ outside of $[-x_M',x_M]$ is the same as the computation of $\hat{P}_C$ in 
\eqref{eq:ptildeoutofeq}. Hence, for all $x\in [-1,1], \hat{P}'(x)=\frac{x}{2}$.
So, if $(X',P(X'))$ is an equilibrium of $\mathcal{K}(C)$ such that $X'$ is non-decreasing, $X_C$ and $X'$ 
maximise the same objective $\psi_C$ over $[-1;1]$. Since the maximisers agree outside of a countable set,
so do $X_C$ and $X'$. In turn, we have $P(X')=P_C$. Hence, $(X_C,P_C)$ and $(X',P(X'))$ are the
same equilibrium, which establishes uniqueness.
\end{proof}

\subsection{Examples of equilibria}
\label{sec:examplesequilibria}
\noindent In this section, we use Theorem \ref{theorem:existenceuniqueness} in order to understand how the presence of penalties affects the trading strategy of the IT and the pricing function. 

Consistent with intuition, penalties reduce the demand of the IT. By how much $X(v)$ is reduced depends on the functional form of the cost $C$ and the realisation of $v$. This leads in general to a non-linear demand schedule. In the following examples, we will illustrate some important determinants of the IT demand. 

The price function can be very flat in some regions and increase sharply in others. In particular, the price impact of a marginal uninformed trade $\frac{d}{du}P(X(v)+u)$ strongly depends on both the realisations of $u$ and $v$. By constrast, in the mimicking equilibrium of the model without penalties, this price impact is constant, regardless of the distributional assumptions on the noise.

We consider three examples of penalty: quadratic, linear, and constant over large trades.

\subsubsection{Quadratic cost}
\noindent In this very particular instance, $X$ remains linear
after the introduction of the penalty. Imposing quadratic costs is akin to increasing
the perceived expected price impact. Since this cost is in $x^2$ while the gross gains
of trading are in $x$, the IT always trade as soon as $v\neq 0$, and the magnitude
of the trade increases with the absolute value of $v$. Note that the result that $X$ is linear can also obtain
in a one-period Kyle model with Gaussian noises when one makes one of the following assumptions:
(i) there is a quadratic penalty on trading,
(ii) the insider is risk-averse instead of risk-neutral,
(iii) the insider observes a signal imperfectly
correlated with $v$ instead of observing $v$ directly.

\begin{figure}[H]
\begin{center}
\includegraphics[width=220pt]{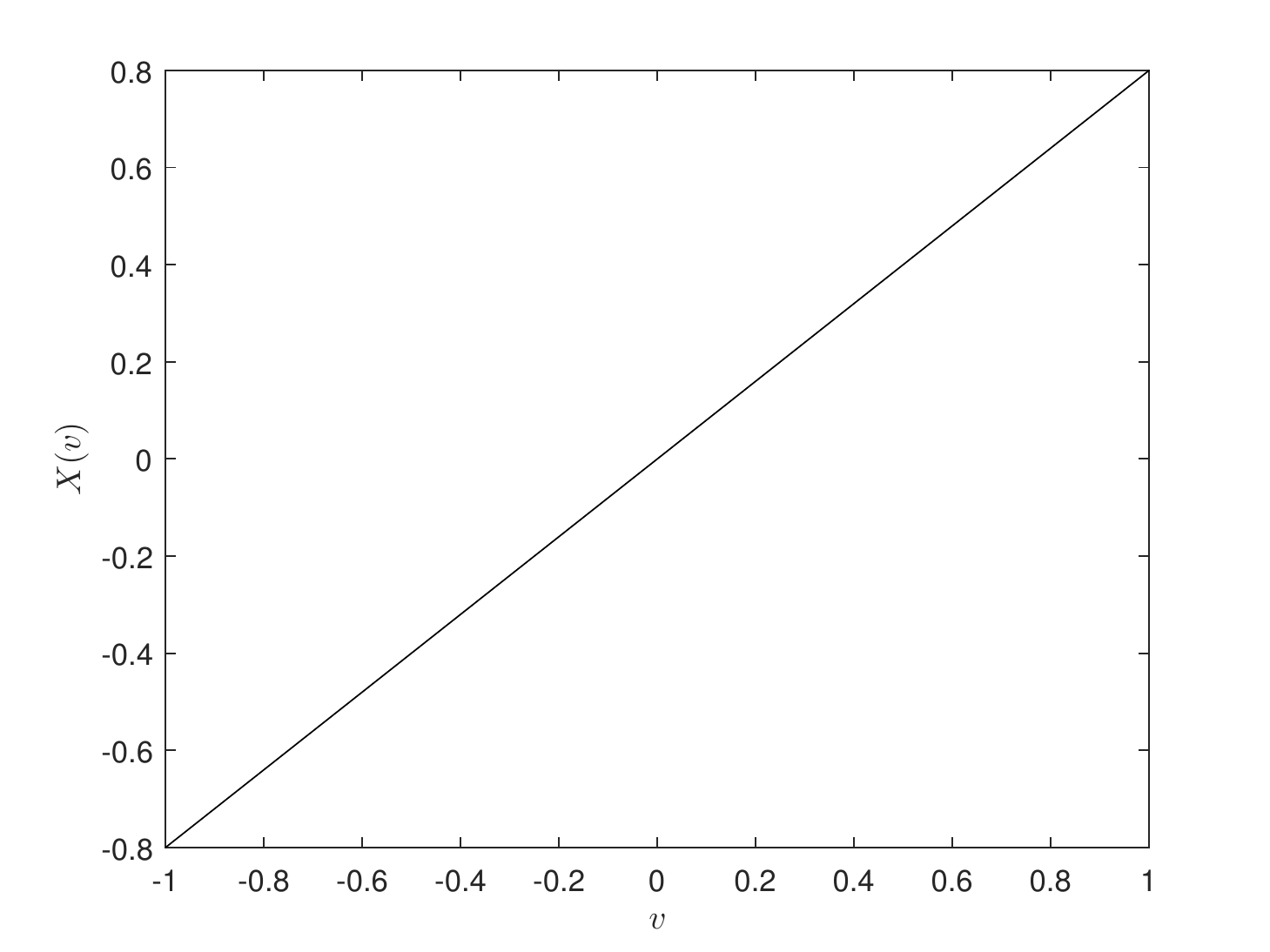} \hspace{20pt}\includegraphics[width=220pt]{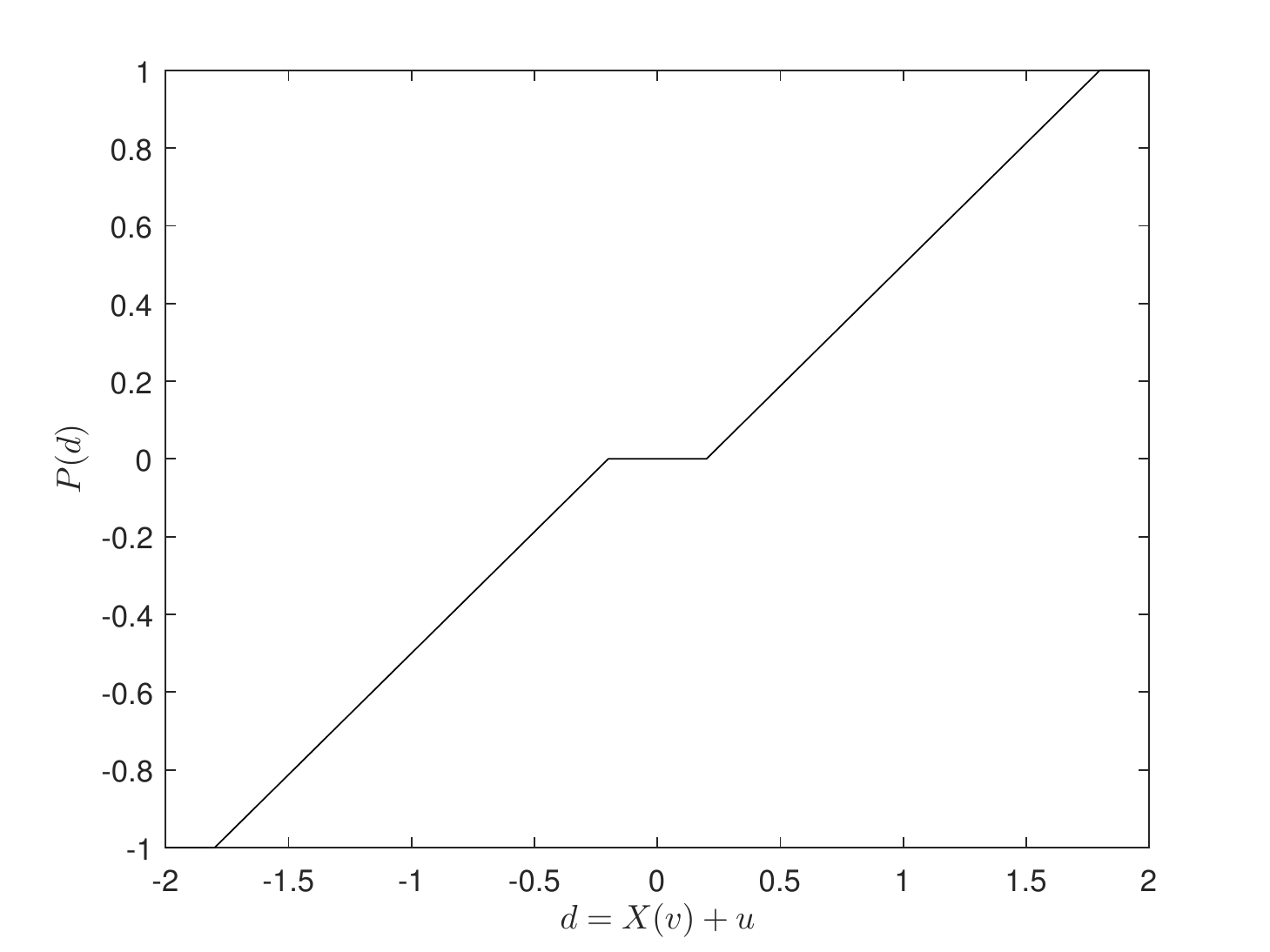}

\caption{Insider's demand and pricing under quadratic penalty}
\label{fig:xpquadraticpenaltyunif}

\

\noindent $C(x)=\alpha x^2$, $\alpha=0.125$. Left panel: IT demand $X$. Right panel: price function $P$.
\end{center}
Due to the presence of the penalty, the insider trades less than in the linear mimicking equilibrium,
so that $X(1)=x_M<1$ ($=0.8$ in this example).
\end{figure}

\noindent When $\lvert d\lvert\leq 1-x_M(=0.2)$, any demand of the IT
is compatible with the observed aggregate order, so all values $v$ remain
equally likely, as explained in section \ref{sec:intuititionlemmacs}. No information
is incorporated and the price remains at the initial expected value of the asset: 0.
When $d>1-x_M$, one knows that $v$ has not realized at a very low value.
This provides a lower bound on $v$ and the price becomes positive. As $d$ increases,
so do the lower bound and the price, until $d=1+x_M(=1.8)$. In that case, one knows for sure
that the IT has placed an order $x_M$, which means that $v=1$, and $P$ reaches 1.
The situation is symmetrical for values of $d$ below $x_M-1(=-0.2)$.

\subsubsection{Linear cost}
\begin{figure}[H]
\begin{center}
\includegraphics[width=220pt]{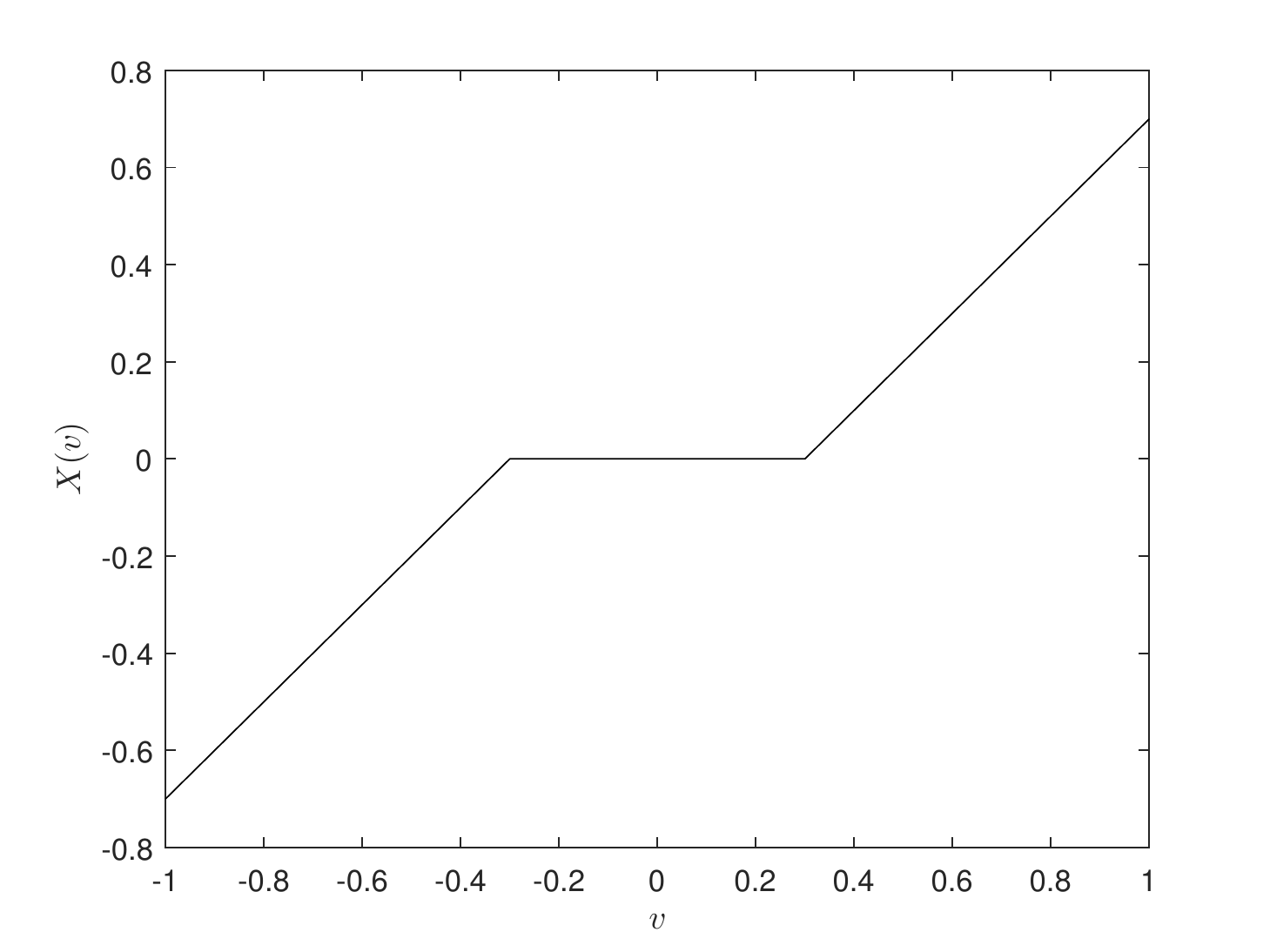} \hspace{20pt}\includegraphics[width=220pt]{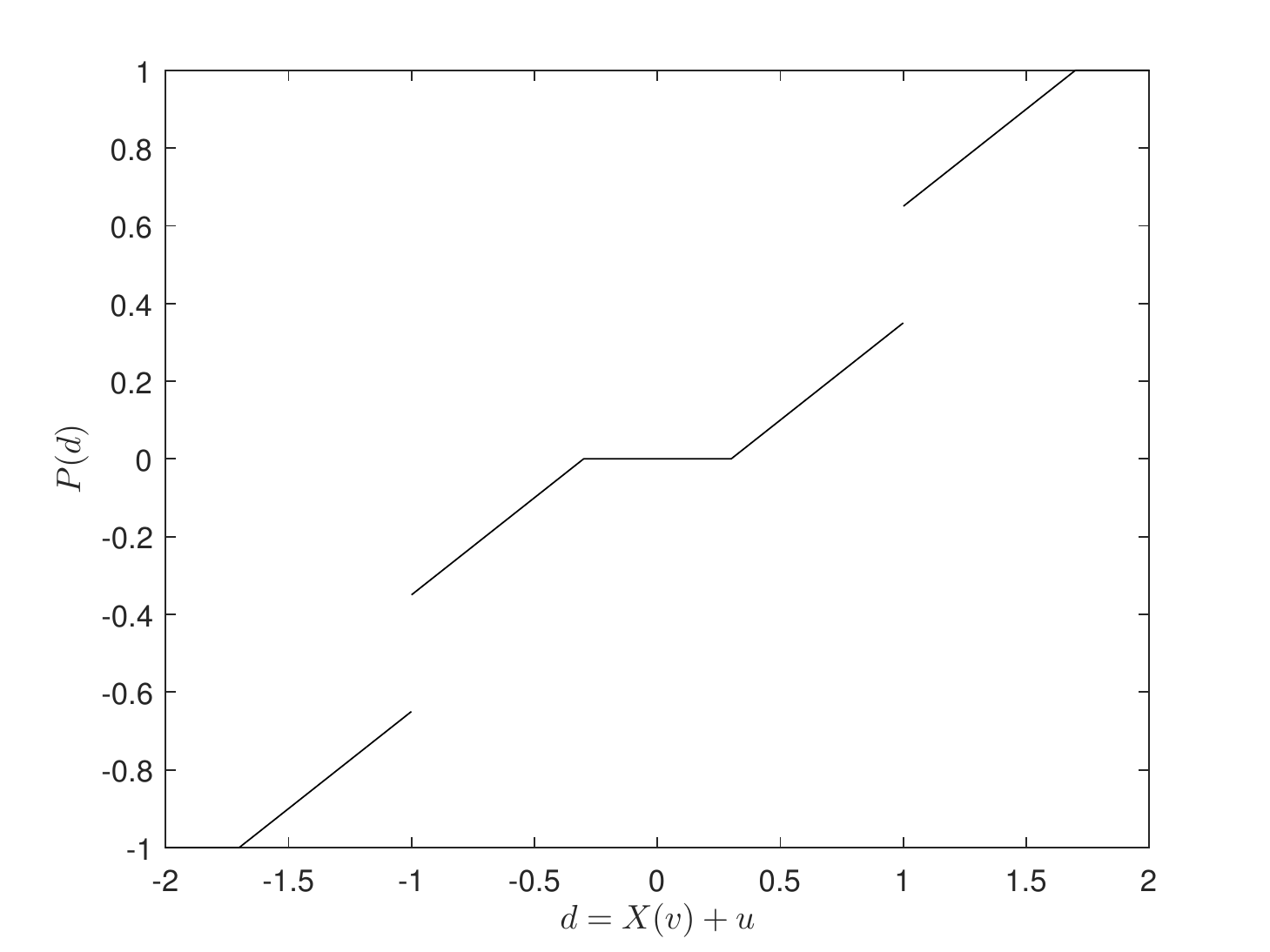}

\caption{Insider's demand and pricing under linear penalty}
\label{fig:xplinearpenaltyunif}

\

\noindent  $C(x)=\alpha \lvert x \lvert$, $\alpha= 0.3$. Left panel: IT demand $X$. Right panel: price function $P$.

\end{center}
\end{figure}

\noindent When the penalty is linear, $C(x)=\alpha \lvert x \lvert$,
the maximisation program of the IT can be rewritten as
\begin{equation*}
\max x\left((v-\alpha)-\frac{x}{2}\right).
\end{equation*}
If $v\geq\alpha$, one sees that a linear cost has the same effect 
as reducing the value of the fundamental $v$ by an amount $\alpha$,
and having no cost. Therefore, the strategy of the IT for values $v\in[\alpha,1]$
is a translation of the linear mimicking strategy over $v\in [0,1-\alpha]$. 
Similarly, the strategy of the IT for values $v\in[-1,-\alpha]$
is a translation of the linear mimicking strategy over $v\in [\alpha-1,0]$. 
This creates the two increasing linear segments in the left panel of Figure \ref{fig:xplinearpenaltyunif}.
In the flat middle section,
$v$ is not sufficient to cover the expected penalty and the IT does not trade.

\

\noindent The price function depicted in the right panel of Figure \ref{fig:xplinearpenaltyunif}
exhibits a flat section in the center surrounded by increasing linear segments. The intuition
is exactly the same as in the quadratic penalty case: when the magnitude of $d$ is small ($\lvert d\lvert\leq \alpha (=0.3)$),
all values of $v$ remain (equally) possible and no information is incorporated. As $d$ grows,
a lower bound on $v$ can be deduced and the price increases. The key difference
with the quadratic penalty case is that the price function jumps at $d=\pm 1$.
Indeed, when $d>1$, the market maker knows for sure that the insider has placed
a positive order. But the IT only does so when $v>\alpha$. By contrast, if $d=1^-$,
$X(v)=0$ remains possible, so we can only deduce that $v>-\alpha(=-0.3)$. In terms of information
incorporation, there is a huge difference between $d=1^+$ and $d=1^-$.

\subsubsection{Constant cost on trades of magnitude larger than $x_0$}
\begin{figure}[H]
\begin{center}
\includegraphics[width=220pt]{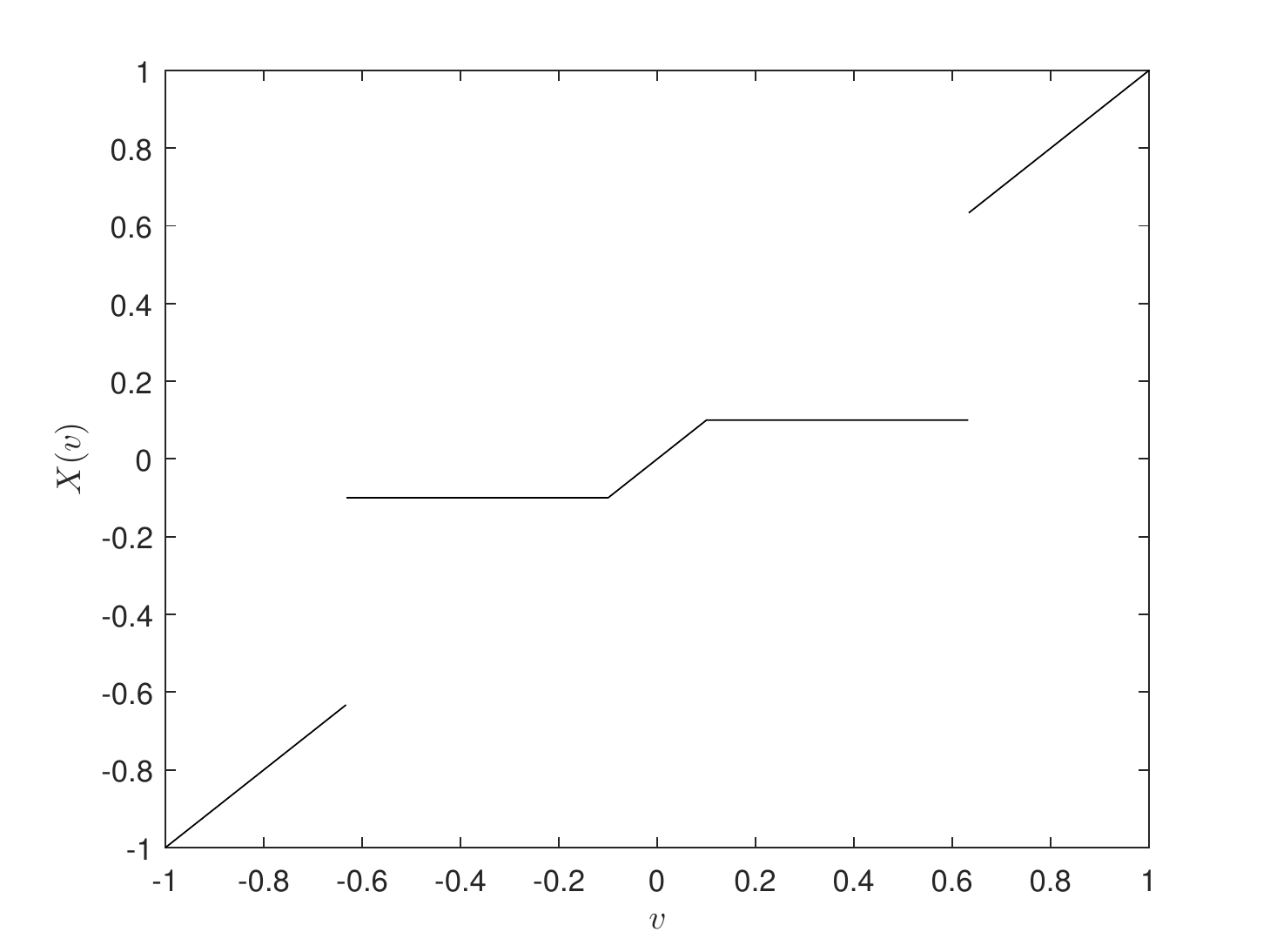}\hspace{20pt}\includegraphics[width=220pt]{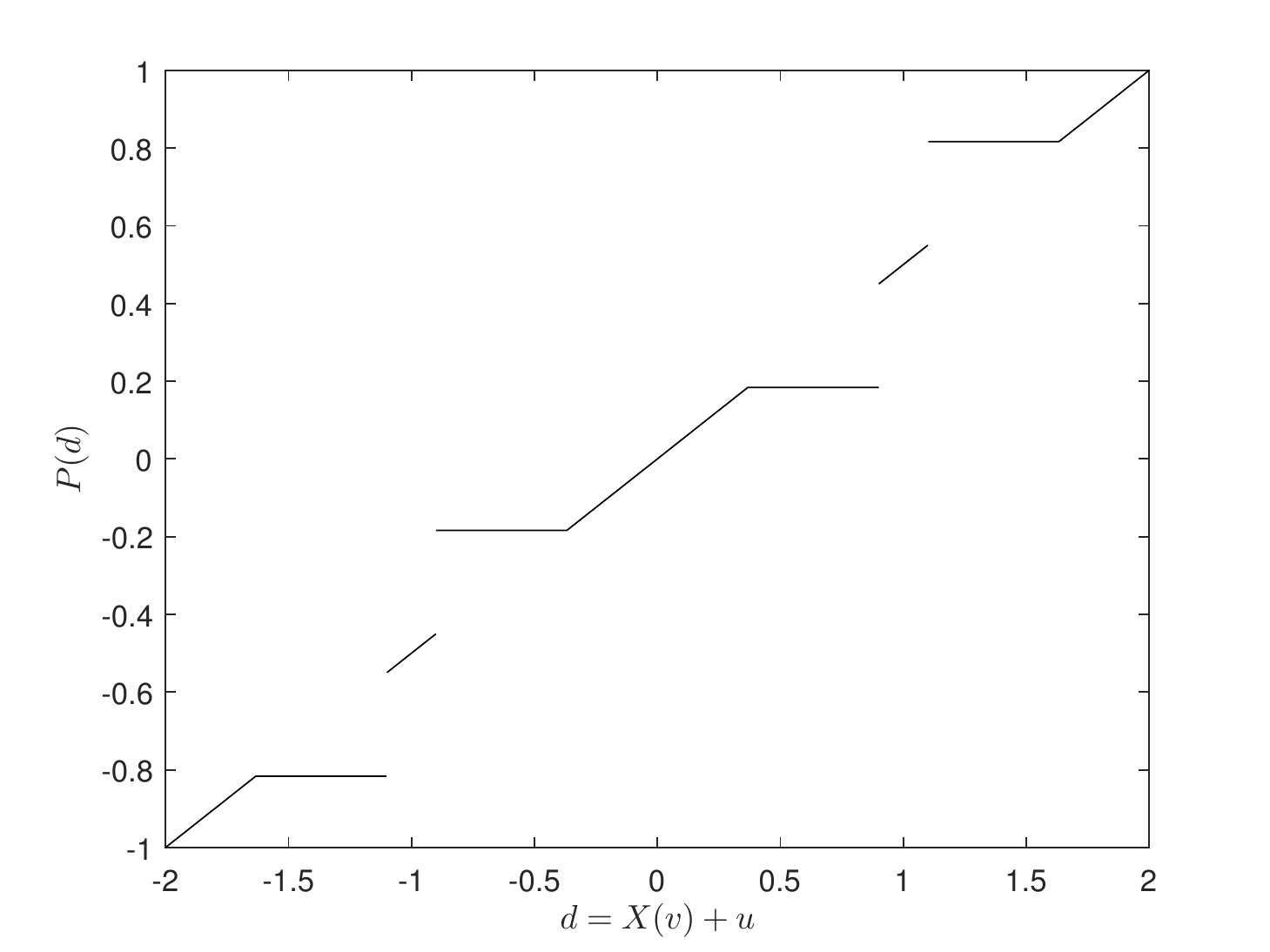} 

\caption{Insider's demand and pricing under constant penalty on large trades}
\label{fig:xpconstantpenaltyunif}

\

\noindent  $C(x)=K\mathbb{I}_{\lvert x \lvert > x_0}$, $K=0.2$, $x_0=0.1$. 

\noindent Left panel: IT demand $X$. Right panel: price function $P$. 

\end{center}

\end{figure}

\noindent Absent penalties, the IT picks $X(v)=v$. Hence, if she is sanctionned only
for trades of magnitude larger than $x_0$, she will not change her demand
as long a $\lvert v \lvert \leq x_0$: this corresponds to the increasing linear section
in the middle of Figure \ref{fig:xpconstantpenaltyunif}. For intermediate values of $v$,
the IT prefers to block her demand at the value $x_0$ (or $-x_0$) in order
to avoid the penalty: this corresponds to the flat sections in Figure \ref{fig:xpconstantpenaltyunif}.
When $v$ becomes large enough ($\lvert v\lvert>\sqrt{2K}(\approx 0.63)$), the penalty
is recouped in expectation by using the strategy that prevails in the absence of costs:
it appears as a sunk cost and the IT selects again the demand $X(v)=v$. This
corresponds to the increasing linear sections at the left and right of Figure \ref{fig:xpconstantpenaltyunif}.

\noindent The price function jumps at $d=\pm (1-x_0) (= \pm 0.9)$ and $d=\pm (1+x_0) (= \pm 1.1)$.
The intuition is as in the linear penalty case. When $d$ exceeds $1-x_0$, the MM knows that 
the demand of the IT was larger than $-x_0$ which rules out all values of $v$ at the left of $-\sqrt{2K}$, the 
left jump of $X$. Similarly, when $d$ exceeds $1+x_0$, the MM knows that the demand of the IT was larger
than $x_0$, which rules out all values of $v$ at the left of $\sqrt{2K}$, the right jump of $X$. 

\

\noindent A robustness exercise in the case of Gaussian noise is conducted
in Appendix \ref{sec:robustnessxp} and shows that most of the effects described above qualitatively subsist.

\newpage
\section{Efficient frontier without a budget constraint}
\label{sec:solutionregulator}
\noindent We now solve the regulatory problem laid out in section \ref{sec:theregulatorsproblem}
by proving the following theorem:

\begin{theorem}
\label{theorem:efficientfrontier}
The equation of the efficient frontier is
\begin{equation*}
S=\frac{1}{\sqrt{3}}(1+2G), \quad -\frac{1}{6}\leq G\leq 0.
\end{equation*}
The set of regulations that implements the efficient frontier is exactly the class $\mathcal{O}$ defined as
\begin{eqnarray}
\label{eq:defmathcalo}
\mathcal{O}=\biggl\{ C \in \mathcal{C}, \, \exists K\in \left[0,1/2\right], \, &&C(x) \geq x\left(\sqrt{2K}-\frac{x}{2}\right) \, \, \, \, \text{for} \, \, 0\leq x\leq \sqrt{2K},\nonumber\\
&& C(x) = K \, \quad \quad \quad \quad \quad \quad \text{for} \, \, \sqrt{2K}<x\leq 1\biggr\}.
\end{eqnarray}

\noindent When $C\in\mathcal{O}$, the demand of the insider writes
\begin{equation*}
X_K(v)=
\begin{cases}
0 \quad \quad \, \lvert v\lvert\leq\sqrt{2K} \\
v \quad \quad \, \lvert v\lvert>\sqrt{2K}
\end{cases}
\end{equation*}
for the $K\in [0,1/2]$ associated with $C$. 
\end{theorem}

\begin{figure}[H]
\begin{center}
\includegraphics[width=\textwidth]{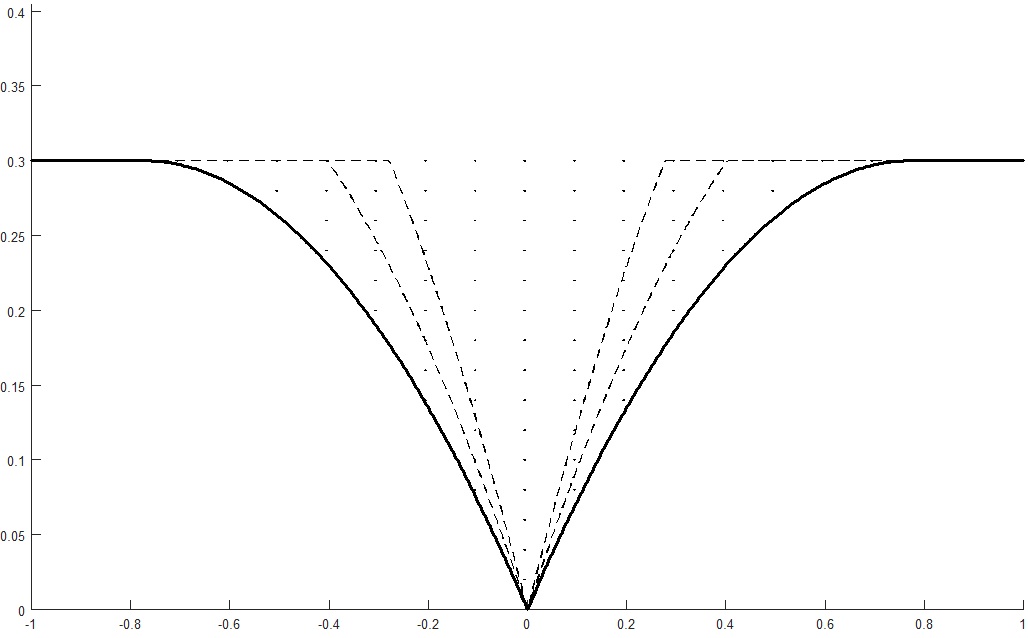}
\end{center}
\caption{Some penalty functions in $\mathcal{O}$.}
\label{fig:optimalpenalties}

\

The thick line represents the lower bound in the definition of $\mathcal{O}$
when $K=0.3$ (then, $\sqrt{2K}\approx 0.77$) :
any penalty in $\mathcal{O}$ must be above this line. Given that a penalty
is symmetrical and non-decreasing over $[0,1]$, the graph of a function in $\mathcal{O}$
must be included in the dotted area. The two dashed lines represent two such functions.
\end{figure}

\noindent Figure \ref{fig:optimalpenalties} gives a graphical representation of functions in $\mathcal{O}$.

\

\noindent If two penalties in $\mathcal{O}$ are associated with the same $K$,
they implement the same demand schedule $X_K$. Moreover, it is easy to see
that any point in the efficient frontier is implemented by $X_K$ for exactly one value of $K$.\footnote{A direct calculation shows that the P\&\,L $G$ of the uninformed traders under the demand $X_K$ is $-\frac{1}{6}\left(1-(2K)^{3/2}\right)$. Hence, the value of $K$ that implements the point $(G,S)$ of the efficient frontier is the solution to $G=-\frac{1}{6}\left(1-(2K)^{3/2}\right)$.}
Therefore, $K$ parametrizes the efficient frontier. Points associated with a small (resp. large) $K$ are selected by
a regulator who puts more weight on information incorporation (resp. on restricting the uninformed traders' losses).

\

\noindent Any regulator that puts nonzero weight on both objectives must at least somewhat
reduce insider trading, but not totally. As we shall detail later, the optimal solution is to allow
some large trades for large realisations of $\lvert v\lvert$, because they incorporate a lot of information;
more precisely, the regulator wants to implement $X(v)=v$ for large values of $\lvert v\lvert$.
The cutoff point $\sqrt{2K}$ in the schedule $X_K$ then appears as the solution
to the equation $\frac{v^2}{2}=K.$ (Recall that $\frac{v^2}{2}$ is the profit of the IT when there is no penalty). This characterizes the magnitude of $v$ above which
the penalty appears as a sunk cost to the insider, who then effectively optimizes as if there
was no penalty and selects the mimicking demand $X(v)=v$.

\

\noindent With noise $u\sim U(-a,a)$ and $v\sim U(b,c)$ with $a>0$ and $b<c$,
one can conduct a similar reasoning. By identifying the points where the mimicking strategy
exactly compensates for the penalty $K$, we find that the cutoff points $\pm\sqrt{2K}$ become 
\begin{equation*}
\frac{b+c}{2}\pm\sqrt{\frac{c-b}{a}K}.
\end{equation*}
Moreover, the maximal $K$ we need to consider is the smallest one that suppresses
net profits even at the extreme realizations $v\in\{b,c\}$; so
$K$ is now varying in the interval $\left[0,\frac{1}{4}a(c-b)\right]$.
Details can be found in Appendix \ref{sec:normalizationsupports}.

\subsection{Preliminary results on the regulator's objective}
\noindent Before characterizing the efficient frontier,
we need to derive some useful formulas.

\

\noindent The {\em expected net profit} of the insider trader in state $v$ is
\begin{equation}
\label{eq:profitbrut}
\pi^N(v) := X(v)(v-\hat{P}(X(v)))-C(X(v)).
\end{equation}
Note that in terms of the profit function $\psi_C$, the net profit is
$\pi^N(v)=\psi_C(X(v),v)$.

\

\noindent The overall  {\em expected net profit} (after fine, if any) is
\begin{equation}
\label{eq:profit}
\Pi^N := \mathbb{E}_v[\pi^N(v)].
\end{equation}
The {\em expected penalty} that the insider undergoes is
\begin{equation*}
F:=\mathbb{E}[C(X(v))].
\end{equation*}
The overall  {\em expected gross profit} (before fine, if any) is
\begin{equation}
\label{eq:profitbrut}
\Pi^T := \Pi^N + F = \lvert G \lvert.
\end{equation}
\noindent Observe that we can write
\begin{equation}
\label{eq:profitdistancel2}
\lvert G\vert = \int_0^1 X(v)\left(v-\frac{X(v)}{2}\right) \ud v = \underbrace{\int_0^1 \frac{v^2}{2}\ud v}_{1/6}-\frac{1}{2}\int_0^1 (v-X(v))^2 \ud v.
\end{equation}
This way of seeing the expected losses of the uninformed traders as (an affine transformation of) the $L^2$
distance between $X$ and the identity will be useful in section \ref{sec:nonpecuniary}. We continue by 
providing some convenient expressions of the quantities defined above.

\begin{lemma}
\label{lemma:envelope}
In equilibrium, the net profits satisfy
\begin{eqnarray}
\pi^N(v) &=& \int_0^v X(s) \ud s,\label{eq:evp}\\
\Pi^N &=& \int_0^1 (1-v) X(v) \ud v.\label{eq:evpintegree}
\end{eqnarray}
\end{lemma}

\begin{proof} Consider the parametrized objective function 
\begin{equation*}
\psi_C: [0,1]\times [0,1] \rightarrow \mathbb{R}
\end{equation*}
defined in \eqref{eq:profitpsi}.
Notice that (i) $\psi_C(x,.)$ is linear in $v$ and therefore absolutely continuous,
(ii) $\lvert\partial_v\psi_C(x,v)\lvert=\lvert x\lvert\leq 1$. (i) and (ii) guarantee that
the assumptions of 
Theorem $2$ in
\cite{MilgromSegal}
are satisfied. In the present case, this theorem tells us that we can write:
\begin{eqnarray*}
\pi^N(v) &=& \pi^N(0)+\int_0^v \partial_2 \psi_C(X(s),s) \ud s \\
&=& \int_{0}^{v} X(s) \ud s,
\end{eqnarray*}
since the insider does not make any profit when the fundamental $v$ is $0$.
Finally,
\begin{eqnarray*}
\Pi^N= \frac{1}{2} \int_{-1}^{1}\pi^{N}(v) \ud v &=& \frac{1}{2}\int_{-1}^{1}\int_0^v X(y) \ud y \ud v\\
&=&\int_{0}^{1}\int_0^v X(y) \ud y \ud v\\
&=&\int_0^{1}(1-v)X(v) \ud v.
\end{eqnarray*}
\end{proof}

\

\noindent Lemma \ref{lemma:expstd} expresses the expected post-trade standard deviation as a function of the demand profile $X$.
One consequence of this Lemma is that large orders associated with large values of the fundamental
are the ones that contribute the most to incorporating information into prices. Indeed, the values
of $v$ such that the product $vX(v)$ is large have the strongest negative impact on $S$,
as can be seen from \eqref{eq:formulastd}. This provides intuition on why $\mathcal{O}$
is the class of optimal penalties: when $C\in\mathcal{O}$, the regulator knows that the IT
will trade large quantities should $v$ realize at a large value because penalties are flat
for large $v$. Although costly for the
uninformed traders, these orders are those that contribute the most to reducing
uncertainty about $v$, making the regulator unwilling to prevent them.

\begin{lemma}
\label{lemma:expstd}
The expected post-trade standard deviation satisfies
\begin{equation}
\label{eq:formulastd}
S= \frac{1}{\sqrt{3}}\left(1-\int_0^1 vX(v) \ud v\right).
\end{equation}
\end{lemma}

\begin{proof} By the proof of Lemma \ref{lemma:cs}, $v\lvert d$ is uniform
over $$I_X(d)\equiv[(X^{-1}_{\ell}\left((d-1)\vee(-x_M)\right);X^{-1}_r\left((d+1)\wedge x_M\right))].$$
Since the standard deviation of a uniform variable over $[a;b]$ equals $\frac{1}{2\sqrt{3}}(b-a)$,
Lemma \ref{lemma:expstd} is an immediate consequence of the following result: \emph{if $X$ is an odd
non-decreasing function from $[-1;1]$ to $[-x_M;x_M]$, then the expected length of the interval
$I_X(X(v)+u)$ equals $2\left(1-\int_0^1 v X(v) \ud v\right)$ }, which we must now prove. 
\

For  $v\in [-1;1]$, define
\begin{eqnarray*}
Y_v &=& X_r^{-1}\left((X(v)+u+1)\wedge x_M\right)\\ 
Z_v &=& X_l^{-1}\left((X(v)+u-1)\vee (- x_M)\right).
\end{eqnarray*} 
What we need to prove is that $\mathbb{E}_{v,u}[Y_v-Z_v]=2\left(1-\int_0^1 v X(v) \ud v\right)$. By symmetry, $\mathbb{E}_{v,u}[Z_v]=-\mathbb{E}_{v,u}[Y_v]$, thus, it remains to prove that: 
\begin{align*}
\mathbb{E}_{v,u}[Y_v]= 1-\int_0^1 vX(v) \ud v.
\end{align*}
Let us consider $v$ fixed. The random variable $Y_v$ takes values in $[-1,1]$: using Fubini theorem, 
\begin{align*}
\mathbb{E}[Y_v] = \mathbb{E}\left[\int_{-1}^{1} \mathbb{I}_{-1\leq y\leq Y_v} \ud y\right] -1 =\int_{-1}^{1}\mathbb{P}\left(y\leq Y_v\right) \ud y -1. 
\end{align*}
By definition of $X_r^{-1}$, if $X(y) \leq (X(v)+u+1)\wedge x_M$ then $y \leq Y_v$. Besides, if $y < Y_v$, then using the fact that $X$ is non decreasing, $X(y) \leq (X(v)+u+1)\wedge x_M$. Thus: 
\begin{align*}
\{y \leq Y_v\} \setminus \{X(y) \leq (X(v)+u+1)\wedge x_M \} \subset \{ y=Y_v\}. 
\end{align*}
Let us remark that $Y_v = y$ can hold for two different values of $u$ if and only if $X$ is discontinuous at $y$ or $y=1$. In particular, 
\begin{align*}
\{y\neq 1 | \mathbb{P}(y =Y_v) > 0\} \subset \{y | X(y^{-}) \neq X(y^{+})\}. 
\end{align*}
It follows from this discussion that :  
\begin{align*}
\left\lvert \mathbb{E}[Y_v] - \int_{-1}^{1} \mathbb{P}(X(y) \leq X(v)+u+1) \ud y + 1 \right\lvert \leq \int_{-1}^1 \mathbb{P}(Y_v=y) \ud y \leq \mu \left(  \{y | X(y^{-}) \neq X(y^{+})\} \right),
\end{align*}
where $\mu$ is the Lebesgue measure on $[-1,1]$. Since $X$ is non-decreasing, it has a countable number of discontinuity points. In particular $\mu \left(  \{y | X(y^{-}) \neq X(y^{+})\} \right)=0$ and:   
\begin{align*}
\mathbb{E}[Y_v] = \int_{-1}^{1} \mathbb{P}(X(y) \leq X(v)+u+1) \ud y - 1. 
\end{align*}
Now,
\begin{eqnarray*}
\mathbb{P}(X(y) \leq X(v)+u+1) &=&\mathbb{P}(u \geq X(y)-X(v)-1)\\
&=& 1 + \left( \frac{1}{2}(X(v)-X(y)) \wedge 0\right).
\end{eqnarray*}
Going back to the expression of $\mathbb{E}[Y_v]$, we obtain
\begin{equation*}
\mathbb{E}[Y_v]=1-\frac{1}{2}\int_v^1 X(y)\ud y+\frac{1}{2}(1-v)X(v).
\end{equation*}
Integrating over $v$:
\begin{eqnarray*}
\mathbb{E}_{v,u}[Y_v]&=&1-\frac{1}{4}\int_{-1}^1\int_v^1 X(y)\ud y \ud v+\frac{1}{4}\int_{-1}^1(1-v)X(v)\ud v\\
&=&1-\frac{1}{4}\int_{-1}^1 (v+1)X(v)\ud v+\frac{1}{4}\int_{-1}^1(1-v)X(v)\ud v\\
&=&1-\frac{1}{2}\int_{-1}^1 vX(v)\ud v\\
&=&1-\int_0^1 vX(v) \ud v,
\end{eqnarray*}
where in line $3$, we used the fact that $X$ is odd. This concludes the proof. 
\end{proof}

\subsection{Characterization of the efficient frontier}
\subsubsection{Shape of the efficient frontier and efficient demand functions}
\noindent In this section, we give the shape of the efficient frontier and
explain what demand schedules are compatible with it. We call these schedules
\emph{efficient demand functions}.
\begin{lemma}
\label{lemma:lowerbound}
Let $C$ be a penalty function in $\mathcal{C}$. In the equilibrium of $\mathcal{K}(C)$,
\begin{equation*}
S\geq\frac{1}{\sqrt{3}}(1+2G)
\end{equation*}
with equality if and only if there is $v^*\in [0,1]$ such that $X(v)=0$ for $\lvert v\lvert<v^*$ and $X(v)=v$ for
$\lvert v\lvert>v^*$.
\end{lemma}
\begin{proof} Due to Lemma \ref{lemma:expstd}, what we need to show
is that
\begin{equation*}
-\int_0^1 v X(v) \ud v\geq -2\int_0^1 X(v)\left(v-\frac{X(v)}{2}\right) \ud v.
\end{equation*}
This is equivalent to
\begin{equation*}
\int_0^1 vX(v) dv\geq \int_0^1 X(v)^2 \ud v,
\end{equation*}
or
\begin{equation}
\label{eq:mainineqlowerbound}
\int_0^1 X(v)(v-X(v)) \ud v\geq 0
\end{equation}
which holds because $0\leq X(v)\leq v$ for $v\in [0;1]$.

\

\noindent For the equality to hold, it is necessary and sufficient to have $X(v)=0$
or $X(v)=v$ almost everywhere. Since $X$ is non-decreasing, it is equivalent
to $X(v)=0$ for $\lvert v\lvert<v^*$ and $X(v)=v$ for
$\lvert v\lvert>v^*$, where $v^*=\sup\{v, \, X(v)=0\}.$
\end{proof} 

\

\noindent Equation \eqref{eq:mainineqlowerbound} is particularly convenient
because it immediately indicates what type of demand function is needed to implement
the efficient frontier. Of course,  $X$ is an endogenous outcome: what remains to
be seen is what regulations implement the efficient demand functions.

\subsubsection{Implementation of the efficient demand functions}
\label{sec:implementationefficientdemand}

\begin{lemma}
\label{lemma:implementation}
The efficient demand functions derived in Lemma \ref{lemma:lowerbound}
are implemented exactly by the penalties $C\in\mathcal{O}$.
\end{lemma}

\noindent By construction, penalties in $\mathcal{O}$ are flat for large values
of $\lvert v\lvert$ and increase quickly as $\lvert v\lvert$ departs from 0
(see Figure \ref{fig:optimalpenalties}). Intuitively, this is what is required
to implement the efficient demand functions. Indeed, when $\lvert v \lvert$ realizes
at a small value, the marginal impact of increasing demand on the expected penalty
is large, and the IT prefers to refrain from trading. For $\lvert  v\lvert$ large, however,
the penalty schedule being flat on large demands, a large order allows to cover
the expected fine, which appears as a sunk cost. The IT then optimizes as in the
linear mimicking equilibrium and demands $X(v)=v$. The proof of the Lemma can be found in Appendix \ref{app:proofs}.

\subsubsection{Proof of Theorem \ref{theorem:efficientfrontier}, illustrations and discussions}
\label{sec:prooftheoremefficientfrontier}
\noindent The proof of Theorem \ref{theorem:efficientfrontier}
is complete:
Lemma \ref{lemma:lowerbound} characterizes the efficient frontier
and due to Lemma \ref{lemma:implementation}, achieving the efficient frontier can only be done by selecting
a cost $C\in\mathcal{O}$, characterized by a $K\in[0,1/2]$.

Note that by varying $K$
between 0 and $\frac{1}{2}$, one clearly covers the full efficient frontier. 
As $K$ increases, the losses ($-G$) of the uninformed traders decrease from $\int_0^1\frac{v^2}{2} \ud v=\frac{1}{6}\approx 0.167$
to 0, while the expected post-trade standard deviation increases from $\frac{1}{\sqrt{3}}(1-2/6)=\frac{2}{3\sqrt{3}}\approx 0.385$
to $\frac{1}{\sqrt{3}}\approx 0.577$.

\begin{figure}[H]

\begin{center}
\includegraphics[width=1\textwidth]{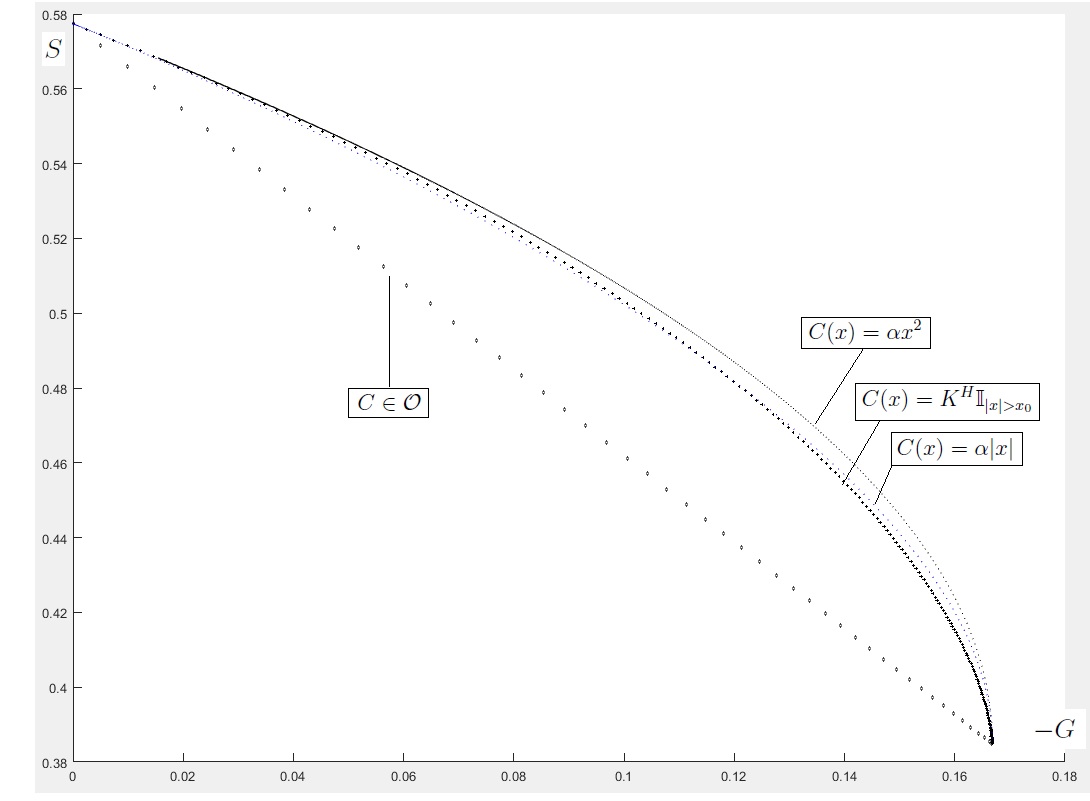}
\caption{Locus of $(S,-G)$ for some penalty functions.}
\label{fig:fig1}
\end{center}
\end{figure}

Each point of Figure \ref{fig:fig1} corresponds to a penalty function $C$;
it represents the outcomes $(S,-G)$ in the unique equilibrium of $\mathcal{K}(C)$.
The losses of the uninformed traders, $-G$, read on the $x$-axis. The expected
post-trade standard deviation, $S$, reads on the $y$-axis. For a fixed $y$-coordinate
(a fixed $S$)
the preferred option of the regulator is to select a point with the smallest $x$-coordinate
(that minimises $-G$).

Consistent with Theorem \ref{theorem:efficientfrontier}, penalties in $\mathcal{O}$
achieve the efficient frontier, which is linear as indicated by Lemma \ref{lemma:lowerbound}.

Outcomes $(S,-G)$ corresponding to quadratic and linear penalties ($C(x)=\alpha x^2$, $C(x)=\alpha\lvert x\lvert$
for varying $\alpha\geq 0$) are also reported in Figure \ref{fig:fig1}. As one can see, they perform significantly
worse than penalties $C\in\mathcal{O}$. This is also the case of penalties with no cost on small trades
and big costs on large trades, $C(x)=K^H\mathbb{I}_{\lvert x \lvert >x_0}$. Here $K^H$ is a constant
large enough so that the insider never chooses to trade more than $x_0$. The fact that these particular
penalty functions perform poorly compared to penalties in $\mathcal{O}$ is consistent with the intuition
given above Lemma \ref{lemma:expstd}. Indeed, they imply that $X(v)=v$ for $\lvert v\lvert$ small 
and $X(v)=0$ for $\lvert v\lvert$ large (the opposite of the demand functions implied by $C\in\mathcal{O}$),
so that the reduction of the expected standard deviation, measured by the term $\int_0^1 vX(v) \ud v$ (see Lemma \ref{lemma:expstd}), is low.

\noindent Figure \ref{fig:fig1} shows that quadratic costs are the most inefficient
among the considered costs. In fact, they have the worst performance among
\emph{all penalty functions}:

\begin{proposition}
\label{prop:quadraticareworst}
Quadratic penalties implement the upper frontier of the locus of outcomes $(S,G)$ generated
by all penalty functions in $\mathcal{C}$, i.e. they induce the highest possible expected 
post-trade standard deviation for a given P\,\&\,L of the uninformed traders.
\end{proposition}

\begin{proof}
See Appendix \ref{app:proofs}.
\end{proof}

\

\noindent In Appendix \ref{sec:robustnessfig1}, we repeat numerically the construction of Figure \ref{fig:fig1} in the case of Gaussian
noise: $u,v\sim N(0,1)$ and obtain similar results.

\newpage
\section{Efficient frontiers under a budget constraint}
\label{sec:efficientfrontiersunderabudgetconstraint}
\noindent So far, by imposing virtually no restriction on the set
of admissible penalties, our analysis potentially assumes away a 
real-world constraint on the regulator: investigation costs. 
Conducting investigations requires time, financial and human resources.
How do the regulator's efficient policies change
in that case? 

\

\noindent In section \ref{sec:nonpecuniary}, we consider the case of non-pecuniary
penalties: the regulator cannot balance its budget by collecting fines. This translates into
a bound on the investigation probability, which in turn caps the maximal expected
penalty that can be imposed an insider trades. In section \ref{sec:pecuniary},
we study pecuniary fines. In that case, the regulator needs to collect at least some fines
to balance its budget. This constraint forces the regulator to select ``intermediate'' levels of
penalties: if $C$ is too small, not enough fines are collected, but the same holds if $C$ is too large,
as this induces insider traders to refrain from trading. 

\subsection{Non-pecuniary penalties}
\label{sec:nonpecuniary}
\noindent We maintain the assumption that investigation occurs 
with a constant probability, $\alpha$, leaving the analysis of the case where $\alpha$ is a function
of an observable (e.g. the aggregate order) for future research.
We also suppose that the regulator cannot use fines
to relax its budget constraint. This is the case as soon as penalties are non-pecuniary,
e.g. an imprisonment sentence.

\subsubsection{Setup and Characterization of the efficient frontier}

\noindent With an investigation cost $\kappa$, since the expected expenses
of the regulator are given by $\alpha \kappa$, and denoting the alloted budget $B$, we consider
the constraint
\begin{equation}
\label{eq:budgetalphakappa}
\alpha\kappa\leq B.
\end{equation}
Note that the insider trader optimizes under an expected
penalty schedule $C = \alpha\tilde{C}$, where $\tilde{C}$ is the actual sanction
conditional on investigation success. Absent a cap on $\tilde{C}$, the regulator could
trivially get around its budget constraint by reducing $\alpha$ and increasing $\tilde{C}$.
We would be back to the case studied in section \ref{sec:solutionregulator}.
There are, however, several reasons that justify the existence of a bound on $\tilde{C}$.
The first one is simply that the worst possible sanction, say lifetime imprisonment,
does not provide $-\infty$ utility. Another rationale comes from the fact
that the stronger a sanction, the harder it is to implement it,
as the legally required amount of evidence increases. For instance, in the Netherlands at the
end of the XXth century, a very strong penalization of insider trading was enforced, which
actually led to a quasi-impossibility to convict people of insider trading.\footnote{This is documented in \cite{SECspeech}.}
From \eqref{eq:budgetalphakappa}, with a cap $\tilde{C}^M$ on $\tilde{C}$, the insider trader faces an expected penalty 
\begin{equation}
\label{eq:expectedpenalty}
C=\alpha\tilde{C}\leq K:=\frac{B}{\kappa}\tilde{C}^M.
\end{equation}

\

The constraint on $C$, equation \eqref{eq:expectedpenalty},
means that we now work with a restricted set of admissible penalties:\footnote{
Of course, we could obtain the same constraint by ignoring investigation costs, setting $\alpha=1$
and assuming that the cap  $\tilde{C}^M$ on $\tilde{C}=C$ is below 1/2. The idea here is that if investigation
was systematic, the bound $\tilde{C}^M$ would likely be non-binding. It only becomes binding
because investigation is costly, which reduces the expected penalty that the IT faces. The extent to which it binds depends on the budget-relevant parameters $B$ and $\kappa$: see \eqref{eq:expectedpenalty}.} 

\begin{definition}
In the non-pecuniary case, the set of admissible penalties with a budget constraint is
\begin{equation*}
\mathcal{C}_K = \{C\in\mathcal{C}, \, C(1)\leq K \}.
\end{equation*}
\end{definition}

\noindent Note that $C(1)\leq K$ is equivalent to \eqref{eq:expectedpenalty} because
any penalty in $C$ is symmetrical and non-decreasing over $[0,1]$. Moreover,
the budget constraint is an actual constraint for $K\in [0,1/2)$; for $K\geq 1/2$, $\mathcal{C}_K=\mathcal{C}$.

What happens when one restricts the set of
admissible penalties? First, some previously efficient points
may no longer be feasible. Second, some points that were
not previously efficient may no longer be dominated by any point
still implementable under the budget constraint. We recast Definition
\ref{def:dominatedpoints} in this new setting:

\begin{definition}
In the non-pecuniary case, the efficient frontier under a budget
constraint is the set of points $(G,S)$ implementable by a penalty in $\mathcal{C}_K$
that are not dominated by any point implementable by a penalty in $\mathcal{C}_K$.
\end{definition}

\begin{figure}[H]
    \centering
    \begin{subfigure}[b]{0.5\textwidth}
        \centering
        \includegraphics[width=230pt]{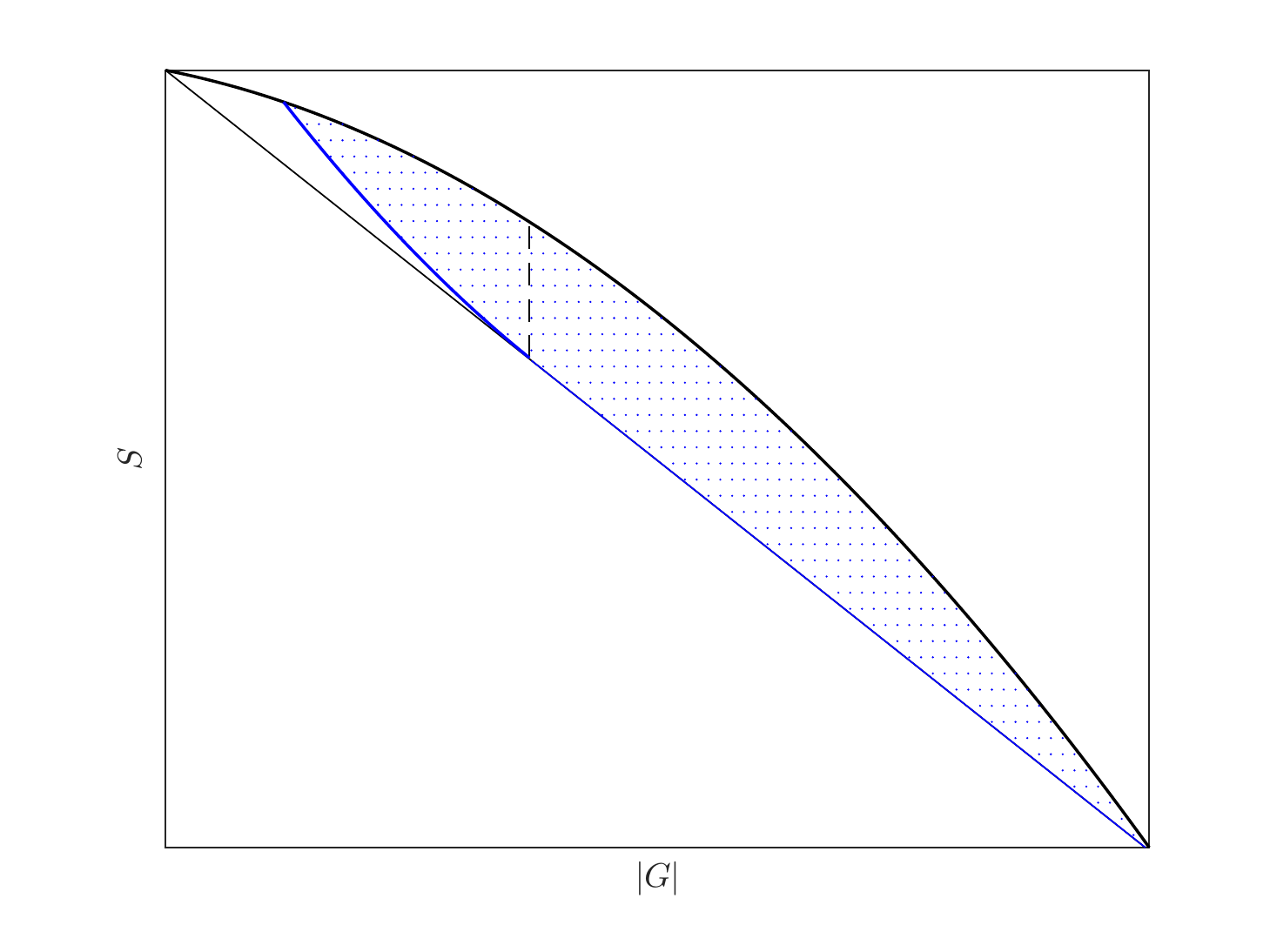}
        \caption{}
    \end{subfigure}%
    ~ 
    \begin{subfigure}[b]{0.5\textwidth}
        \centering
        \includegraphics[width=230pt]{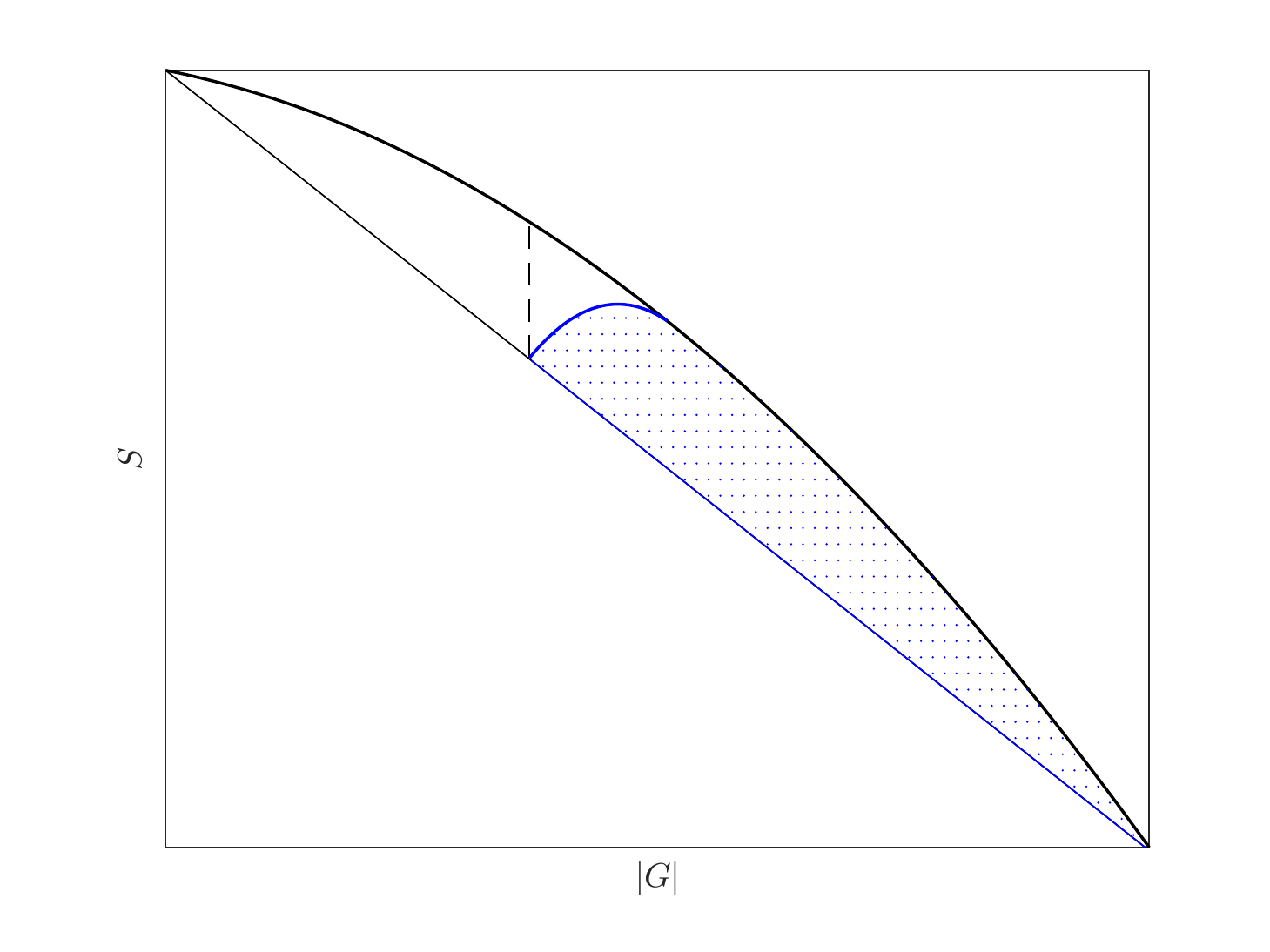}
        \caption{}
    \end{subfigure}
    \caption{Introduction of a constraint: two possible scenarios.}
\label{fig:twotypesofconstrainedfrontiers}
\end{figure}

As discussed above, the introduction of a constraint on the set
of admissible penalties should in general make new efficient points appear.
Consider for instance the situation depicted in panel (a) of Figure 
\ref{fig:twotypesofconstrainedfrontiers}. The dotted region represents
the set of feasible points under the constraint. Points of the previously
efficient frontier (oblique straight line) at the right of the dashed line
are still feasible and therefore still efficient. Those at the left on the dashed
line are not implementable anymore. The lower frontier of the blue area
is the new efficient frontier. In particular, new efficient points appear
at the left of the dashed line.

By contrast, in panel (b), there are no feasible points at the left
of the dashed line: the efficient frontier is truncated.

\noindent It is a priori quite unclear in which situation we are.
Denote

\begin{equation*}
\mathcal{O}_K =  \mathcal{O} \cap \mathcal{C}_K.
\end{equation*}

\noindent $\mathcal{O}_K$ is the set of efficient penalties derived in section \ref{sec:implementationefficientdemand},
that are still feasible under the budget constraint. These penalties are still
efficient under the budget constraint. Moreover, by direct computation, we obtain that
as $C$ varies in $\mathcal{O}_K$, $\vert G\vert$ describes the interval 
\begin{equation*}
\left[\vert G\vert_{\text{min}}(K),\frac{1}{6}\right],
\end{equation*}
where
\begin{equation}
\label{eq:gmink}
\vert G\vert_{\text{min}}(K):=\frac{1}{6}\left(1-(2K)^{3/2}\right).
\end{equation}

\

\noindent The truncature of the previously
efficient frontier at the right (in the $(\vert G\vert,S)$ plane) of $\vert G\vert_{\text{min}}(K)$ is part of
the efficient frontier under the budget constraint. In light of the discussion above, the key question is to know what happens
at the left of $\vert G\vert_{\text{min}}(K)$.
Theorem \ref{lemma:gmaxsouscontrainte} shows that no penalty
in $\mathcal{C}_K$ can implement $\lvert G\lvert<\vert G\vert_{\text{min}}(K)$
(i.e. we are in the situation of panel (b)).

This immediately
implies the characterisation of the constrained efficient frontier: 

\begin{theorem}
\label{theorem:efficientfrontiernonpecuniary}
The efficient frontier under the constraint $C\leq K$ is the truncature $\vert G\vert\geq \vert G\vert_{\text{min}}(K)$
of the efficient frontier of Theorem \ref{theorem:efficientfrontier} and is implemented exactly by penalties in $\mathcal{O}_K$.
\end{theorem}

\

\noindent Theorem \ref{theorem:efficientfrontiernonpecuniary} is a consequence of the following:
\begin{theorem}
\label{lemma:gmaxsouscontrainte}
Let $K\leq 1/2$.
Under the constraint $C\leq K$, the expected losses of the uninformed traders are at least
\begin{equation*}
\lvert G\lvert\geq \vert G\vert_{\text{min}}(K).
\end{equation*}
This lower bound is attained by the demand schedules $X_{\alpha}$ 
for $0\leq\alpha \leq 1-\sqrt{2K}$ and by the $X_{\alpha}$  only, where

\begin{equation*}
X_{\alpha}(v)=
\begin{cases}
v \quad \quad \, 0\leq v\leq\alpha\\
\alpha \quad \quad \alpha<v \leq \alpha+\sqrt{2K}\\
v \quad \quad \, v > \alpha+\sqrt{2K}\\
-X_{\alpha}(-v) \quad \quad v<0.
\end{cases}
\end{equation*}

\noindent These demand schedules are implemented by the penalties $C_{\alpha}$
where
$C_{\alpha}(x)=K\mathbb{I}_{\lvert x\lvert>\alpha}$.
\end{theorem}

\noindent Theorem \ref{lemma:gmaxsouscontrainte} shows that one 
cannot implement $\vert G\vert<\vert G\vert_{\text{min}}(K)$ with $C\in\mathcal{C}_K$
and provides penalty functions $C_{\alpha}$ that achieve $\vert G\vert=\vert G\vert_{\text{min}}(K)$.
While the $C_{\alpha}$ imply the same expected losses of the uninformed traders, they all
imply different expected post-trade standard deviations. In particular, all the $C_{\alpha}$ for $\alpha\neq 0$
are \emph{not} efficient penalties.

\

\noindent We supplement the proof with several discussions, and therefore present it
in a separate section.

\subsubsection{Proof of Theorem \ref{lemma:gmaxsouscontrainte} and intuition}
\noindent\textit{Step 1}: transformation of the problem into a constrained
problem of $L^2$ distance maximisation.

\

\noindent Recall equation \eqref{eq:profitdistancel2}:
\begin{equation*}
\lvert G\vert = \frac{1}{6}-\frac{1}{2}\int_0^1 (v-X(v))^2 \ud v.
\end{equation*}
This means that obtaining the bound of the Theorem is equivalent to showing
\begin{equation}
\label{eq:maximisedistancel2}
\max_{C\in\mathcal{C}_K} \int_0^1 (v-X(v))^2 \ud v = \frac{(2K)^{3/2}}{3},
\end{equation}
subject to the constraint that $X(v)$ maximises the net profit $\psi_C(.,v)$.

\

\noindent Let $g(v)=v-X(v)$, so that we are looking for an upper bound
of $\int_0^1 g^2$. By Lemma \ref{lemma:envelope} and under the constraint $C\leq K$,
we obtain:

\begin{eqnarray}
\label{eq:applicationenveloppe}
\int_0^1 g &=& \int_0^1 v \ud v - \int_0^1 X(v) \ud v\nonumber\\
&=& \frac{1}{2} - \pi^N(1)\nonumber\\
&\leq& K.
\end{eqnarray}
This is because, when $v=1$, the IT can achieve at least a net profit of $\frac{1}{2}-C(1)\geq \frac{1}{2}-K$. Therefore, the maximum in \eqref{eq:maximisedistancel2}
is less or equal to
\begin{equation*}
\sup \int_0^1 g^2
\end{equation*}
subject to the constraints (i) $\int_0^1 g \leq K$, and (ii) $g(0)=0\leq g(v)$ and
$v\mapsto v-g(v)$ is non-decreasing. (i) comes from \eqref{eq:applicationenveloppe},
and (ii) is an immediate consequence of the properties of an optimal demand
schedule $X$. 

\

\noindent Notice how crucial Lemma \ref{lemma:envelope} is, and therefore how
effective the result of \cite{MilgromSegal} is. Once noted
that $C(1)\leq K$ implies a lower bound on the net profit at 1, Lemma \ref{lemma:envelope}
allows (i) to incorporate the constraint the $X$ is a maximiser in a parsimonious way, (ii)
to reduce the two constraints --- $C\leq K$ and $X$ must maximise $\psi_C$ ---, into a single
condition, $\int g\leq K$, which is particularly convenient, as it is a $L^1$ bound in a $L^2$ 
maximisation problem.

\

\noindent Absent the fact that $X$ must be non-decreasing, which translates into
the fact that $v\mapsto v-g(v)$ is non-decreasing, the maximisation of $\int g^2$
subject to $\int g = K$ (and $0\leq g(v)\leq v$) would be standard: to ``spread mass
as unenvenly as possible'', one would pick $g(v)=v\mathbb{I}_{v\geq v^*}$ with $\int_{v^*}^1 v \ud v=K$.
This is not feasible, however, because it violates the monotonicity constraint.
The $g_{\alpha}:v\mapsto v-X_{\alpha}(v)$ are then natural candidate maximisers, as they
are constructed in a similar spirit of variance maximisation, but respect the monotonicity constraint.

\

\noindent The $g_{\alpha}$ all have the same $L^2$ norm, but are away from zero over different
intervals. This hints at the fact that for a general function $g$, when trying to find a bound
on $\int g^2$, we will have no way to know where $g$ must be small or large, and therefore little grip
on $g$. The idea is then to consider the repartition function $\varphi$ of $g$, because (i) one can
reconstruct the moments of $g$ with those of $\varphi$ (see Step 3) and (ii) it does not matter
\emph{where} $g$ is large, only how often it is large. In fact, all the $g_{\alpha}$ have the same
repartition function, which suggests that this is the correct perspective to adopt.
 
\begin{figure}[H]
\begin{center}
\includegraphics[width=0.65\textwidth]{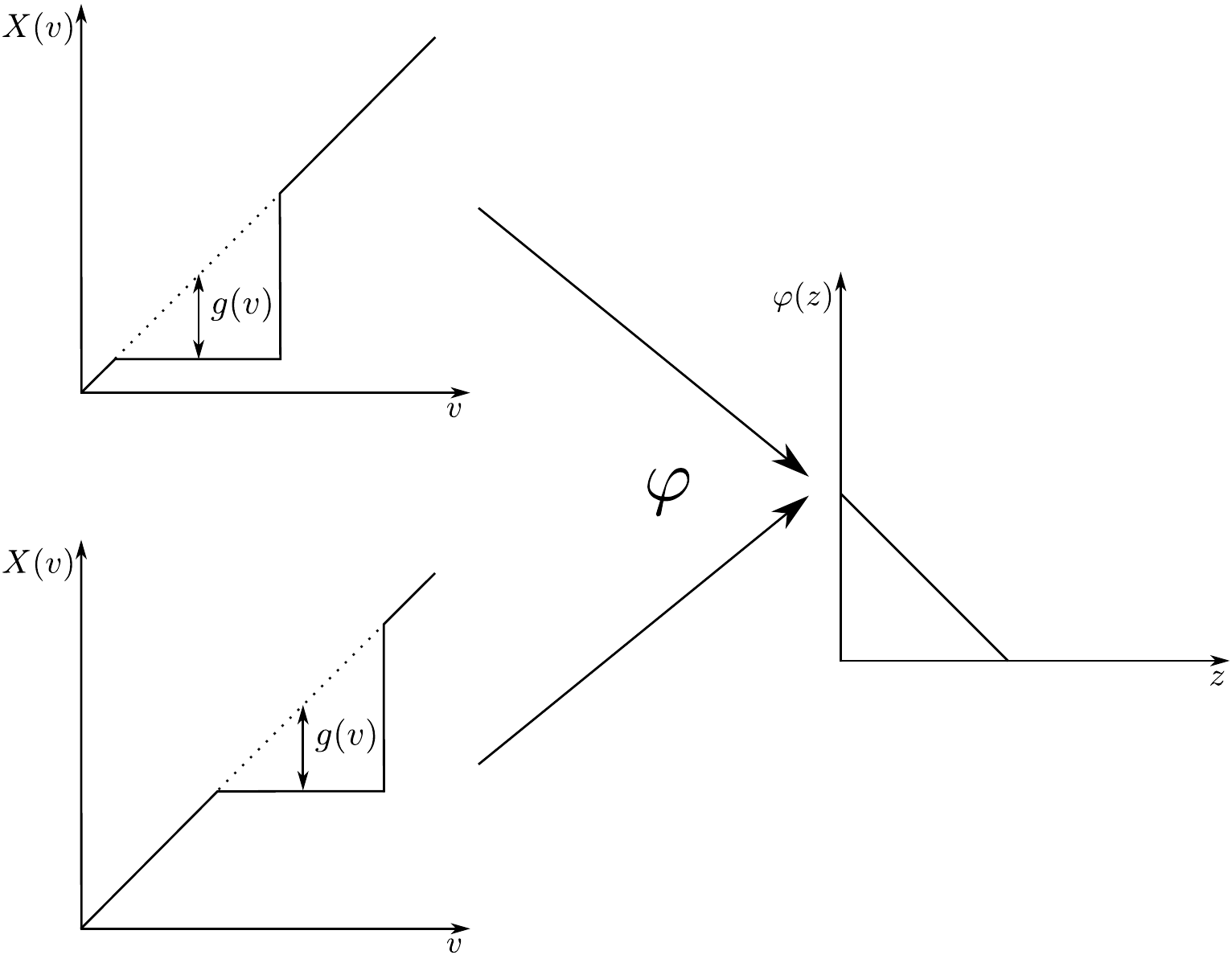}
\caption{Using the repartition function to transform $g$}
\label{fig:transformeerepartition}
\end{center}
\end{figure}

\

\noindent For any function $f$ and $x\neq y$, let
\begin{equation*}
\tau_{x,y}f = \frac{f(y)-f(x)}{y-x}.
\end{equation*}
Since $X$ is non-decreasing, we have
\begin{equation}
\label{eq:tauxbelowone}
\tau_{x,y}g\leq 1
\end{equation}
for all $x\neq y$. Now, define
\begin{equation*}
\varphi(z) = \mu\left(\left\{x, \, g(x)\geq z\right\}\right).
\end{equation*}

\noindent \textit{Step 2}: \eqref{eq:tauxbelowone} implies
\begin{equation}
\label{eq:tauxbelowminusone}
\tau_{x,y}\varphi\leq -1
\end{equation}
for all $x<y$ such that $\varphi(y)>0$.

\

\noindent $g$ is subject to a monotonicity constraint
(namely $v\mapsto v-g(v)$ must be non-decreasing), which we need
to transform into a constraint for $\varphi$. Clearly, if $g$ increases at speed 1,
$\varphi$ decreases at speed 1. What we show here is that if $g$ increases at speed less
than 1 then $\varphi$ decreases at speed larger than 1.

\

\noindent Since $y>0$, the set $\{u, \, g(u)\geq y \}$
is nonempty, so we can consider
\begin{equation*}
u^+ = \inf \{u, \, g(u)\geq y \}.
\end{equation*}
Since $g(0)=0\leq x$ we can also define
\begin{equation*}
u^- = \sup\{u \leq u^+, \, g(u)\leq x \}.
\end{equation*}
Because of \eqref{eq:tauxbelowone}, the function $g$ can not jump upwards,
hence $g(u^-)=x$ and $g(u^+)=y$. By construction of $u^-$ and $u^+$, we have:
\begin{align}
\label{eq:inclusionumoinsuplus}
[u^-,u^+)\subset \{u, \, g(u) \in [x,y) \}.
\end{align} 
Since $\tau_{u^-,u^+}g \leq 1$, we have:
\begin{align}
\label{eq:integaliteumoinsuplus}
 u^+-u^- \geq g(u^+)-g(u^-) = y-x, 
\end{align}
We can now obtain \eqref{eq:tauxbelowminusone}:
\begin{eqnarray*}
\tau_{x,y}\varphi&=&\frac{\mu\left(\left\{u, \, g(u)\geq y\right\}\right)-\mu\left(\left\{u, \, g(u)\geq x\right\}\right)}{y-x}\\
&=&-\frac{\mu\left(\left\{u, \, g(u)\in [x,y)\right\}\right)}{y-x}\\
&\leq&-\frac{\mu\left([u^-,u^+)\right)}{y-x}\\
&\leq& -1.
\end{eqnarray*}
Line 3 uses \eqref{eq:inclusionumoinsuplus} and Line 4 
is a consequence of \eqref{eq:integaliteumoinsuplus}.

\

\noindent \textit{Step 3}: expression of the moments of $g$ as a function of the moments of $\varphi$.

\

\noindent Recall that
\begin{eqnarray}
\int_0^1 g &=& \int_0^1 \varphi \nonumber\\
\int_0^1 g^2 &=& 2\int_0^1 y \varphi(y) \ud y. \label{eq:transfolineaire}
\end{eqnarray}

\

\noindent Indeed, 
\begin{eqnarray*}
\int_0^1 g^2(y) \ud y &=& \int_0^1\int_0^1 \mathbb{I}_{0\leq s\leq g^2(y)} \ud s \ud y\\
&=&\int_0^1 \mu\left(\left\{u, \, g^2(u)\geq s\right\}\right) \ud s\\
&=&\int_0^1 \mu\left(\left\{u, \, g(u)\geq \sqrt{s}\right\}\right) \ud s\\
&=&2\int_0^1 y \varphi(y) \ud y,
\end{eqnarray*}
by using the change of variable $y=\sqrt{s}$. 
The other equality in \eqref{eq:transfolineaire} is proven similarly.

\

\noindent \textit{Step 4}: translation into a functional maximisation 
problem with respect to the transform $\varphi$.

\

\noindent Using the previous discussion, 
\begin{eqnarray}
\label{eq:transformationlineaire}
\sup_{C\in\mathcal{C}_K} \int_0^1 (v-X(v))^2 \ud v &\leq& 2 \sup_{\varphi \in \Phi_K^{\leq}} \int_0^1 y \varphi(y) \ud y\nonumber\\
&\leq& 2 \sup_{\varphi \in \Phi_K} \int_0^1 y \varphi(y) \ud y
\end{eqnarray}
where $\Phi_K^{\leq}$ is the set of measurable functions  $\left\{ \varphi: [0,1]\to [0,1], \, \underset{x,y}{\sup} \, \tau_{x,y} \varphi \leq -1, \, \int_0^1\varphi(y)\ud y \leq K \right\}$
and $\Phi_K = \{ \varphi\in \Phi_K^{\leq},\int_0^1 \varphi = K\}.$
Clearly, in \eqref{eq:transformationlineaire} the right-hand-side of Line 1 equals the term in Line 2.

\

\noindent Define $\varphi_K(z)=\max\left\{ \sqrt{2K}-z,\, 0 \right\}$ for $0\leq z \leq 1$.
Note that $\varphi_K\in\Phi_K$.
If $\varphi\in\Phi_K$, $\varphi(0)\geq \varphi_K(0)$. Otherwise, using the fact that $\tau_{0,y} \varphi \leq -1$, 
\begin{align*}
\varphi(y) \leq \varphi(0)-y < \varphi_K(0) - y \leq \varphi_K(y).
\end{align*}Hence, $\int_0^1 \varphi(y) \ud y$ would be strictly less than $K=\int_0^1  \phi_K(y) \ud y$.  

\noindent Define $\Delta = \varphi - \varphi_K$: we proved that $\Delta (0) > 0$. Besides, by construction, $\int_0^1 \Delta (y) \ud y=0$. Define
\begin{align*}
y_0 = \inf\left \{y, \, \Delta(y)\leq 0 \right\}.
\end{align*}
Because $\tau_{y_0,y}\varphi\leq -1$, we have $\Delta(y)\leq 0$ 
for $y>y_0$ and $\Delta(y)\geq 0$ for  $y< y_0$. Hence: 
\begin{align*}
\int_0^1 y\varphi(y) \ud y -\int_0^1 y\varphi_K(y) \ud y &= \int_0^1 y \Delta(y) \ud y\\
&=\int_0^{y_0} y \Delta(y) \ud y + \int_{y_0}^1 y \Delta(y) \ud y \\
& \leq y_0  \int_0^{y_0} \Delta(y) \ud y + y_0 \int_{y_0}^1 \Delta(y) \ud y \leq 0. 
\end{align*}

\begin{figure}[H]
\begin{center}
\includegraphics[width=0.55\textwidth]{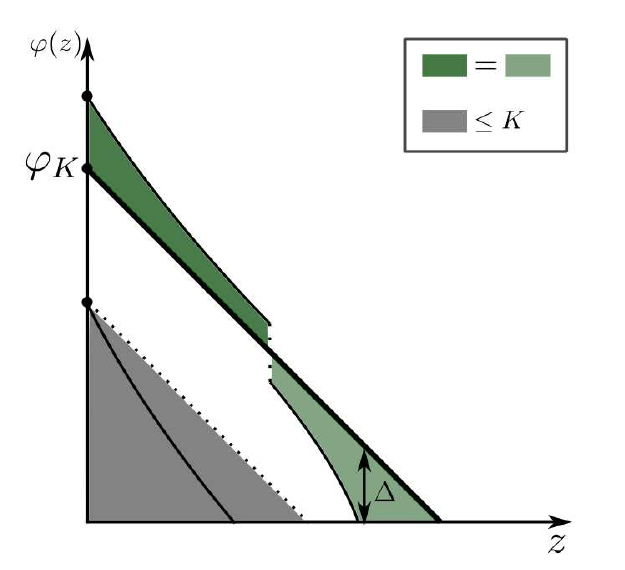}
\caption{The transform $\varphi$ of a maximiser $g$ must be $\varphi_K$.}
\label{fig:preuveairesphik}
\end{center}
(i) Starting from a point $\varphi(0)<\varphi_K(0)$ (lowest thick dot on the $y$-axis),
$\varphi$ (solid black curved line) remains below the dotted line and its integral is therefore
smaller than the area of the grey region, itself below $K$. (ii) After crossing $\varphi_K$,
$\varphi$ must remain below $\varphi_K$. Here, the crossing occurs through a downwards jump
of $\varphi$.
\end{figure}

\noindent Thus, the supremum in \eqref{eq:transformationlineaire} is attained only by the function $\varphi_K$ and equal to
\begin{eqnarray*}
2\int_0^1 y\varphi_K(y) \ud y &=& \int_0^{\sqrt{2K}}y(\sqrt{2K}-y) \ud y\\
&=&\frac{(2K)^{3/2}}{3},
\end{eqnarray*}
which establishes the bound of the Theorem. 

\

\noindent \textit{Step 5}: The maximum in \eqref{eq:maximisedistancel2} is attained exclusively by the demand schedules $\left( X_\alpha\right)_{\alpha \in [0,1-\sqrt{2K}]}$ defined in the Theorem. 

\

\noindent First, it is easy to see that these demand schedules achieve the maximum in \eqref{eq:maximisedistancel2}. It remains to show that they are the only one to do so. Let $X$ be a demand schedule obtained under a penalty $C \in \mathcal{C}$, $C\leq K$. Let us suppose that it achieves the maximum in \eqref{eq:maximisedistancel2}. Consider, as in step $2$, the function $\varphi$ associated with $g(v)=v-X(v)$. The function $\varphi$ is then a supremum of \eqref{eq:transformationlineaire} and by step $3$, $\varphi = \varphi_K$. Since 
\begin{align*}
\sup_x g(x) \geq \sup \{x , \, \varphi(x) > 0 \} = \sup \{x , \, \varphi_K(x) > 0 \}=\sqrt{2K},
\end{align*}
the supremum of $g(v)$ is at least $\sqrt{2K}$. Let us remark that: 
\begin{align*}
\sup_{v} g(v) = \sup_v \sup_{s \in [0,v]} g(s). 
\end{align*}
Since $\tau_{.,.}g \leq 1$, the function $\overline{g}(v)= \sup_{s \in [0,v]} g(s)$ is continuous: the supremum of $\overline{g}(v)$ and thus of ${g}(v)$ is attained at a point $v_0$. Since $\tau_{.,.}g \leq 1$, $v_0\geq \sqrt{2K}$ and for
$v\in [v_0-\sqrt{2K},v_0]$, $g(v)\geq v-v_0+\sqrt{2K}$. Since $g\geq 0$, we obtain
\begin{eqnarray*}
\int_0^1 g &\geq&\int_{v_0-\sqrt{2K}}^{v_0}g\\
&\geq&\int_{v_0-\sqrt{2K}}^{v_0}(v-v_0+\sqrt{2K}) \ud v\\
&\geq& K
\end{eqnarray*}
with equality if and only if $g=0$ outside $[v_0-\sqrt{2K},v_0]$
and $g(v)= v-v_0+\sqrt{2K}$ over $[v_0-\sqrt{2K},v_0]$.
But there must be equality because $g\in\Phi_K$. 
Hence $g$ has the above form, and the demand function $X$, given by
$X(v)=v-g(v)$, is equal to $X_{\alpha}$ as stated in the Theorem,
with $\alpha=v_0-\sqrt{2K}$.

\

\noindent \textit{Step 6}: It is easy to see that the demand schedules $X_\alpha$ are implemented by the penalties $C_\alpha$. This allows to conclude the proof of the Theorem.  

\

\noindent One consequence of Theorem \ref{theorem:efficientfrontiernonpecuniary}
is that it is not possible to infer from a regulator's choice of penalty whether she is
constrained or not. In the non-pecuniary case, a regulator subject to a binding budget constraint 
effectively behaves like an unconstrained regulator that would assign less
weight to curtailing the losses of the uninformed traders. In the next section,
we study the case of pecuniary penalties and show that, by contrast to the 
previous result, the introduction of the constraint creates new efficient points.
In theory, observing that the regulator has selected one of these points
would imply that she is constrained.

\subsection{Pecuniary penalties}
\label{sec:pecuniary}
\noindent We now consider pecuniary penalties,
collected by the regulator. For simplicity, we maintain the assumption
of a constant $\alpha$ and assume that a potential cap on $\tilde{C}$ does not bind.
We suppose that the regulator must have a balanced budget in expectation.
The budget constraint \eqref{eq:budgetalphakappa} transforms into
\begin{equation}
\label{eq:budgetpecuniary}
\alpha\kappa\leq B + \mathbb{E}[C(X(v))].
\end{equation}

\noindent If $B\geq\alpha\kappa$, since we assume that a potential cap
on $\tilde{C}$ is not binding, there is no constraint, and we are back to the case studied in section \ref{sec:solutionregulator}. The interesting case is therefore $B<\alpha\kappa$.

\

\begin{definition}
The efficient surface $\Sigma$ is the locus of points $(G,S,F)$
generated by any $C\in\mathcal{C}$ such that no $C'\in\mathcal{C}$ can weakly (i) increase $G$,
(ii) decrease $S$, (iii) increase $F$ with at least one among (i), (ii) or (iii) being in fact strictly. 

\

\noindent Recall that $G$, $S$ and $F$ denote respectively the P\,\& L of the uninformed traders,
the expected post-trade standard deviation and the expected collected fine. When convenient,
we use the notations $G(X)$, $S(X)$ or $F(X)$ to say that the quantities are implied by the demand
schedule~$X$.
\end{definition}

\subsubsection{Characterization of the efficient surface}

\noindent Let $J$ be the set of indices 
\begin{equation*}
J:=\left\{(x,y), \, 0\leq \frac{y}{1+y}\leq x\leq y\leq 1\right\}.
\end{equation*}
\begin{theorem}
\label{theorem:efficientsurface}
A parametric equation of the efficient surface $\Sigma$ in the space $(G,S,F)$ is 
\begin{equation*}
\left\{\left(\frac{1}{6}(v_1^2v_2-1);\frac{1}{\sqrt{3}}\left(\frac{2}{3}+\frac{1}{6}(v_1^2v_2+v_1v_2^2)\right);\frac{v_1v_2}{6}\left(3-2v_1-v_2\right)\right)\right\}_{(v_1,v_2)\in J}
\end{equation*}
and it is achieved exactly by the demand schedules
$\left(X_{v_1,v_2}\right)_{(v_1,v_2)\in J}$ where
\begin{equation*}
X_{v_1,v_2}(v) =
\begin{cases}
0 \quad \quad \quad \quad \quad \quad v\in [0,v_1] \\
\frac{v_2}{v_2-v_1}(v-v_1) \quad v\in (v_1,v_2] \\
v \quad \quad \quad \quad \quad  \quad v\in (v_2,1] \\
-X_{v_1,v_2}(-v) \quad \, v<0.
\end{cases}
\end{equation*}
These demand functions can be implemented by the penalties $(C_{v_1,v_2})_{(v_1,v_2)\in J}\in\mathcal{C}$ where
\begin{equation*}
C_{v_1,v_2}(x) =
\begin{cases}
v_1\lvert x\lvert - \frac{v_1}{2v_2}x^2 \quad \lvert x\lvert\leq v_2 \\
\frac{v_1v_2}{2} \quad \lvert x\lvert>v_2.
\end{cases}
\end{equation*}

\end{theorem}
\begin{figure}[H]
\begin{center}
\includegraphics[width=0.7\textwidth]{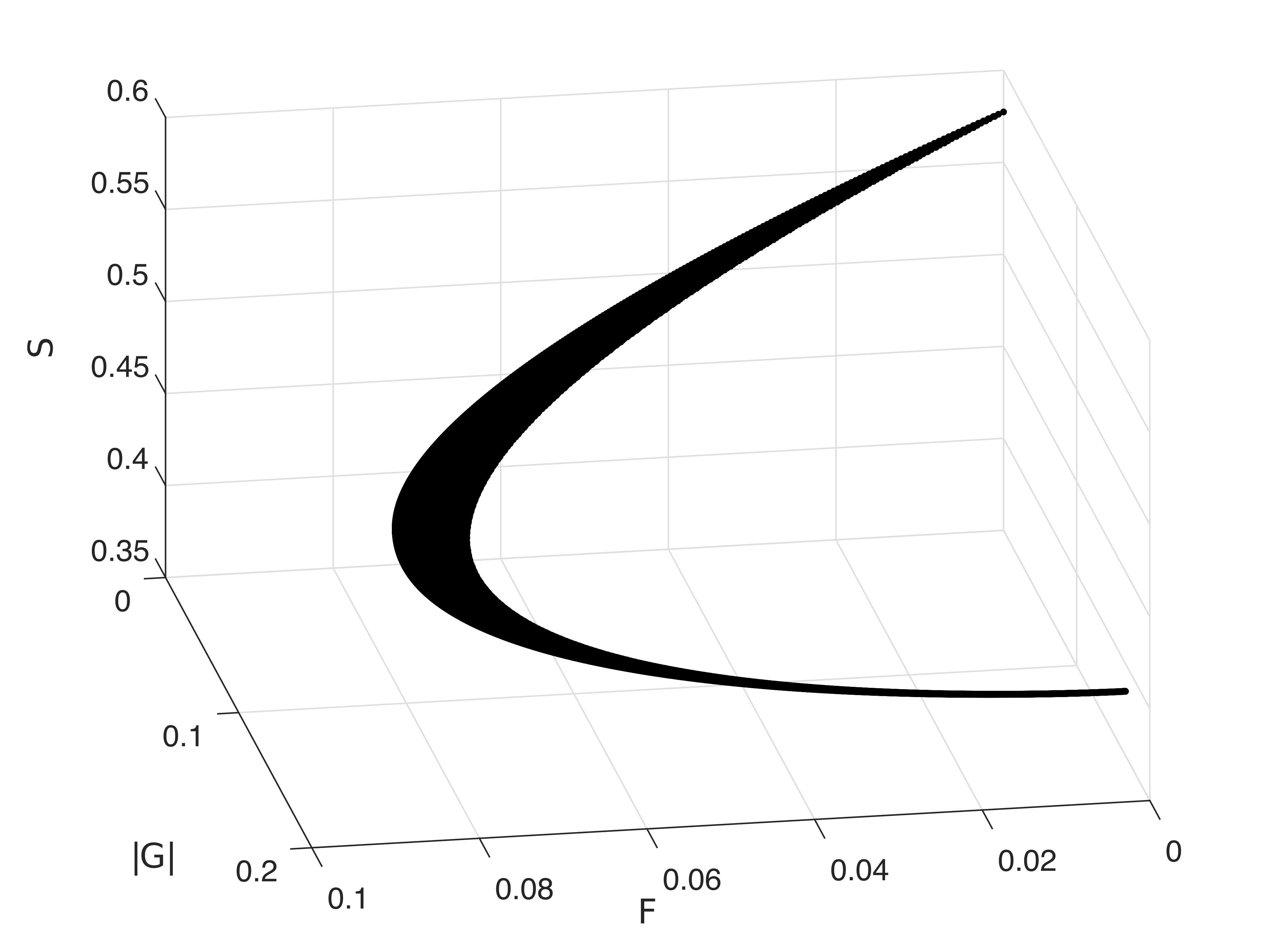}
\caption{The efficient surface $\Sigma$.}
\label{fig:surfaceefficiente}
\end{center}
\end{figure}
\begin{figure}[H]
\begin{center}
\includegraphics[width=0.9\textwidth]{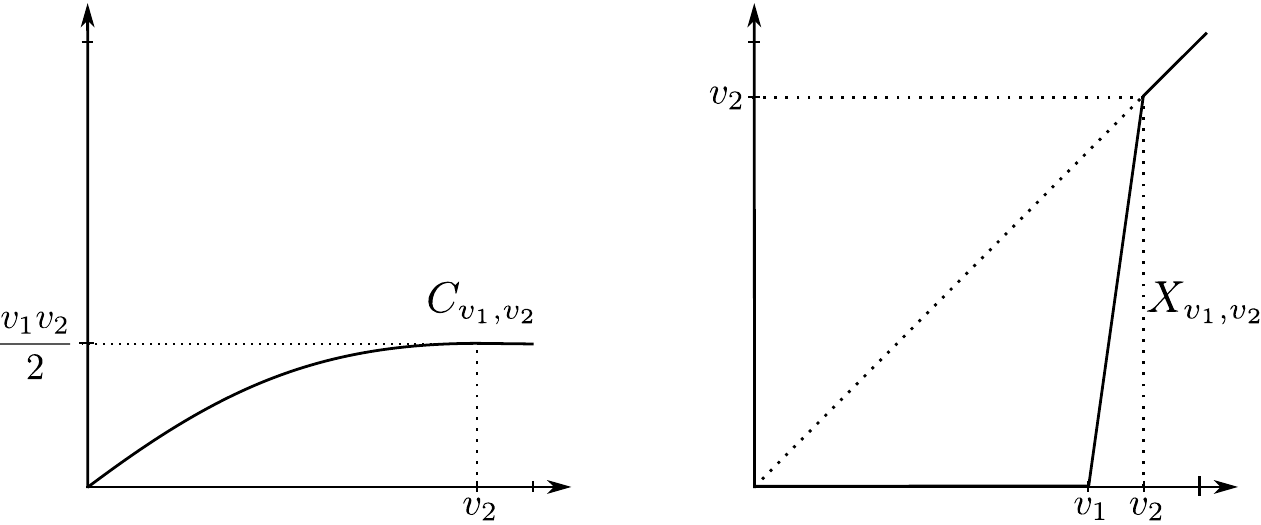}
\caption{Efficient demand schedule and penalty function under a budget constraint with pecuniary fines.}
\label{fig:optimalxcv1v2}
\end{center}
\end{figure}
\noindent There is a key difference in proving Theorem \ref{lemma:gmaxsouscontrainte}
and Theorem \ref{theorem:efficientsurface}. Here, the most natural candidate optimiser of the weighted objective,
i.e. the pointwise minimiser, turns out to be an implementable demand schedule. Since pointwise
minimisation is a simple task, the proof of  Theorem \ref{theorem:efficientsurface} is fairly straightforward.
Such an approach was not possible in proving Theorem \ref{lemma:gmaxsouscontrainte}.

\

\begin{proof}
As a consequence of Lemma \ref{lemma:envelope}, in equilibrium the expected fine satisfies
\begin{equation*}
\mathbb{E}[C(X(v))]=\int_0^1 X(v)\left(v-\frac{X(v)}{2}\right) \ud v-\int_0^1 (1-v)X(v) \ud v,
\end{equation*}
and we are working under a constraint $\mathbb{E}[C(X(v))]\geq K_1$. 

By Lemma \ref{lemma:expstd}, an upper bound constraint
on the expected post-trade standard deviation translates into
a constraint
\begin{equation*}
\int_0^1 vX(v) dv\geq K_2.
\end{equation*}

\

\noindent This leads us to consider the following minimisation problem:
\begin{eqnarray*}
\min_X && \int_0^1 X(v)\left(v-\frac{X(v)}{2}\right) \ud v+\gamma\left(K_1-\int_0^1 X(v)\left(v-\frac{X(v)}{2}\right) \ud v+\int_0^1 (1-v)X(v) \ud v\right)\\
&+&\eta\left(K_2-\int_0^1 vX(v)\ud v\right),
\end{eqnarray*}
for some weights $\gamma,\eta\geq 0$.
Gathering terms, we obtain that this program is equivalent to
\begin{equation}
\label{eq:minimisationlambdamu}
\min_X \int_0^1 X(v)\left(\gamma+(1-2\gamma-\eta)v+\frac{\gamma-1}{2}X(v)\right) \ud v
\end{equation}
For $0\leq v\leq 1$, define
\begin{eqnarray*}
P_v:[0,v]&&\rightarrow\mathbb{R}\\
x&&\mapsto x\left(\gamma+(1-2\gamma-\eta)v+\frac{\gamma-1}{2}x\right)
\end{eqnarray*}

\noindent\textit{Case 1:} $\gamma>1$. $P_v$ is the restriction to $[0,v]$
of a second-order polynomial with positive leading coefficient. 
Therefore it reaches its minimum at either 0, $v$, or when the first order condition is satisfied,
say at $x_0(v)$, and $x_0(v)$ achieves the minimum as soon as $0\leq x_0(v)\leq v$. Given that
\begin{equation*}
x_0(v) = \frac{(2\gamma+\eta-1)v-\gamma}{\gamma-1},
\end{equation*}
algebra shows that 
\begin{equation*}
\arg \max \, P_v =
\begin{cases}
0 \quad\quad \quad \quad v\leq \frac{\gamma}{2\gamma+\eta-1}\\
x_0(v) \quad \quad \frac{\gamma}{2\gamma+\eta-1}\leq v\leq\frac{\gamma}{\gamma+\eta}\\
v \quad \quad \quad \quad v>\frac{\gamma}{\gamma+\eta}.
\end{cases}
\end{equation*}
Let $v_1=\frac{\gamma}{2\gamma+\eta-1}$ and $v_2=\frac{\gamma}{\gamma+\eta}$. 
We have obtained that with the function $X_{v_1,v_2}$ given in the Theorem, the equality
\begin{equation*}
\arg \max P_v = X_{v_1,v_2}(v)
\end{equation*}
holds. Direct calculations show that $X_{v_1,v_2}$ is implemented by $C_{v_1,v_2}$. 
This means that we have found an implementable demand schedule that maximises the integral
in \eqref{eq:minimisationlambdamu} pointwise, which implies that $X_{v_1,v_2}$ is a 
minimiser of the program \eqref{eq:minimisationlambdamu}, and it is the only one because
the pointwise minimisation of the integral in \eqref{eq:minimisationlambdamu} has a unique solution.

\

\noindent\textit{Case 2:} $\gamma\leq 1$. $P_v$ is now either linear or with a negative
leading coefficient, meaning that its minimum is attained either at 0 or $v$. Algebra shows
that $\arg \max P_v=v$ (for $0\leq v\leq 1$) if and only if 
\begin{equation}
\label{eq:conditionlambdamu}
\gamma+2\eta\geq 1
\end{equation}
and
\begin{equation*}
v\geq v^*:=\frac{\gamma}{\eta+\frac{3\gamma}{2}-\frac{1}{2}},
\end{equation*}
where, by condition \eqref{eq:conditionlambdamu}, $v^*\in [0,1]$.
With $v_1=v_2=v^*$ we conclude as before that $X_{v_1,v_2}$ is the unique
minimiser of \eqref{eq:minimisationlambdamu}. Finally, if \eqref{eq:conditionlambdamu} is not satisfied,
the minimiser of \eqref{eq:minimisationlambdamu} is identically zero, which corresponds to $X_{1,1}$
defined in the Theorem.

\

\noindent Finally, it is easy to see that the $(v_1,v_2)$ constructed above
describe the set $J$ as
$\gamma,\eta\geq 0$ vary, and $J$ is the family of indices specified in the Theorem.
So any index in $J$ corresponds
to an efficient demand function. This shows that $\left(X_{v_1,v_2}\right)_{(v_1,v_2)\in J}$
is the family of efficient demand functions.

\

\noindent The proof is complete, because the set of maxima we obtain as
$\gamma,\eta\geq 0$ vary is connected, which implies that we have found all the
points of the efficient surface.
\end{proof}

\subsubsection{Efficient $(G,S)$ frontiers for various regulator's budgets}
\noindent Assume that the regulator has budget $B$, which translates into a constraint
\begin{equation*}
F = \mathbb{E}[C(X(v))]\geq F_{\text{min}}:=\alpha\kappa-B.
\end{equation*}
\begin{definition}
\label{definition:frontierepecuniaire}
\noindent The $F_{\text{min}}$-efficient frontier is the set of non-dominated points in
\begin{equation*}
\mathcal{F}(F_{\text{min}}):=\left\{(G(X),S(X)), \, X \text{implemented by some} \,\, C\in\mathcal{C} \, \text{with} \,\, \mathbb{E}[C(X(v))]\geq F_{\text{min}}\right\}.
\end{equation*}
\end{definition}

\noindent We can now construct the $F_{\text{min}}$-efficient frontiers from the 
efficient surface $\Sigma$ (denote $\pi_{GS}:(G,S,F)\mapsto (G,S)$ the projection on the $(G,S)$-plane):

\begin{lemma}
\label{lemma:projectionSigma}
The $F_{\text{min}}$-efficient frontier is the set of points of $\pi_{GS}\left(\Sigma\cap\{F\geq F_{\text{min}}\}\right)$ that
are not dominated in $\pi_{GS}\left(\Sigma\cap\{F\geq F_{\text{min}}\}\right)$.
\end{lemma}

\begin{proof}
See Appendix \ref{app:proofs}.
\end{proof}

\

\noindent This means that to obtain the $F_{\text{min}}$-efficient frontier,
one must first project the relevant points $(G,S,F)$ of $\Sigma$, and then
select those that are efficient in the plane (note that this second step is necessary,
as the projection of a point of the efficient surface will in general not be a point 
of the efficient frontier). $\Sigma$ was found by solving an optimization problem,
from which the $F_{\text{min}}$-efficient frontiers are deduced geometrically: we do not
need to solve again a minimisation problem.

\

\begin{figure}[H]
\begin{center}
\includegraphics[width=0.8\textwidth]{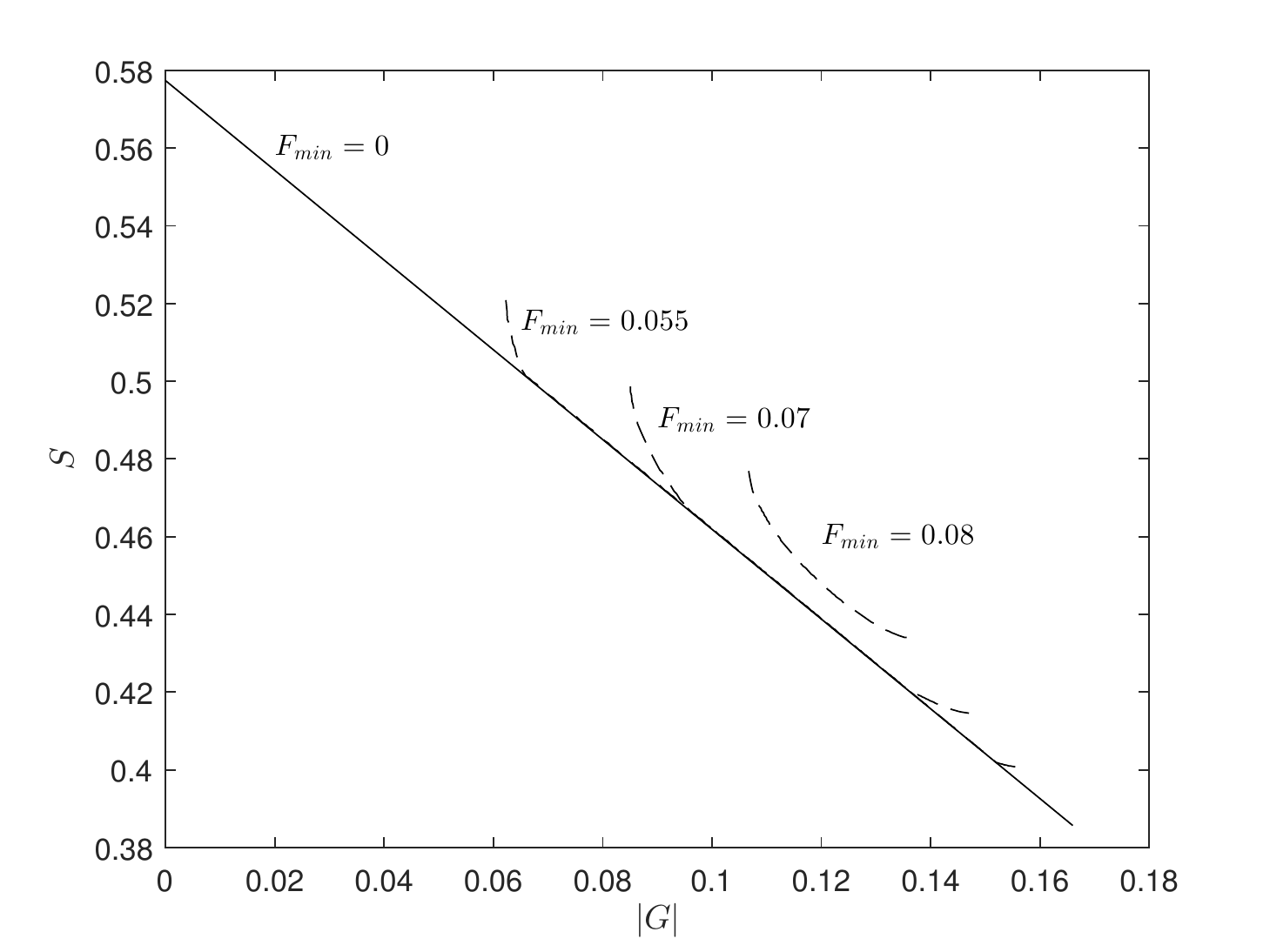}
\caption{Efficient $(\vert G\vert,S)$ frontiers under various constraints $F\geq F_{\text{min}}$.}
\label{fig:frontieresGSsousFgeqFmin}
\end{center}
\end{figure}

\noindent We are now in a position to provide the $F_{\text{min}}$-efficient frontiers:
see Figure \ref{fig:frontieresGSsousFgeqFmin}.

\noindent An important difference emerges with respect to the case
of non-pecuniary penalties studied above. Here, the efficient frontiers
are not truncatures of the frontier that obtains absent the constraint.
Of course, the penalties in $\mathcal{O}$ that implement $F\geq F_{\text{min}}$
are still part of the $F_{\text{min}}$-efficient frontier, but new constrained efficient
points emerge (dotted arcs), which are associated with penalty functions and demand
schedules that were not previously optimal. The $F_{\text{min}}$ efficient
frontier does not even intersect the unconstrained frontier for $F_{\text{min}}$ very large.
To see why, note that the maximal expected fine under a penalty in $\mathcal{O}$ is
\begin{eqnarray*}
\max \{F(X), \, X \, \text{implemented by} \, C\in\mathcal{O}\}&=&\max_{0\leq K\leq 1/2} K\left(1-\sqrt{2K}\right)\\
&=& \frac{2}{27}\approx 0.074, \, \text{attained by} \, K=\frac{2}{9}.
\end{eqnarray*}

\noindent This means that if $F_{\text{min}}>\frac{2}{27}$, no penalty in $\mathcal{O}$ allows
to balance the regulator's budget. In fact, a penalty that provides the highest expected fine
(regardless of $S$ and $G$) is $C_{\frac{1}{2},1}$ (defined in Theorem \ref{theorem:efficientsurface}),
and it gives $F=\frac{1}{12}$.

\

\begin{figure}[H]
\begin{center}
\includegraphics[width=0.8\textwidth]{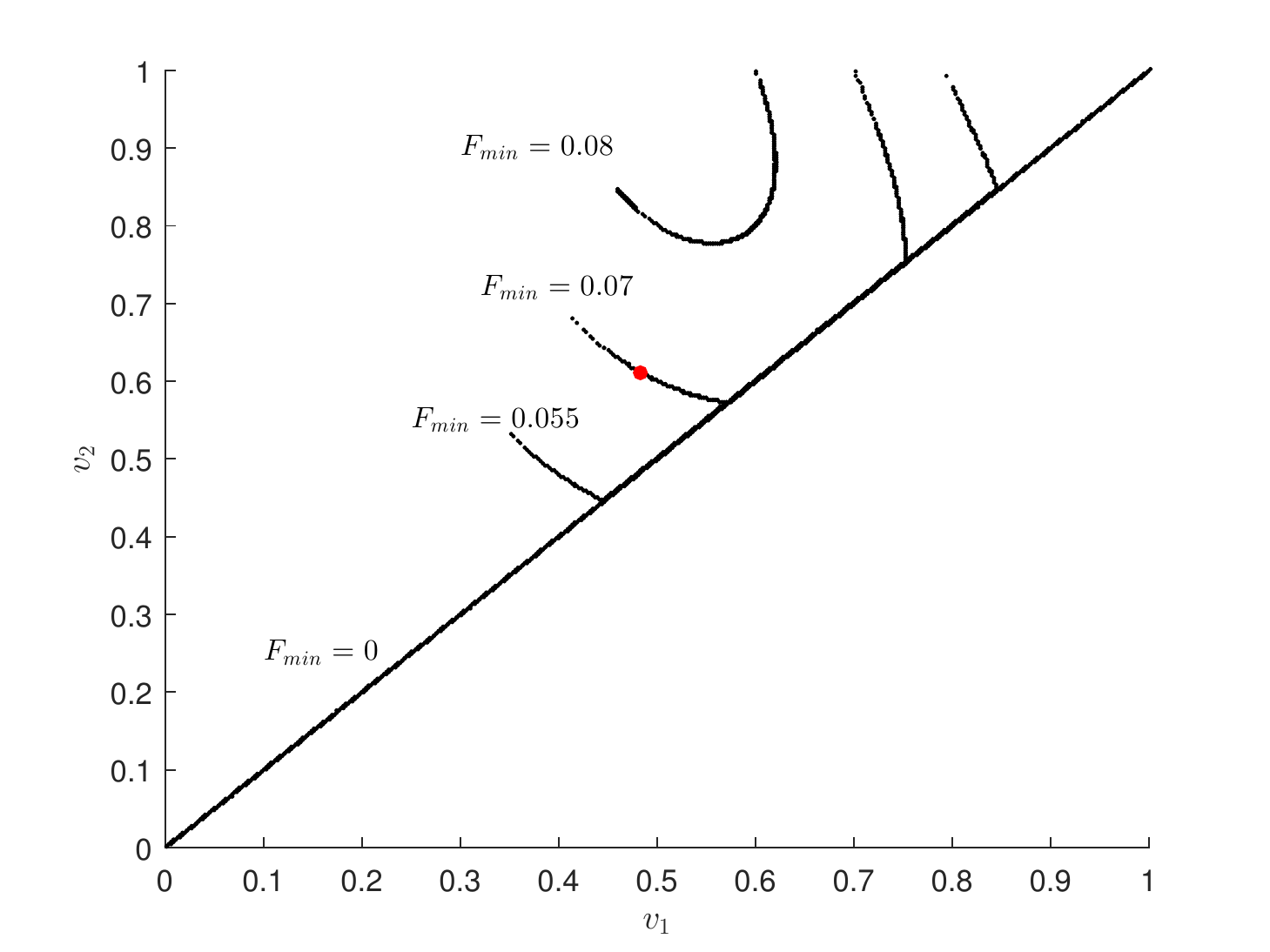}
\caption{Indices $(v_1,v_2)$ of the efficient demand functions $X_{v_1,v_2}$ associated with various constraints
$F\geq F_{\text{min}}$.}
\label{fig:v1v2Fmin}
\end{center}
As an illustration, the red filled dot, which corresponds to $(v_1,v_2)\approx (0.48,0.61)$ represents
the demand schedule $X_{0.48,0.61}$ (where $X_{v_1,v_2}$ is defined in Theorem \ref{theorem:efficientsurface})
and indicates that this demand schedule implements one point of the efficient frontier when the budget
constraint of the regulator is such that $F_{\text{min}} = 0.07$.
\end{figure}

\noindent From Lemma \ref{lemma:projectionSigma}, we know that points of the
$F_{\text{min}}$-efficient frontier correspond to points in $\Sigma$, which means
that they are associated with demand schedules of the form $X_{v_1,v_2}$ defined
in Theorem \ref{theorem:efficientsurface}. To understand how the budget constraint $F\geq F_{\text{min}}$
modifies the nature of the optimal strategies, Figure \ref{fig:v1v2Fmin} plots the
$(v_1,v_2)$ used on the $F_{\text{min}}$-efficient frontier for various values of $F_{\text{min}}$.

\noindent When $F_{\text{min}}=0$, we obtain the line $v_2=v_1$, in which case $X_{v_1,v_2}$
is implemented by a penalty $C\in\mathcal{O}$, consistent with section \ref{sec:implementationefficientdemand}. We observe that as
$F_{\text{min}}$ increases, one needs to widen the gap $v_2-v_1$. The intuition is that the linear
section over $[v_1,v_2]$ of the demand schedule $X_{v_1,v_2}$ best resolves the trade off between large fines and
large trade volumes of the insider trader and allows to collect a relatively high amount of fines in expectation.
As an example, recall that the demand schedule that implies the highest expected fine (1/12) had $v_2-v_1=\frac{1}{2}$.

\

\noindent When the regulator must balance its budget through the collection of pecuniary fines,
some previously optimal strategies are no longer feasible as they do not induce the insider trader
to pay enough fines in expectation. New constrained efficient points appear, and the class
of efficient penalties is modified, as well as the equilibrium demand schedules and price functions.

\begin{figure}[H]
\begin{center}
\includegraphics[width=\textwidth]{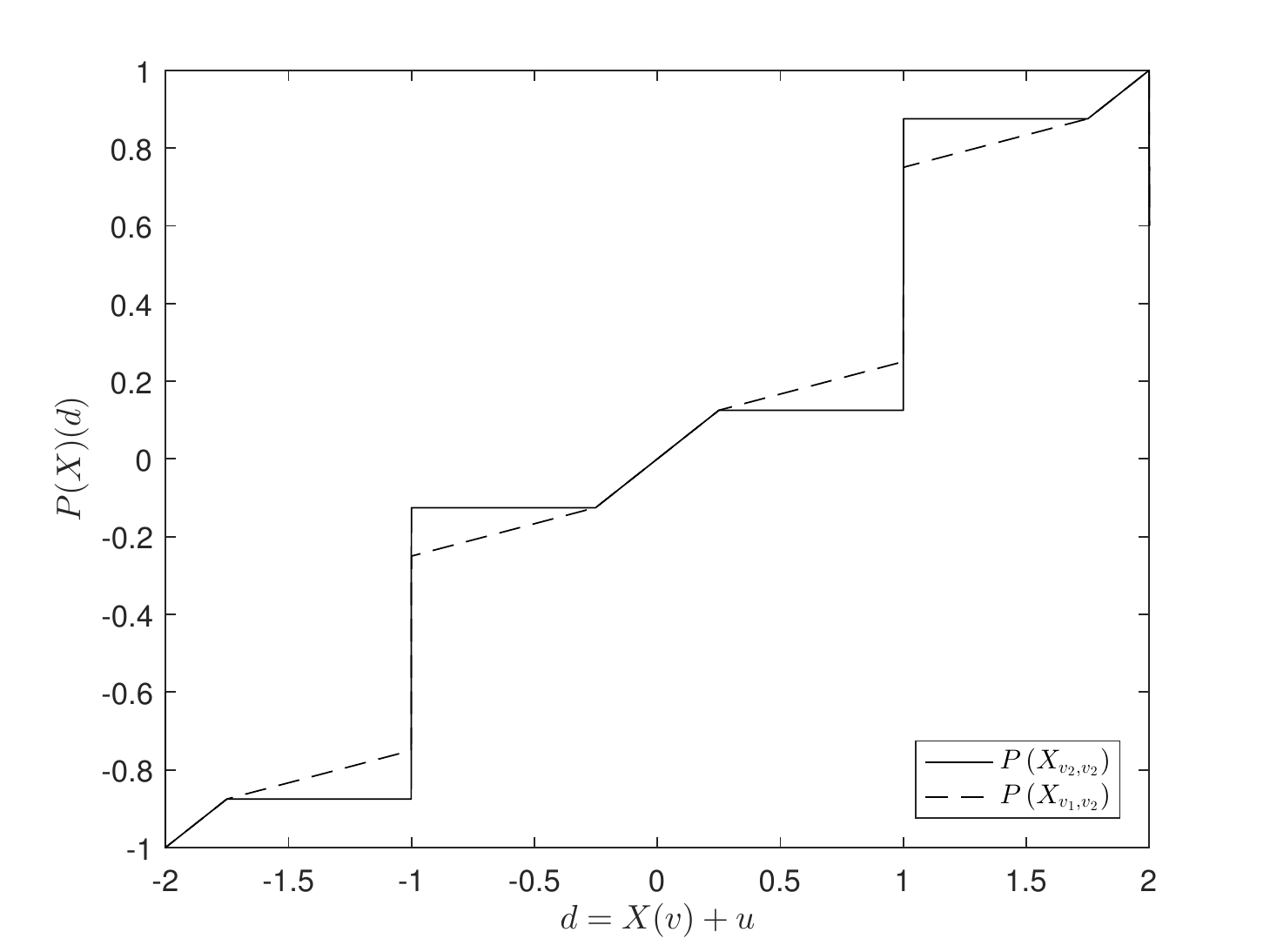}
\caption{New patterns of price functions.}
\label{fig:pricesv1v2}
$v_1=0.5$ and $v_2=0.75$.
\end{center}
\end{figure}
\noindent Figure \ref{fig:pricesv1v2} compares the price functions implied by
a demand schedule efficient absent a budget constraint, $X_{v_2,v_2}$ and a
constrained efficient demand schedule $X_{v_1,v_2}$.
Contrary to $P\left(X_{v_2,v_2}\right)$, $P\left(X_{v_1,v_2}\right)$ has no flat sections and is
everywhere increasing. In particular, in the unconstrained case, the random price is partly discrete: with positive probability, it will be equal to one of the ordinates of the flat sections of $P(X_{v_2,v_2})$. Conversly, in the case of a strong budget constraint, the random price has a continuous density.

\bibliography{insider}

\appendix

\section{Additional Proofs}
\label{app:proofs}

\subsection{Normalization of supports to $[-1,1]$}
\label{sec:normalizationsupports}
\noindent Assume $u\sim U(-a,a)$ and $v\sim U(b,c)$ with $a>0$ and $b<c$.
We want to map an equilibrium with these noise terms and penalty $C$ to an equilibrium
with normalized noises. Let $C^0(x^0)=\frac{1}{\sigma}C(ax^0)$ for $-1\leq x^0\leq 1$.
$C^0$ defines a penalty in $\mathcal{C}$.

\

\noindent Let $(X^0,P^0)$ be an equilibrium of $\mathcal{K}(C^0)$ under uniform noises
distributed over $[-1,1]$, and admissible demands $I^0=[-1,1]$.
Let $\Phi$ be the linear application mapping $[b,c]$ to $[-1,1]$:
\begin{equation*}
\Phi(v) = \frac{2}{c-b}v-\frac{c+b}{c-b}
\end{equation*}
Similar to Lemma \ref{lemma:cs}, the expected price function must be $\hat{P}(x)=m+\frac{\sigma x}{2a}$ where
\begin{eqnarray*}
m &=& \frac{b+c}{2},\\
\sigma &=& \frac{c-b}{2}.
\end{eqnarray*}
For any $v\in [b,c]$, the maximisation program of the IT is
\begin{equation*}
\max_{x\in [-a,a]} x\left(v-m- \frac{\sigma}{2a}x\right)-C(x).
\end{equation*}
This can be rewritten as
\begin{equation*}
\max_{x^0\in [-1,1]} (ax^0)\left(v-m- \frac{\sigma}{2a}(ax^0)\right)-C(ax^0).
\end{equation*}
Or:
\begin{equation*}
(a\sigma) \max_{x^0\in [-1,1]} x^0\left(v/\sigma-m/\sigma- \frac{x^0}{2}\right)-C^0(x^0),
\end{equation*}
By definition of $X^0$, the solution of this program is given by $X^0\left(\frac{v-m}{\sigma}\right)
=X^0\left(\Phi(v)\right)$.
Recalling that the actual demand of the IT is $x=ax^0$, we obtain 
\begin{equation*}
X(v) = a X^0\left(\frac{v-m}{\sigma}\right)=aX^0\left(\Phi(v)\right).
\end{equation*}
We can also express the price function using $P^0$. Since $\Phi$ is linear, we can write
\begin{eqnarray*}
P(d) &=& \mathbb{E}[v\lvert d]\\
&=&\Phi^{-1}\left(\mathbb{E}\left[\Phi(v)\lvert X(v)+u=d\right]\right)\\
&=&\Phi^{-1}\left(\mathbb{E}\left[\Phi(v)\lvert aX^0(\Phi(v))+a(u/a)=d\right]\right)\\
&=&\Phi^{-1}\left(\mathbb{E}\left[v^0\lvert X^0(v^0)+ u^0=d/a\right]\right)\\
&=&\Phi^{-1}\left(P^0(d/a)\right),
\end{eqnarray*}
because $v^0=\Phi(v)$ and $u^0=u/a$ are independent $U(-1,1)$ variables.

\

\noindent So the equilibrium with noises $u\sim U(-a,a)$ and $v\sim U(b,c)$, penalty $C$ and 
admissible demands $I=[-a,a]$ can be mapped to the equilibrium of $\mathcal{K}(C^0)$
with normalized noises and admissible demands $I^0=[-1,1]$. By the 
same procedure, one can do the reverse mapping.

Absent penalties, $X(v)=\frac{2a}{c-b}v-a\frac{c+b}{c-b}$ and the profit at $v$ is given by
\begin{equation*}
X(v)\left(v-\frac{b+c}{2}-\frac{c-b}{4a}X(v)\right)=\frac{a}{c-b}\left(v-\frac{c+b}{2}\right)^2.
\end{equation*}
This recoups a cost $K$ as soon as $v$ is outside
\begin{equation*}
\frac{b+c}{2}\pm\sqrt{\frac{c-b}{a}K},
\end{equation*}
and is maximal when $v\in\{b,c\}$, where it equals $\frac{1}{4}a(c-b)$.

Finally, we note that the model quantities of interest ($S$, $G$ and $F$) are
mapped one-to-one and ranked identically regardless of the chosen supports,
\emph{i.e.} the assertions ``$S<S'$'',``$G<G'$'' or ``$F<F'$'' do not depend on which supports we consider.
Therefore, the choice of $[-1,1]$ as the support of the noises
is without loss of generality once we assume uniform distributions and a centered uninformed traders' demand.

\subsection{Proposition \ref{prop:quadraticareworst}}
\noindent Using Lemma \ref{lemma:expstd}, we can write
\begin{eqnarray}
\label{eq:step1quadratic}
-G &=& \int_0^1 X(v)\left(v-\frac{X(v)}{2}\right) \ud v\nonumber\\
&=& 1-\sqrt{3}S-\frac{1}{2}\int_0^1 X(v)^2 \ud v.
\end{eqnarray}
By Cauchy-Schwarz inequality
\begin{eqnarray}
\label{eq:cauchyschwarz}
\left(\int_0^1 vX(v)dv\right)^2 &\leq&\int_0^1 v^2 dv\int_0^1 X(v)^2 \ud v\\
&\leq&\frac{1}{3}\int_0^1 X(v)^2 \ud v\nonumber\\
-\frac{1}{2}\int_0^1 X(v)^2 \ud v&\leq& -\frac{3}{2}\left(\int_0^1 vX(v)\right)^2 \ud v=-\frac{3}{2}(1-\sqrt{3}S)^2.\nonumber
\end{eqnarray}
Plugging this into \eqref{eq:step1quadratic}, we obtain
\begin{equation}
\label{eq:ineqgs}
G\geq \sqrt{3}S-1+\frac{3}{2}(1-\sqrt{3}S)^2.
\end{equation}
This inequality determines the highest possible $S$ given $G$.
But there is equality in \eqref{eq:ineqgs} if and only if there is equality in the Cauchy-Schwarz
bound \eqref{eq:cauchyschwarz}. This is the case if and only if the two functions in the left-hand side
are colinear, i.e. if $X(v)$ is proportional to $v$: $X(v)=\beta v$. Since $0\leq X(v)\leq 1$ for $0\leq v\leq 1$,
$\beta\in [0;1]$. We conclude by noting that if $\beta\in [0;1]$ and $\gamma\in [0;\infty]$ is defined by
$\gamma=\frac{1}{2\beta}-\frac{1}{2}$, the quadratic penalty $C(x)=\gamma x^2$ implements $X(v)=\beta v$.

\subsection{Lemma \ref{lemma:implementation}}
\noindent We first need to introduce some definitions:

\

\noindent\textit{
Let $f$ be a function defined over $[0,1]$ and $x \in [0,1]$. We define: }
\begin{eqnarray*}
\overline{D}^-f(x) &=& \limsup_{x'\nearrow x} \, \frac{f(x')-f(x)}{x'-x},  \\
\underline{D}^-f(x) &=& \liminf_{x'\nearrow x} \, \frac{f(x')-f(x)}{x'-x},
\end{eqnarray*}
One can define similarly $\overline{D}^+f(x) $ and $\underline{D}^+f(x) $.
Let us recall the first order conditions satisfied by a function at a local maximum. 

\

\noindent\textit{If $x^*$ is a local maximum of $f$, then: }
\begin{align*}
\overline{D}^+ f(x^*) &\leq 0, \\
\underline{D}^- f(x^*) &\geq 0
\end{align*}

\noindent We will also use the following real analysis result:

\begin{lemma}
\label{lemma:leftderiv}
Any continuous function $f$ on $]0,1]$ with a null left derivative is contant. 
\end{lemma}

\noindent Let $C$ be a penalty function such that the strategy of the IT satisfies that for any $v \in [0,1]$, $X(v)$ is either $0$ or $v$. Since the strategy of the IT is non-decreasing, there exists $v_0$ such that $X(v) = 0$ for any $v \in [0,v_0[$ and $X(v) = v$ for any $v \in ]v_0,1]$.

Besides, the penalty function $C$ must be continuous on $]v_0,1]$. Indeed, if $v' > v \geq v_0$, using the fact that $X(v') = v'$, 
\begin{align*}
v \left(v'-\frac{v}{2}\right) - C(v) \leq v' \left(v'-\frac{v'}{2}\right) - C(v'), 
\end{align*}
thus, since $C$ is non-decreasing, 
\begin{align*}
0 \leq {C(v')-C(v)} \leq v'\left(v'-\frac{v'}{2}\right)- v\left(v'-\frac{v}{2}\right). 
\end{align*}
Taking the limit as $v'$ goes to $v$, we see that $C$ is right continuous at $v$. Since by hypothesis it is left continuous on $[0,1]$, the penalty function $C$ is continuous on $]v_0,1]$. 
 
Let us show that $C$ has a null left derivative on $]v_0, 1]$. If $v \in ]v_0,1]$, we know that 
$v$ is a profit maximiser at $v$: $v \in \arg \max_x f_v(x)$:.
Using the first order condition for the lower left derivative $\underline{D}^-$ recalled above,
at $v$, $\underline{D}^- f_v(v) \geq 0$. Since $\underline{D}^- f_v(v) = - \overline{D}^- C(v)$, we obtain $\overline{D}^- C(v) \leq 0$. Yet, $C$ is increasing, so the lower and upper left derivatives must be positive : $0 \leq \underline{D}^- C(v) \leq \overline{D}^- C(v)$. Thus: 
\begin{align*}
\underline{D}^- C(v) = \overline{D}^- C(v) = 0. 
\end{align*}

This means that the cost function $C$ admits a left derivative at any $v \in ]v_0,1]$,
and the value of this left derivative is zero.

Thus $C$ is continuous and has a null left derivative on $]v_0,1]$. Using Lemma \ref{lemma:leftderiv}, we obtain that $C$ is constant on $]v_0,1]$. Let us denote by $K$ the value of $C$ on this interval.

The IT does not trade for $v\in [0,v_0)$. In that case, since we know that $0\leq X(v)\leq v$,
we must have
\begin{equation*}
\forall \, x \, \in [0,v], \quad x\left(v-\frac{x}{2}\right)\leq C(x).
\end{equation*}
By continuity of the left-hand term and the fact that the right-hand term is non-decreasing, we obtain
\begin{equation*}
\forall \, x \, \in [0,v_0], \quad x\left(v_0-\frac{x}{2}\right)\leq C(x).
\end{equation*}
There must be equality for $x=v_0$, because otherwise it would not be optimal
to select $X(v)=v$ on the right neighborhood of $v_0$. For the same reason,
$C$ can not jump at $v_0$. This implies that $v_0\left(v_0-\frac{v_0}{2}\right)=K$,
or $v_0=\sqrt{2K}$ and therefore $C$ must belong to $\mathcal{O}$. 

\

\noindent Assume conversely that $C\in\mathcal{O}$. Then for $0\leq v < v_0$, the insider trader
will make negative expected profits if she trades, so that $X(v)=0$. For $v>v_0$, 
there are two cases to consider. (i) The IT plays $x\geq v_0$. In that case, the expected
penalty $K$ appears as a sunk cost and the best choice is $x=v$, leading to a net profit
of $\frac{v^2}{2}-K$. (ii) The IT plays $x\in [0,v)$. The net profit is then
\begin{eqnarray*}
x\left(v-\frac{x}{2}\right)-C(x)&=& x\left(v_0-\frac{x}{2}\right)-C(x)+x(v-v_0)\\
&\leq& x(v-v_0)\\
&\leq& v_0(v-v_0)
\end{eqnarray*}
where the second line uses the fact that $C\in\mathcal{O}$.
Since
\begin{eqnarray*}
\frac{v^2}{2}-K&=&\frac{v^2}{2}-\frac{v_0^2}{2}\\
&=&\frac{1}{2}(v+v_0)(v-v_0)\\
&>&v_0(v-v_0),
\end{eqnarray*}
choice (i) is always preferred. Hence, if $C\in\mathcal{O}$, $X(v)=0$ for $\lvert v\lvert< v_0$
and $X(v)=v$ for $\lvert v\lvert > v_0$, which concludes the proof.

\subsection{Lemma \ref{lemma:projectionSigma}}
\noindent (i) We first show that 
the $F_{\text{min}}$-efficient frontier is included in the set of points of $$\pi_{GS}\left(\Sigma\cap\{F\geq F_{\text{min}}\}\right)$$ that
are not dominated in $\pi_{GS}\left(\Sigma\cap\{F\geq F_{\text{min}}\}\right)$.

\

\noindent Let $(G,S)$ be in the $F_{\text{min}}$-efficient frontier. 
By definition, there is $X$ implemented by $C\in\mathcal{C}$ such that
$G=G(X)$, $S=S(X)$ and $F:=\mathbb{E}[C(X(v))]\geq F_{\text{min}}$.
Only two cases are possible: (a) $(G,S,F)\in\Sigma$ or (b) $(G,S,F)$ is dominated
by a point $(G',S',F')$ non-dominated in the closure (in $\mathbb{R}^3$) of all the implementable points, which is exactly $\Sigma$. ($(G',S',F')$ is obtained by constructing a sequence $(G_n,S_n,F_n)$ where each point
dominates the previous one and define $(G',S',F')$ as its limit, or as $(G_N,S_N,F_N)$ if the procedure
stops at $N$.) In case (a), we see that $(G,S)=\pi_{GS}(G,S,F)\in\pi_{GS}(\Sigma\cap\{F\geq F_{\text{min}}\})$,
and since it is in the $F_{\text{min}}$-efficient frontier, it cannot be dominated in that space. In case (b),
since $(G,S)$ is in the $F_{\text{min}}$-efficient frontier, we must have $G=G'$ and $S=S'$ and $F'\geq F_{\text{min}}$,
so  $(G,S)=\pi_{GS}(G',S',F')\in\pi_{GS}(\Sigma\cap\{F\geq F_{\text{min}}\})$ and we conclude as in case (a).

\

\noindent (ii) Let us show the other inclusion. It is enough to prove that if a point is dominated in $\mathcal{F}(F_{\min})$, it is dominated in $\pi_{GS}\left(\Sigma\cap\{F\geq F_{\text{min}}\}\right)$.

\

\noindent  Let $(G,S)\in\pi_{GS}(\Sigma\cap\{F\geq F_{\text{min}}\})$ and $F$ be
associated with this point. Assume $(G,S)$ is dominated in $\mathcal{F}(F_{\text{min}})$,
say by $(G',S')$, associated with $F'\geq F_{\text{min}}$. As before, either $(G',S',F')\in\Sigma$ or
$(G',S',F')$ is dominated by a point in $\Sigma$. In both cases, this means that there exists a point
in $\Sigma\cap\{F\geq F_{\text{min}}\}$ whose projection dominates $(G,S)$. This concludes the proof.

\section{Robustness checks: the case of Gaussian noise}
\label{sec:robustnesschecks}
\subsection{Shape of $X$ and $P$ under Gaussian noise}
\label{sec:robustnessxp}
\noindent Which effects of section \ref{sec:examplesequilibria} are peculiar to uniform noises and which effects
are robust to other distributional assumptions?

The qualitative behaviour of the demand function $X$ does not depend
on the distribution of the noise. Consider for
instance a cost $C(x)=K\mathbb{I}_{\lvert x\lvert > x_0}$ with $K, x_0>0$.
When the magnitude of the optimal demand absent penalties is below $x_0$, 
it remains optimal under the penalty $C$. The IT then blocks its demand at $x_0$ in order to avoid
the expected penalty $K$, as long as trading does not allow to recoup $K$ on average. For $v$
sufficiently large ($\lvert v\lvert >v_0$ for some $v_0>0$), the IT switches back to trading.
This creates a jump in the demand function at $\pm v_0$. All these effects are independent of
the assumptions on the noise.

The qualitative behaviour of the price function is robust as far as non-linearity
is concerned. Flat sections in the demand schedule $X$ induce steep sections in the
price function $P$. Indeed, when $X$ increases slowly as a function of $v$, the information
that $X$ is likely to have increased a little (obtained through the observation of $d=X(v)+u$) implies that
$v$ is likely to have increased a lot. Similarly, steep sections of $X$ induce flat sections of $P$.
Since the introduction of penalties produces steep and flat sections for $X$, it produces flat and
steep sections for $P$.

What does not hold in general is the fact that $P$ has discontinuities.
Those are due to the fact
that the uniform distribution has a discontinuous density $\frac{1}{2}\mathbb{I}_{[-1,1]}$.
In general, one must have discontinuities in the density of the noise to obtain
discontinuities in the price function. With a continuous noise density, jumps are
replaced with sections where $P$ increases fast.

To support these arguments, we report the equilibrium $(X,P)$ for 
the model with Gaussian noise ($u,v\sim N(0,1)$) and penalty $C$.\footnote{We are not able to prove formally
existence (and even less uniqueness) in the case of Gaussian noise. What we do
is run a fixed-point algorithm on equations \eqref{eq:maximisationIT} and \eqref{eq:breakevenMM}
and assume that the functions to which it converges indeed correspond to an exact 
equilibrium.} 
We consider the same penalties $C$
as above: quadratic, linear and constant on large trades. 

\begin{figure}[h!]
\begin{center}
\includegraphics[width=185pt]{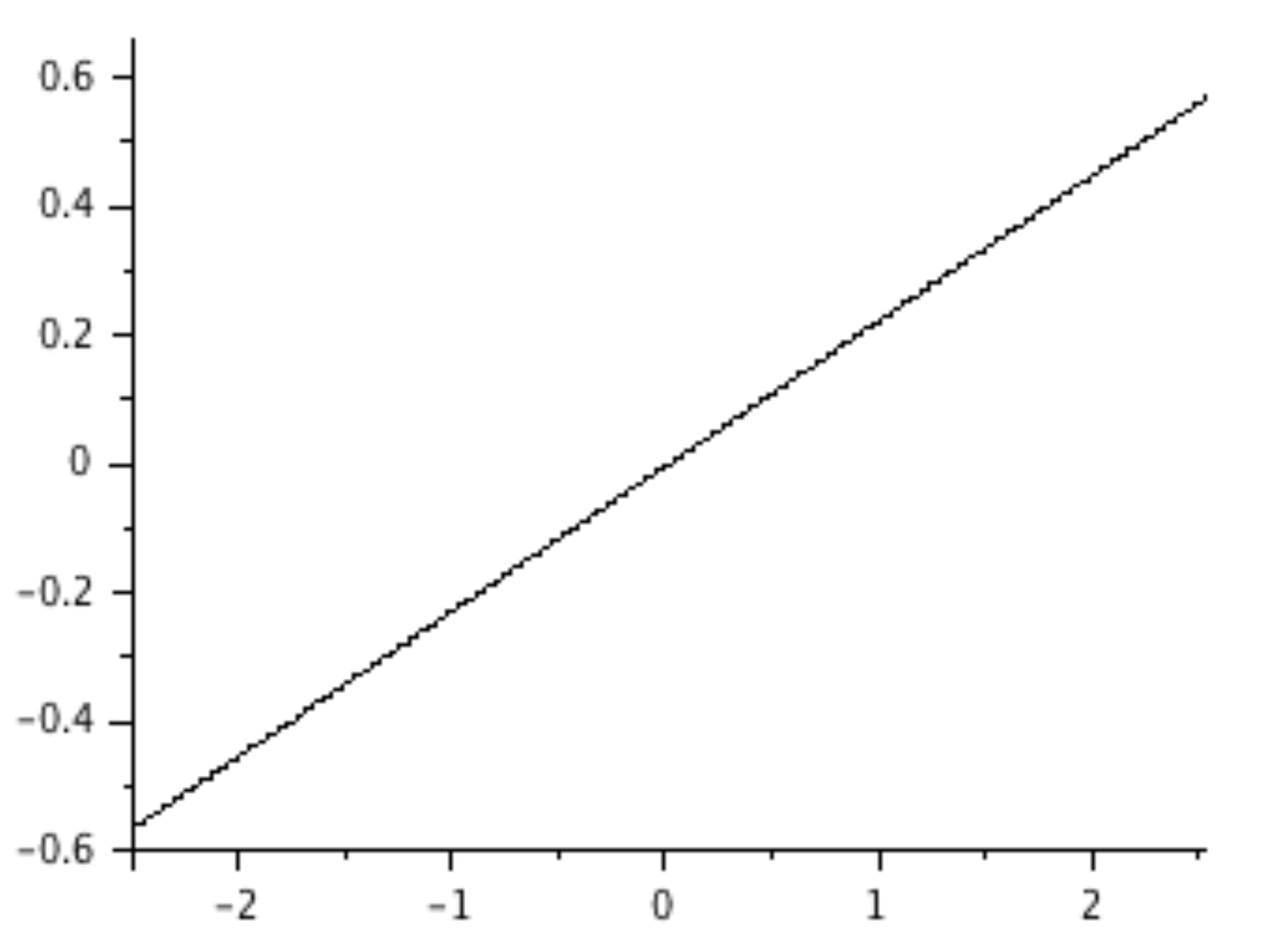} \hspace{50pt}\includegraphics[width=190pt, height=115pt]{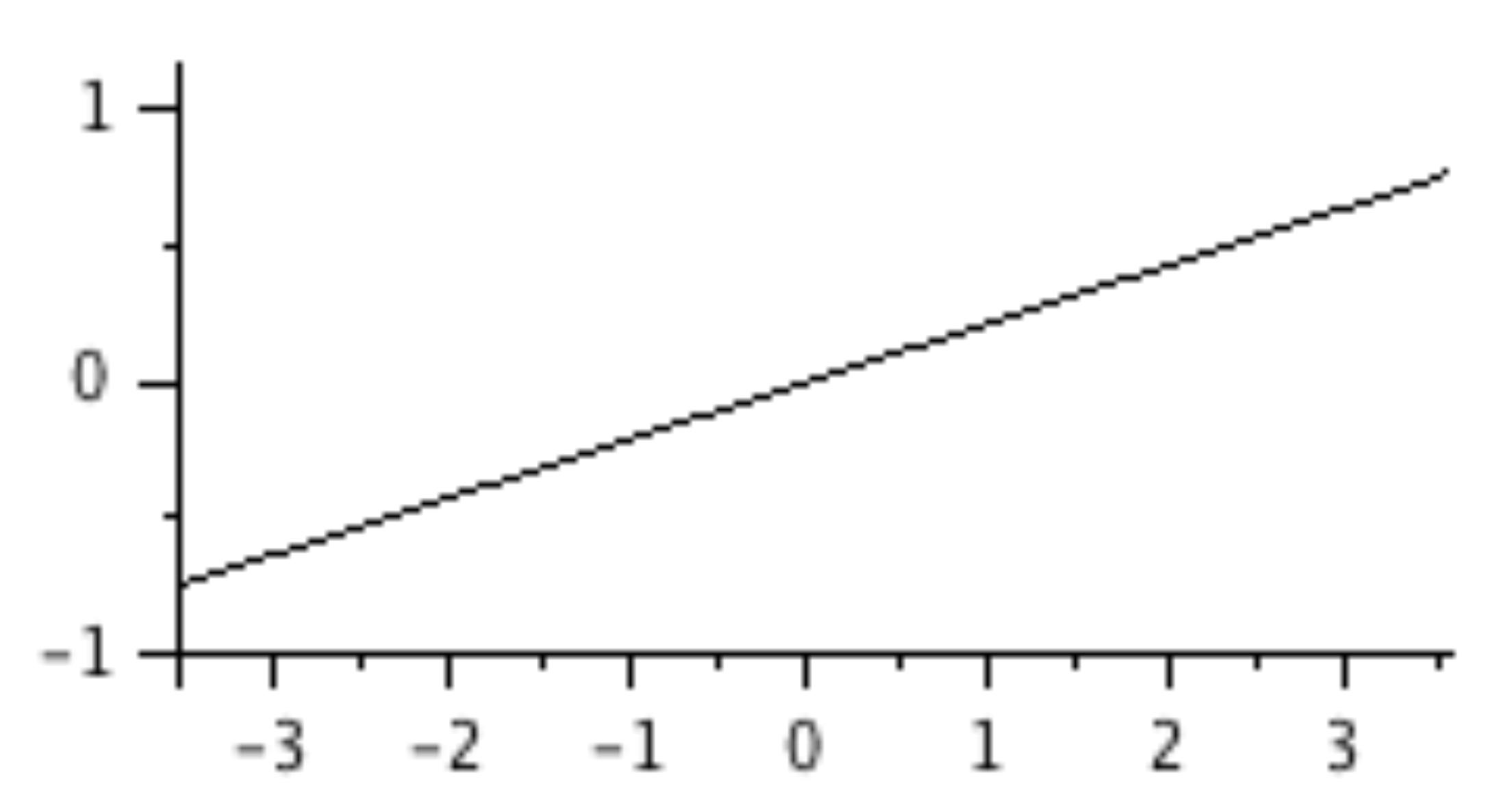}
\caption{IT demand and pricing under quadratic penalty, Gaussian case}
\label{fig:fig3-gauss}

\noindent $C(x)=\alpha x^2$, $\alpha=2$. Left panel: IT demand $X$. Right panel: price function $P$.
\end{center}

\end{figure}

\begin{figure}[h!]

\begin{center}
\includegraphics[width=185pt]{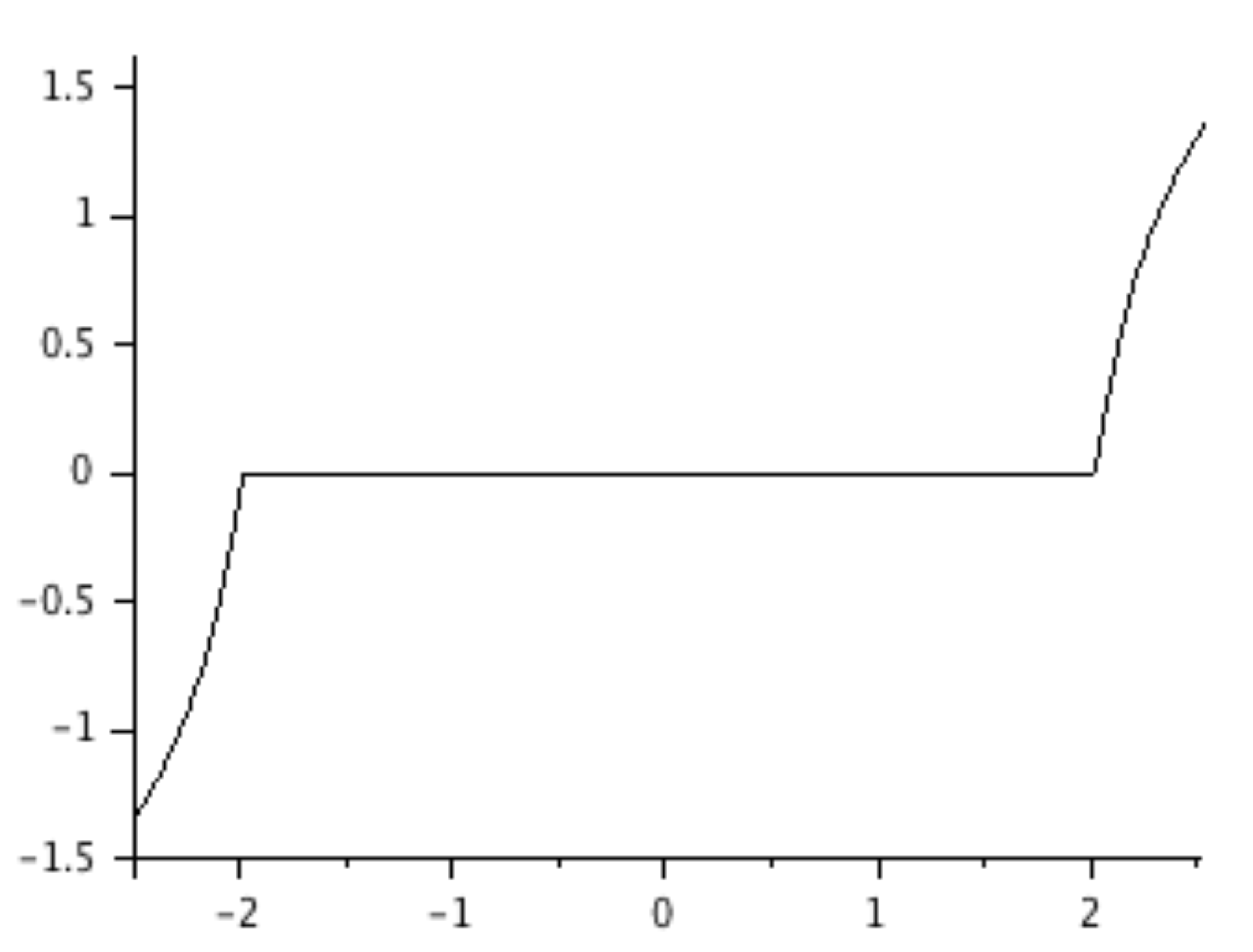} \hspace{50pt}\includegraphics[width=190pt, height=110pt]{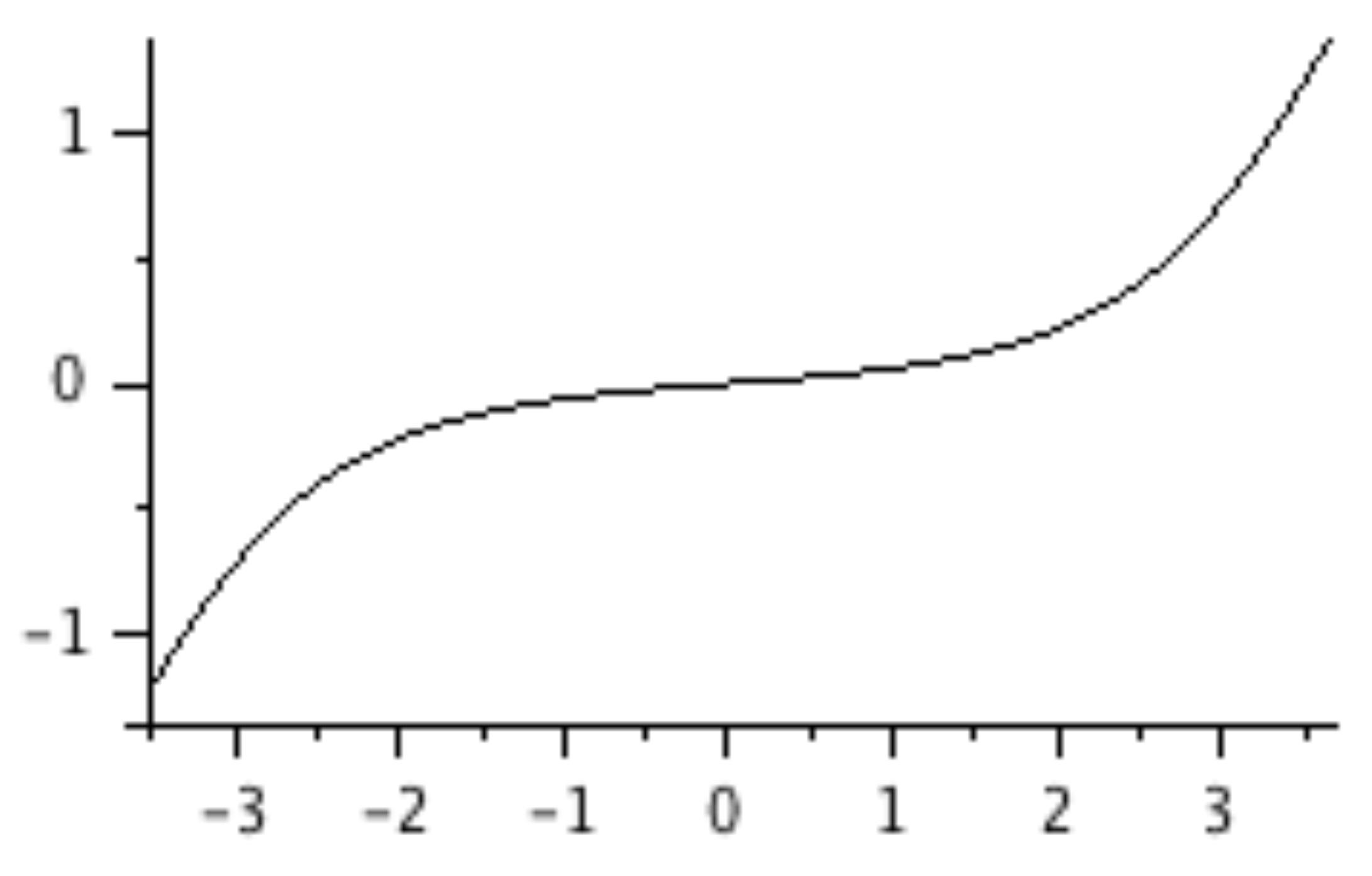}
\caption{IT demand and pricing under linear penalty, Gaussian case}
\label{fig:fig2-gauss}

\noindent $C(x)=\alpha \lvert x\lvert$, $\alpha=2$. Left panel: IT demand $X$. Right panel: price function $P$.
\end{center}
\end{figure}

\begin{figure}[h!]

\begin{center}
\includegraphics[width=185pt]{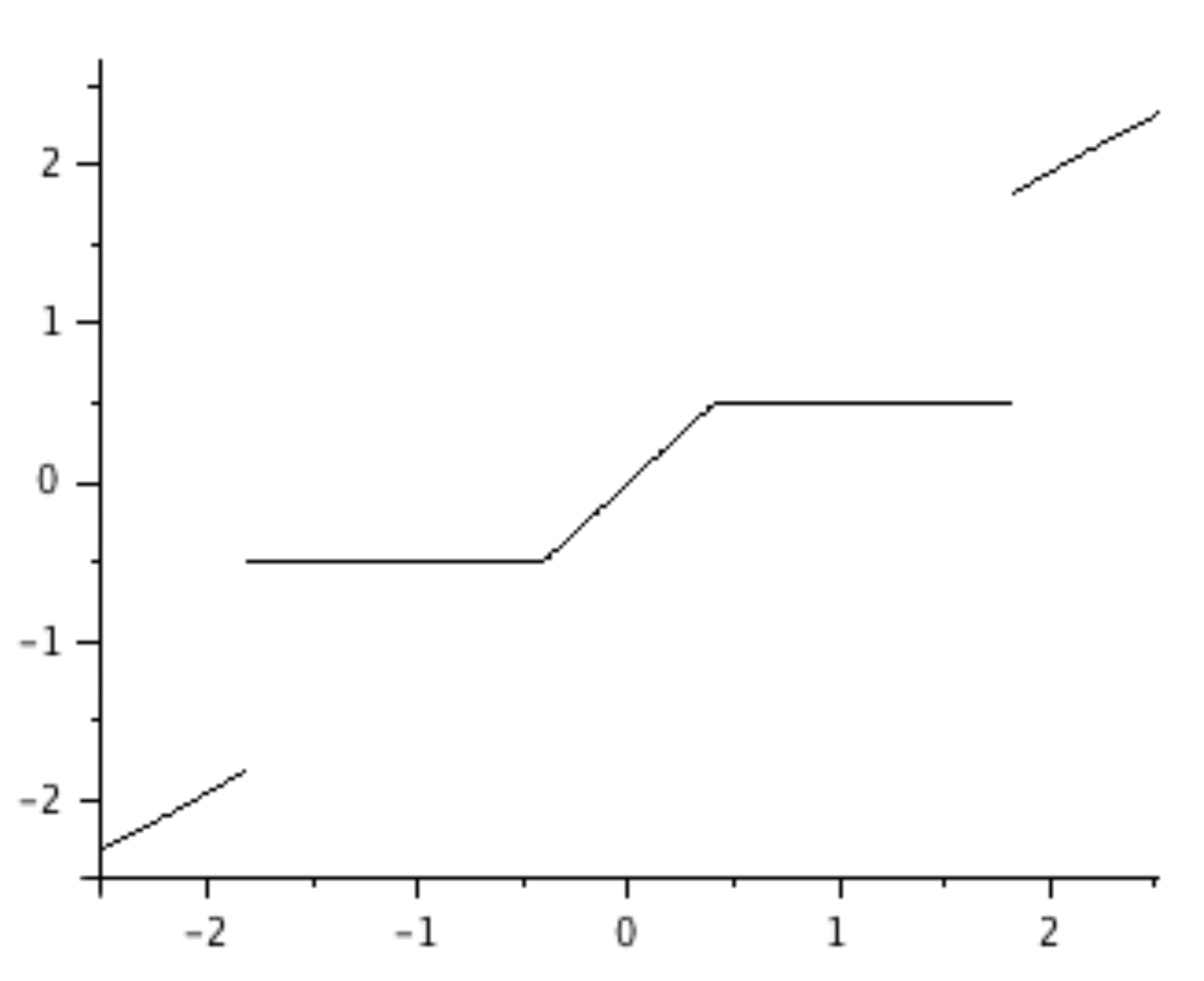} \hspace{50pt}\includegraphics[width=210pt, height=150pt]{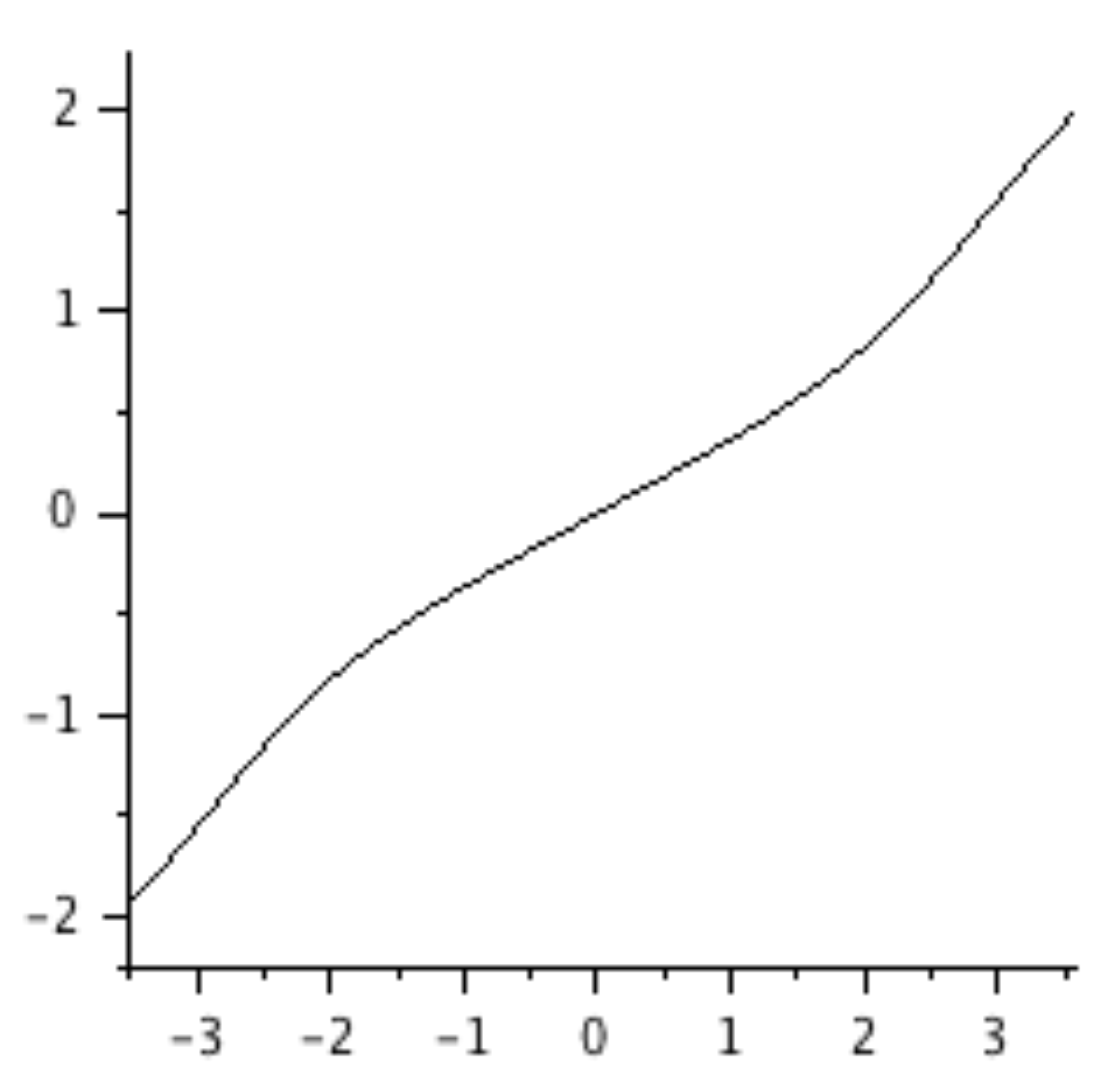}
\caption{IT demand and pricing under constant penalty, Gaussian case}
\label{fig:fig1-gauss}

\noindent $C(x)=K\mathbb{I}_{\lvert x\lvert>x_0}$, $K=1$, $x_0=0.5$. 

\noindent Left panel: IT demand $X$. Right panel: price function $P$.
\end{center}
\end{figure}

\subsection{Figure \ref{fig:fig1} under Gaussian noise}
\label{sec:robustnessfig1}
\begin{figure}[H]
\begin{center}
\includegraphics[width=0.75\textwidth]{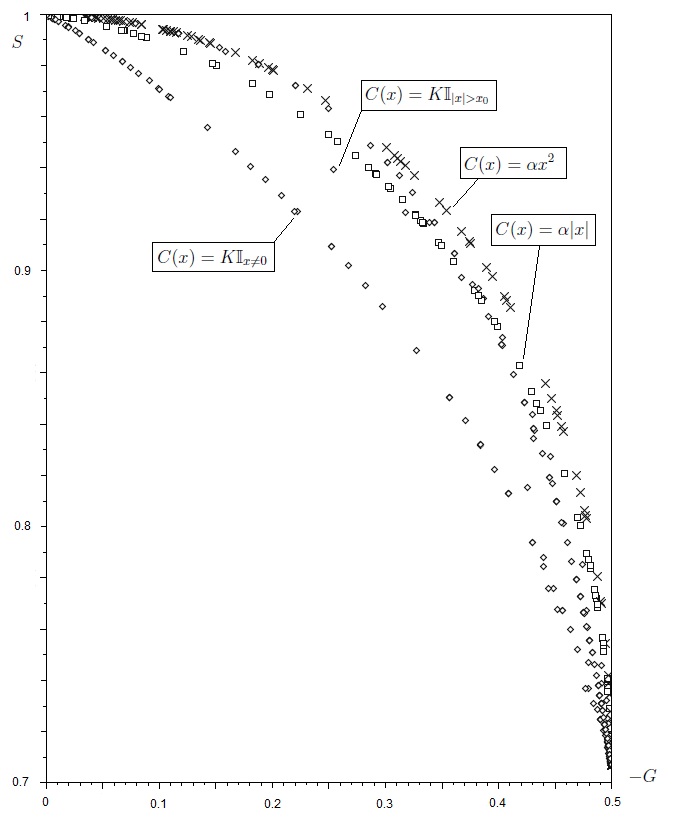}
\caption{Locus of $(S,-G)$ for different penalty functions - Gaussian noise.}
\label{fig:gaussianfrontier}
\end{center}
\end{figure}

\noindent We repeat the construction of Figure \ref{fig:fig1} by assuming Gaussian noise:
$u,v\sim N(0;1)$. We obtain Figure \ref{fig:gaussianfrontier}. The constant
costs upon nonzero trades $C(x)=K\mathbb{I}_{x\neq 0}$ are doing best
among the penalty functions considered. This is consistent with the results
in the uniform noise case. Other penalties are suboptimal, as before, and the locus of points $(S,-G)$ they
generate is very similar in shape.

\section{Discussion: \cite{BVH}}
\label{sec:discussionBVH}
\noindent Consider a static Kyle model where the NT trade with noise $u$ and the fundamental
has distribution $v$. We provide an informal discussion with two results that complement those
of \cite{BVH}. Result 1 is new. Result 2 provides an alternative proof for a particular case of their work.
Recall that the model is said to be with \emph{individual orders} when the market maker observes the
set $\{x;u\}$ where $x$ is the order of the insider trader, and to be with \emph{aggregate orders}
when she observes $x+u$ instead.

\

\noindent Let $v_0=\mathbb{E}[v]$, which we do not necessarily assume to be 0. We focus on increasing demand schedules $X$. We have the following\newline\newline\noindent\textbf{Result 1 (individual orders)} \textit{A mimicking equilibrium strategy $X$ must be affine in $v$.}\newline\newline\noindent$\bullet$ Since $q$ and $u$ are indistinguishable, the price function is given by
\begin{equation*}
P(\{q,u\})=\frac{1}{2}(X^{-1}(u)+X^{-1}(q)).
\end{equation*}
Therefore, the maximisation program of the IT is
\begin{equation*}
\max_q q\left(v-\mathbb{E}_u\left[\frac{1}{2}X^{-1}(u)\right]-\frac{1}{2}X^{-1}(q)\right).
\end{equation*}
Since $X(v)=u$ in distribution, $v=X^{-1}(u)$ in distribution and the program reduces to
\begin{equation*}
\max_q q\left(v-\frac{v_0}{2}-\frac{1}{2}X^{-1}(q)\right).
\end{equation*}
Since $X$ is an equilibrium strategy, the derivative of this expression evaluated at $q=X(v)$ must be zero:
\begin{eqnarray*}
0&=&v-\frac{v_0}{2}-\frac{X^{-1}(q)}{2}-\frac{q}{2}\frac{1}{X'(X^{-1}(q))}\\
&=&\frac{v-v_0}{2}-\frac{X(v)}{2X'(v)}.
\end{eqnarray*}
Therefore, $X$ must satisfy the ODE
\begin{equation*}
X'(v)(v-v_0)=X(v),
\end{equation*}
i.e. $X$ is affine (in fact linear in $v-v_0$).\newline\newline\noindent We already see that if it is impossible to mimick the noise in an affine manner, we can't have a mimicking equilibrium. If, however, this is possible, we automatically have an equilibrium:\newline\newline\noindent\textbf{Result 2 (individual orders and aggregate orders)} \textit{If there exists $X$ increasing and linear in $v-v_0$ such that $X(v)=u$ in distribution, and $P$ is the corresponding pricing function, $(X,P)$ is an equilibrium}.\newline\newline\noindent$\bullet$ In the case of individual orders, this result is immediate from the arguments above: indeed $X^{-1}$ is affine so the first order condition, which is satisfied for $q=X(v)$ indeed characterizes a global maximum.

\

\noindent Then, we note that the result extends to the case of aggregate orders. Indeed since $X(v)$ and $u$ are indistinguishable, by symmetry
\begin{eqnarray*}
\mathbb{E}[X(v)\lvert X(v)+u]&=&\mathbb{E}[u\lvert X(v)+u]\\
&=&\frac{1}{2}\mathbb{E}[X(v)+u\lvert X(v)+u]\\
&=&\frac{X(v)+u}{2}
\end{eqnarray*}
so, since $X$ is linear, $P(x+u)=\frac{X^{-1}(x)+X^{-1}(u)}{2}$ as before, therefore $X(v)$ is still an optimal demand.

\end{document}